\documentclass{ptephy_v1}

\usepackage{graphics} 

\usepackage{dcolumn}
\usepackage{bm}
\usepackage{amsfonts}
\usepackage{latexsym}
\usepackage{amssymb}
\usepackage{verbatim}
\usepackage{amsthm}
\usepackage{amsmath}
\usepackage{mathrsfs}
\usepackage{wrapft}

\newcommand{\FF}{{{\mathbb F}}}

\newcommand{\HF}{{{\mathbb H}}}

\newcommand{\RF}{{{\mathbb R}}}

\newcommand{\scrF}{{{\mathscr F}}}
\newcommand{\scrG}{{{\mathscr G}}}
\newcommand{\scrH}{{{\mathscr H}}}

\newcommand{\scrM}{{{\mathscr M}}}

\newcommand{\scrQ}{{{\mathscr Q}}}

\newcommand{\scrT}{{{\mathscr T}}}

\newcommand{\scrX}{{{\mathscr X}}}
\newcommand{\scrY}{{{\mathscr Y}}}

\newcommand{\FrakG}{{{\mathfrak G}}}
\newcommand{\FrakH}{{{\mathfrak H}}}

\newcommand{\FrakR}{{{\mathfrak R}}}

\newtheorem{theorem}{Theorem}[section]

\newtheorem{conjecture}{Conjecture}[section]
\newtheorem{proposal}{Proposal}[section]

\begin{document}


\numberwithin{equation}{section}

\title{
  Comparing a gauge-invariant formulation and a ``conventional
  complete gauge-fixing approach'' for $l=0,1$ mode perturbations on the
  Schwarzschild background spacetime
}
\author{
  Kouji Nakamura
  \footnote{E-mail address: dr.kouji.nakamura@gmail.com}
}
\address{
  Gravitational-Wave Science Project,
  National Astronomical Observatory of Japan,\\
  2-21-1, Osawa, Mitaka, Tokyo 181-8588, Japan
}
\date{\today}
\begin{abstract}
  Comparison of the gauge-invariant formulation for $l=0,1$-mode
  perturbations on the Schwarzschild background spacetime proposed in
  [K.~Nakamura, Class. Quantum Grav. {\bf 38} (2021), 145010.] and a
  ``conventional complete gauge-fixing approach'' in which we use the
  spherical harmonic functions $Y_{lm}$ as the scalar harmonics from
  the starting point is discussed.
  Although it is often said that ``gauge-invariant formulations in
  general-relativistic perturbations are equivalent to complete
  gauge-fixing approaches,'' as the result of this comparison, we conclude
  that the derived solutions through the proposed gauge-invariant
  formulation and those through a ``conventional complete gauge-fixing
  approach'' are different.
  It is pointed out that there is a case where the boundary conditions
  and initial conditions are restricted in a conventional complete
  gauge-fixing approach.
\end{abstract}

\maketitle


\section{Introduction}
\label{sec:introduction}


Through the ground-based gravitational-wave detectors~\cite{LIGO-home-page,Virgo-home-page,KAGRA-home-page,LIGO-INDIA-home-page},
many events of gravitational waves, mainly from black hole-black hole
coalescences, have now been detected.
We are now at the stage where there is no doubt about the existence of
gravitational waves due to their direct observations.
One of the future directions of gravitational-wave astronomy will be a
precise science through the statistics of many events.
Toward further development of gravitational-wave science, the projects
of future ground-based gravitational-wave
detectors~\cite{ET-home-page,CosmicExplorer-home-page} are also
progressing to achieve more sensitive detectors and some projects of
space gravitational-wave antenna are also
progressing~\cite{LISA-home-page,DECIGO-PTEP-2021,TianQin-PTEP-2021,Taiji-PTEP-2021}.
Although there are many targets of these detectors, the
Extreme-Mass-Ratio-Inspiral (EMRI), which is a source of gravitational
waves from the motion of a stellar mass object around a supermassive
black hole, is a promising target of the Laser Interferometer Space
Antenna~\cite{LISA-home-page}.
Since the mass ratio of this EMRI is very small, we can describe the
gravitational waves from EMRIs through black hole
perturbations~\cite{L.Barack-A.Pound-2019}.
Furthermore, the sophistication of higher-order black hole
perturbation theories is required to support gravitational-wave
physics as a precise science.
The motivation of our series of papers,
Refs.~\cite{K.Nakamura-2021-CQG,K.Nakamura-2021-LHEP,K.Nakamura-2021-PartI,K.Nakamura-2021-PartII,K.Nakamura-2021-PartIII}
and this paper, is in this theoretical sophistication of black hole
perturbation theories toward higher-order perturbations.


Although realistic black holes have their angular momentum and we have
to consider the perturbation theory of a Kerr black hole for direct
applications to the EMRI, we may say that further sophistications are
possible even in perturbation theories on the Schwarzschild background
spacetime.
From the pioneering works by Regge and
Wheeler~\cite{T.Regge-J.A.Wheeler-1957} and
Zerilli~\cite{F.Zerilli-1970-PRL,F.Zerilli-1970-PRD},
there have been many studies on the perturbations in the Schwarzschild
background
spacetime~\cite{H.Nakano-2019,V.Moncrief-1974a,V.Moncrief-1974b,C.T.Cunningham-R.H.Price-V.Moncrief-1978,Chandrasekhar-1983,Gerlach_Sengupta-1979a,Gerlach_Sengupta-1979b,Gerlach_Sengupta-1979c,Gerlach_Sengupta-1980,T.Nakamura-K.Oohara-Y.Kojima-1987,Gundlach-Martine-Garcia-2000,Gundlach-Martine-Garcia-2001,A.Nagar-L.Rezzolla-2005-2006,K.Martel-E.Poisson-2005}.
In these works, perturbations are decomposed through the spherical
harmonics $Y_{lm}$ because of the spherical symmetry of the background
spacetime, and $l=0,1$ modes should be separately treated.
These modes correspond to the monopole and dipole perturbations.
Due to these separate treatments, ``{\it gauge-invariant}'' treatments
for $l=0$ and $l=1$ modes were unclear.


Owing to this situation, in the previous
papers~\cite{K.Nakamura-2021-CQG,K.Nakamura-2021-LHEP,K.Nakamura-2021-PartI,K.Nakamura-2021-PartII,K.Nakamura-2021-PartIII},
we proposed the strategy of the gauge-invariant treatments of these
$l=0,1$ mode perturbations, which is declared as
Proposal~\ref{proposal:treatment-proposal-on-pert-on-spherical-BG}
in Sec.~\ref{sec:Brief_rev_gauge-inv_l=01modes}
of this paper below.
One of the important premises of our gauge-invariant perturbations is
the distinction between the first-kind gauge and the second-kind
gauge.
The first-kind gauge is essentially the choice of the coordinate
system on the single manifold, and we often use this first-kind gauge
when we predict or interpret the measurement results of experiments
and observations.
On the other hand, the second-kind gauge is the choice of the
point identifications between the points on the physical spacetime
$\scrM_{\epsilon}$ and the background spacetime $\scrM$.
This second-kind gauge has nothing to do with our physical spacetime
$\scrM_{\epsilon}$.
Although this difference is extensively explained in the Part I
paper~\cite{K.Nakamura-2021-PartI}, we also emphasize this difference
in Sec.~\ref{sec:Brief_rev_gauge-inv_l=01modes} of this paper.
The proposal in the Part I paper~\cite{K.Nakamura-2021-PartI} is a
part of our developments of the general formulation of a higher-order
gauge-invariant perturbation theory on a generic background spacetime
toward unambiguous sophisticated nonlinear general-relativistic
perturbation theories~\cite{K.Nakamura-2003,K.Nakamura-2005,K.Nakamura-2011,K.Nakamura-IJMPD-2012,K.Nakamura-2013,K.Nakamura-2014}.
Although we have been applied this general framework to cosmological
perturbations~\cite{K.Nakamura-2006,K.Nakamura-2007,K.Nakamura-2008,K.Nakamura-2009a,K.Nakamura-2009b,K.Nakamura-2010,A.J.Christopherson-K.A.Malik-D.R.Matravers-K.Nakamura-2011,K.Nakamura-2020},
we applied it to black hole perturbations in the series of papers,
i.e.,
Refs.~\cite{K.Nakamura-2021-CQG,K.Nakamura-2021-LHEP,K.Nakamura-2021-PartI,K.Nakamura-2021-PartII,K.Nakamura-2021-PartIII}
and this paper.
Even in cosmological perturbation theories, the same problem as the
above $l=0,1$-mode problem exists as gauge-invariant treatments of
homogeneous modes of perturbations.
In this sense, we can expect that the proposal in the previous
paper~\cite{K.Nakamura-2021-PartI} will be a clue to the same problem
in gauge-invariant perturbation theory on the generic background
spacetime.


In the Part I paper~\cite{K.Nakamura-2021-PartI}, we also derived the
linearized Einstein equations in a gauge-invariant manner following
Proposal~\ref{proposal:treatment-proposal-on-pert-on-spherical-BG}.
Perturbations on the spherically symmetric background spacetime are
classified into even- and odd-mode perturbations.
In the same paper~\cite{K.Nakamura-2021-PartI}, we also gave the strategy
to solve the odd-mode perturbations including $l=0,1$ modes.
Furthermore, we also derived the formal solutions for the $l=0,1$
odd-mode perturbations to the linearized Einstein equations following
Proposal~\ref{proposal:treatment-proposal-on-pert-on-spherical-BG}.
In the Part II paper~\cite{K.Nakamura-2021-PartII}, we gave the
strategy to solve the even-mode perturbations including $l=0,1$ modes,
and we also derive the formal solutions for the $l=0,1$ even-mode
perturbations following
Proposal~\ref{proposal:treatment-proposal-on-pert-on-spherical-BG}.
In the Part III paper~\cite{K.Nakamura-2021-PartIII}, we find the fact
that the derived solutions in the Part II
paper~\cite{K.Nakamura-2021-PartII} realize the linearized version of
two exact solutions: one is the Lema\^itre-Tolman-Bondi (LTB)
solution~\cite{L.Landau-E.Lifshitz-1962} and the other is the
non-rotating
C-metric~\cite{W.Kinnersley-M.Walker-1970,J.B.Griffiths-P.Krtous-J.Podolsky-2006}.
Due to this fact, we conclude that the solutions for even-mode
perturbations derived in the Part II
paper~\cite{K.Nakamura-2021-PartII} are physically reasonable.
This series of papers is the full paper version of our short
paper~\cite{K.Nakamura-2021-CQG}.
Furthermore, brief discussions on the extension to the higher-order
perturbations are given in the short
paper~\cite{K.Nakamura-2021-LHEP}.


On the other hand, it is well-known the fact that we cannot construct
gauge-invariant variables for $l=0,1$ modes in a similar manner
to $l\geq 2$ modes if we decompose the metric perturbations through
the spherical harmonics as the scalar harmonics from the starting
point.
For this reason, we usually use gauge-fixing approaches.
Furthermore, it is often said that ``gauge-invariant formulations in
general-relativistic perturbations are equivalent to complete
gauge-fixing approaches.''
In this paper, we check this statement through the comparison of our
proposed gauge-invariant formulation, in which we introduce singular
harmonics at once and regularize them after the derivation of the
mode-by-mode Einstein equation, and a ``conventional complete
gauge-fixing approach'', in which we use the spherical harmonic
functions $Y_{lm}$ as the scalar harmonics $S_{\delta}$ from the
starting point.
As the result of this comparison, we conclude that our gauge-invariant
formulation and the above ``conventional complete gauge-fixing
approach'' are different, though these two formulations lead similar
solutions for $l=0,1$-mode perturbations.
More specifically, there is a case where the boundary conditions and
initial conditions are restricted in a ``conventional complete
gauge-fixing approach'' where we use the decomposition of the metric
perturbation by the spherical harmonics $Y_{lm}$ from the starting point.


The organization of this paper is as follows.
In Sec.~\ref{sec:Brief_rev_gauge-inv_l=01modes}, we briefly review the
premise of our series of
papers~\cite{K.Nakamura-2021-CQG,K.Nakamura-2021-LHEP,K.Nakamura-2021-PartI,K.Nakamura-2021-PartII,K.Nakamura-2021-PartIII},
which are necessary for the ingredients of this paper.
We also emphasize the difference between the concepts of the first-
and the second-kind gauges and summarize the linearized Einstein
equations for $l=0,1$ modes and their formal solutions in
Sec.~\ref{sec:Brief_rev_gauge-inv_l=01modes}.
In Sec.~\ref{sec:Rule_of_comparison}, we specify the rule of our
comparison between our gauge-invariant formulation and a conventional
gauge-fixing approach and summarize the gauge transformation rules for
the metric perturbations of $l=0,1$ modes, because the above statement
 ``gauge-invariant formulations in
general-relativistic perturbations are equivalent to complete
gauge-fixing approaches'' includes some ambiguous.
In Sec.~\ref{sec:l=1-odd-ConventionalGaugeFixing}, we discuss the
linearized Einstein equations for $l=1$ odd-modes and their solutions
in the conventional gauge-fixing approach.
In Sec.~\ref{sec:l=1-even-ConventionalGaugeFixing}, we discuss the
linearized Einstein equations for $l=1$ even-modes and their solutions
in the conventional gauge-fixing approach.
In Sec.~\ref{sec:l=0ConventionalGaugeFixing}, we derive the solution
to the linearized Einstein equations for $l=0$ modes through the
complete gauge-fixing and discuss the comparison with linearized LTB
solution.
Sec.~\ref{sec:Summary_and_Discussions} is devoted to the summary of
this paper and discussions based on our results.


We use the notation used in the previous
papers~\cite{K.Nakamura-2021-CQG,K.Nakamura-2021-LHEP,K.Nakamura-2021-PartI,K.Nakamura-2021-PartII,K.Nakamura-2021-PartIII}
and the unit $G=c=1$, where $G$ is Newton's constant of gravitation
and $c$ is the velocity of light.


\section{Brief review of a gauge-invariant treatment of $l=0,1$ modes}
\label{sec:Brief_rev_gauge-inv_l=01modes}


In this section, we briefly review the premises of our series of
papers~\cite{K.Nakamura-2021-CQG,K.Nakamura-2021-LHEP,K.Nakamura-2021-PartI,K.Nakamura-2021-PartII,K.Nakamura-2021-PartIII}
which are necessary for the ingredients of this paper.
In Sec.~\ref{sec:general-framework-GI-perturbation-theroy}, we review
our framework of the gauge-invariant perturbation
theory~\cite{K.Nakamura-2003,K.Nakamura-2005}.
This is an important premise of the series of our
papers~\cite{K.Nakamura-2021-CQG,K.Nakamura-2021-LHEP,K.Nakamura-2021-PartI,K.Nakamura-2021-PartII,K.Nakamura-2021-PartIII}
and this paper.
In Sec.~\ref{sec:spherical_background_case}, we review the
gauge-invariant perturbation theory on spherically symmetric
spacetimes which includes our proposal in
Refs.~\cite{K.Nakamura-2021-CQG,K.Nakamura-2021-LHEP,K.Nakamura-2021-PartI,K.Nakamura-2021-PartII,K.Nakamura-2021-PartIII}.
In Sec.~\ref{sec:l=1-odd_Einstein_equations_solutions}, we summarize
the $l=1$ odd-mode linearized Einstein equations and their ``formal''
solutions.
In Sec.~\ref{sec:l=01-even_Einstein_equations_solutions}, we summarize
the $l=0,1$ even-mode linearized Einstein equations and their
``formal'' solutions.
The equations and their solutions in
Sec.~\ref{sec:l=1-odd_Einstein_equations_solutions} and
Sec.~\ref{sec:l=01-even_Einstein_equations_solutions} are derived
based on our proposal in Refs.~\cite{K.Nakamura-2021-CQG,K.Nakamura-2021-LHEP,K.Nakamura-2021-PartI,K.Nakamura-2021-PartII,K.Nakamura-2021-PartIII}.
These are necessary for the arguments within this paper.


\subsection{General framework of gauge-invariant perturbation theory}
\label{sec:general-framework-GI-perturbation-theroy}


In any perturbation theory, we always treat two spacetime manifolds.
One is the physical spacetime $(\scrM_{{\rm ph}},\bar{g}_{ab})$,
which is identified with our nature itself, and we want to describe
this spacetime $(\scrM_{{\rm ph}},\bar{g}_{ab})$ by perturbations.
The other is the background spacetime $(\scrM,g_{ab})$,
which is prepared as a reference by hand.
Note that these two spacetimes are distinct.
Furthermore, in any perturbation theory, we always write equations
for the perturbation of the variable $Q$ as follows:
\begin{equation}
  \label{eq:variable-symbolic-perturbation}
  Q(``p\mbox{''}) = Q_{0}(p) + \delta Q(p).
\end{equation}
Equation (\ref{eq:variable-symbolic-perturbation}) gives a
relation between variables on different manifolds.
Actually, $Q(``p\mbox{''})$ in
Eq.~(\ref{eq:variable-symbolic-perturbation}) is a variable on
the physical spacetime $\scrM_{\epsilon}=\scrM_{\rm ph}$, whereas
$Q_{0}(p)$ and $\delta Q(p)$ are variables on the background spacetime
$\scrM$.
Because we regard Eq.~(\ref{eq:variable-symbolic-perturbation}) as
a field equation, Eq.~(\ref{eq:variable-symbolic-perturbation})
includes an implicit assumption of the existence of a point
identification map $\scrX_{\epsilon}$ $:$
$\scrM\rightarrow\scrM_{\epsilon}$ $:$
$p\in\scrM\mapsto ``p\mbox{''}\in\scrM_{\epsilon}$.
This identification map is a {\it gauge choice} in
general-relativistic perturbation theories.
This is the notion of the {\it second-kind gauge} pointed out by
Sachs~\cite{R.K.Sachs-1964,J.M.Stewart-M.Walker-1974,J.M.Stewart-M.Walker-1990,J.M.Stewart-1990}.
Note that this second-kind gauge is a different notion from the
degree of freedom of the coordinate transformation on a single
manifold, which is called the {\it first-kind
  gauge}~\cite{K.Nakamura-2010,K.Nakamura-2020,K.Nakamura-2021-PartI}.
This distinction between the first- and the second-kind of gauges
extensively explained in the Part I paper~\cite{K.Nakamura-2021-PartI} and
is also important to understand the ingredients of this paper.


To compare the variable $Q$ on $\scrM_{\epsilon}$ with its
background value $Q_{0}$ on $\scrM$, we use the pull-back
$\scrX_{\epsilon}^{*}$ of the identification map
$\scrX_{\epsilon}$ $:$ $\scrM\rightarrow\scrM_{\epsilon}$ and
we evaluate the pulled-back variable $\scrX_{\epsilon}^{*}Q$ on the
background spacetime $\scrM$.
Furthermore, in perturbation theories, we expand the pull-back
operation $\scrX_{\epsilon}^{*}$ to the variable $Q$ with respect
to the infinitesimal parameter $\epsilon$ for the perturbation as
\begin{eqnarray}
  \scrX_{\epsilon}^{*}Q
  =
  Q_{0}
  + \epsilon {}^{(1)}_{\;\scrX}Q
  + O(\epsilon^{2})
  .
  \label{eq:perturbative-expansion-of-Q-def}
\end{eqnarray}
Equation~(\ref{eq:perturbative-expansion-of-Q-def}) are evaluated on
the background spacetime $\scrM$.
When we have two different gauge choices $\scrX_{\epsilon}$ and
$\scrY_{\epsilon}$, we can consider the {\it gauge transformation},
which is the change of the point-identification
$\scrX_{\epsilon}\rightarrow\scrY_{\epsilon}$.
This gauge transformation is given by the diffeomorphism
$\Phi_{\epsilon}$ $:=$
$\left(\scrX_{\epsilon}\right)^{-1}\circ\scrY_{\epsilon}$
$:$ $\scrM$ $\rightarrow$ $\scrM$.
Actually, the diffeomorphism $\Phi_{\epsilon}$ induces a pull-back from
the representation $\scrX_{\epsilon}^{*}\!Q_{\epsilon}$ to the
representation $\scrY_{\epsilon}^{*}\!Q_{\epsilon}$ as
$\scrY_{\epsilon}^{*}\!Q_{\epsilon}(q)=\Phi_{\epsilon}^{*}\scrX_{\epsilon}^{*}\!Q_{\epsilon}(q)$
at any point $q\in\scrM$.
From general arguments of the Taylor
expansion~\cite{M.Bruni-S.Matarrese-S.Mollerach-S.Sonego-1997,M.Bruni-S.Sonego-CQG1999,S.Sonego-M.Bruni-1998},
the pull-back $\Phi_{\epsilon}^{*}$ is expanded as
\begin{eqnarray}
  \scrY_{\epsilon}^{*}\!Q_{\epsilon}(q)
  &=&
  \scrX_{\epsilon}^{*}\!Q_{\epsilon}(q)
  + \epsilon {\pounds}_{\xi_{(1)}} \scrX_{\epsilon}^{*}\!Q_{\epsilon}(q)
  + O(\epsilon^{2}),
  \label{eq:Bruni-46-one}
\end{eqnarray}
where $\xi_{(1)}^{a}$ is the generator of $\Phi_{\epsilon}$.
From Eqs.~(\ref{eq:perturbative-expansion-of-Q-def}) and
(\ref{eq:Bruni-46-one}), the gauge transformation for the first-order
perturbation ${}^{(1)}Q$ is given by
\begin{eqnarray}
  \label{eq:Bruni-47-one}
  {}^{(1)}_{\;\scrY}\!Q(q) - {}^{(1)}_{\;\scrX}\!Q(q) &=&
  {\pounds}_{\xi_{(1)}}Q_{0}(q)
\end{eqnarray}
at any point $q\in\scrM$.
We also employ the {\it order by order gauge invariance} as a
concept of gauge invariance~\cite{K.Nakamura-2003,K.Nakamura-2005}.
We call the $k$th-order perturbation ${}^{(k)}_{\scrX}\!Q$ as
gauge invariant if and only if
\begin{eqnarray}
  \label{eq:orderbyorder-gauge-inv-def}
{}^{(k)}_{\;\scrX}\!Q(q) = {}^{(k)}_{\;\scrY}\!Q(q)
\end{eqnarray}
for any gauge choice $\scrX_{\epsilon}$ and $\scrY_{\epsilon}$ at any
point of $q\in\scrM$.


Based on the above setup, we proposed a formulation to construct
gauge-invariant variables of higher-order
perturbations~\cite{K.Nakamura-2003,K.Nakamura-2005}.
First, we expand the metric on the physical spacetime
$\scrM_{\epsilon}$, which was pulled back to the background
spacetime $\scrM$ through a gauge choice $\scrX_{\epsilon}$ as
\begin{eqnarray}
  \scrX^{*}_{\epsilon}\bar{g}_{ab}
  &=&
  g_{ab} + \epsilon {}_{\scrX}\!h_{ab}
  + O(\epsilon^{2}).
  \label{eq:metric-expansion}
\end{eqnarray}
Although the expression (\ref{eq:metric-expansion}) depends
entirely on the gauge choice $\scrX_{\epsilon}$, henceforth,
we do not explicitly express the index of the gauge choice
$\scrX_{\epsilon}$ in the expression if there is no
possibility of confusion.
The important premise of our formulation of higher-order
gauge-invariant perturbation theory was the following
conjecture~\cite{K.Nakamura-2003,K.Nakamura-2005} for the linear
metric perturbation $h_{ab}$:
\begin{conjecture}
  \label{conjecture:decomposition-conjecture}
  If the gauge transformation rule for a perturbative pulled-back
  tensor field $h_{ab}$ from the physical spacetime $\scrM_{\epsilon}$
  to the background spacetime $\scrM$ is given by ${}_{\scrY}\!h_{ab}$
  $-$ ${}_{\scrX}\!h_{ab}$ $=$ ${\pounds}_{\xi_{(1)}}g_{ab}$ with the
  background metric $g_{ab}$, there then exist a tensor field
  $\scrF_{ab}$ and a vector field $Y^{a}$ such that $h_{ab}$ is
  decomposed as $h_{ab}$ $=:$ $\scrF_{ab}$ $+$ ${\pounds}_{Y}g_{ab}$,
  where $\scrF_{ab}$ and $Y^{a}$ are transformed as
  ${}_{\scrY}\!\scrF_{ab}$ $-$ ${}_{\scrX}\!\scrF_{ab}$ $=$ $0$ and
  ${}_{\scrY}\!Y^{a}$ $-$ ${}_{\scrX}\!Y^{a}$ $=$ $\xi^{a}_{(1)}$
  under the gauge transformation, respectively.
\end{conjecture}
We call $\scrF_{ab}$ and $Y^{a}$ as the
{\it gauge-invariant} and {\it gauge-dependent} parts
of $h_{ab}$, respectively.


The proof of Conjecture~\ref{conjecture:decomposition-conjecture} is
highly
nontrivial~\cite{K.Nakamura-2011,K.Nakamura-IJMPD-2012,K.Nakamura-2013},
and it was found that the gauge-invariant variables are essentially
non-local.
Despite this non-triviality, once we accept
Conjecture~\ref{conjecture:decomposition-conjecture},
we can decompose the linear perturbation of an arbitrary tensor field
${}_{\scrX}^{(1)}\!Q$, whose gauge transformation is given by
Eq.~(\ref{eq:Bruni-47-one}), through the gauge-dependent part $Y_{a}$ of
the metric perturbation in
Conjecture~\ref{conjecture:decomposition-conjecture} as
\begin{eqnarray}
  \label{eq:arbitrary-Q-decomp}
  {}_{\scrX}^{(1)}\!Q = {}^{(1)}\!\scrQ + {\pounds}_{{}_{\scrX}\!Y}Q_{0},
\end{eqnarray}
where ${}^{(1)}\!\scrQ$ is the gauge-invariant part of the perturbation
${}_{\scrX}^{(1)}\!Q$.
As examples, the linearized Einstein tensor
${}_{\scrX}^{(1)}G_{a}^{\;\;b}$ and the linear perturbation of the
energy-momentum tensor ${}_{\scrX}^{(1)}T_{a}^{\;\;b}$ are also
decomposed as
\begin{eqnarray}
  \label{eq:Gab-Tab-decomp}
  {}_{\scrX}^{(1)}\!G_{a}^{\;\;b}
  =
  {}^{(1)}\!\scrG_{a}^{\;\;b}\left[\scrF\right] + {\pounds}_{{}_{\scrX}\!Y}G_{a}^{\;\;b}
  ,
  \quad
  {}_{\scrX}^{(1)}\!T_{a}^{\;\;b}
  =
  {}^{(1)}\!\scrT_{a}^{\;\;b} + {\pounds}_{{}_{\scrX}\!Y}T_{a}^{\;\;b}
  ,
\end{eqnarray}
where $G_{ab}$ and $T_{ab}$ are the background values of the Einstein
tensor and the energy-momentum tensor,
respectively~\cite{K.Nakamura-2005,K.Nakamura-2009a}.
Using the background Einstein equation
$G_{a}^{\;\;b}=8\pi T_{a}^{\;\;b}$, the linearized Einstein equation
${}_{\scrX}^{(1)}\!G_{ab}=8\pi{}_{\scrX}^{(1)}\!T_{ab}$ is
automatically given in the gauge-invariant form
\begin{eqnarray}
  \label{eq:einstein-equation-gauge-inv}
  {}^{(1)}\!\scrG_{a}^{\;\;b}\left[\scrF\right]
  =
  8 \pi
  {}^{(1)}\!\scrT_{a}^{\;\;b}\left[\scrF,\phi\right]
\end{eqnarray}
even if the background Einstein equation is nontrivial.
Here, ``$\phi$'' in Eq.~(\ref{eq:einstein-equation-gauge-inv})
symbolically represents the matter degree of freedom.


Finally, we comment on the coordinate
transformation induced by the gauge transformation
$\Phi_{\epsilon}$ of the
second-kind~\cite{K.Nakamura-2010,K.Nakamura-2021-CQG}.
To see this, we introduce the coordinate system
$\{O_{\alpha},\psi_{\alpha}\}$ on the background spacetime $\scrM$,
where $O_{\alpha}$ are open sets on the background spacetime and
$\psi_{\alpha}$ are diffeomorphisms from $O_{\alpha}$ to $\RF^{4}$
($4=\dim\scrM$).
The coordinate system $\{O_{\alpha},\psi_{\alpha}\}$ is the set of
collections of the pair of open sets $O_{\alpha}$ and diffeomorphism
$\psi_{\alpha}:O_{\alpha}\mapsto\RF^{4}$.
If we employ a gauge choice $\scrX_{\epsilon}$ of the second kind, we
have the correspondence of the physical spacetime
$\scrM_{\epsilon}$ and the background spacetime $\scrM$.
Together with the coordinate system $\psi_{\alpha}$ on $\scrM$,
this correspondence $\scrX_{\epsilon}$ between $\scrM_{\epsilon}$ and
$\scrM$ induces the coordinate system on $\scrM_{\epsilon}$.
Actually, $\scrX_{\epsilon}(O_{\alpha})$ for each $\alpha$ is an
open set of $\scrM_{\epsilon}$.
Then, $\psi_{\alpha}\circ\scrX_{\epsilon}^{-1}$ becomes a
diffeomorphism from an open set $\scrX_{\epsilon}(O_{\alpha})\subset
\scrM_{\epsilon}$ to $\RF^{4}$.
This diffeomorphism $\psi_{\alpha}\circ\scrX_{\epsilon}^{-1}$
induces a coordinate system of an open set on $\scrM_{\epsilon}$.
When we have two different gauge choices $\scrX_{\epsilon}$ and
$\scrY_{\epsilon}$ of the second kind,
$\psi_{\alpha}\circ\scrX_{\epsilon}^{-1}$ $:$
$\scrM_{\epsilon}$ $\mapsto$ $\RF^{4}$ ($\{x^{\mu}\}$) and
$\psi_{\alpha}\circ\scrY_{\epsilon}^{-1}$ $:$ $\scrM_{\epsilon}$
$\mapsto$ $\RF^{4}$ ($\{y^{\mu}\}$) become different coordinate systems
on $\scrM_{\epsilon}$.
We can also consider the coordinate transformation from the coordinate
system $\psi_{\alpha}\circ\scrX_{\epsilon}^{-1}$ to another
coordinate system $\psi_{\alpha}\circ\scrY_{\epsilon}^{-1}$.
Because the gauge transformation
$\scrX_{\epsilon}\rightarrow\scrY_{\epsilon}$ is induced by the
diffeomorphism $\Phi_{\epsilon}$ $:=$
$\left(\scrX_{\epsilon}\right)^{-1}\circ\scrY_{\epsilon}$, this
diffeomorphism $\Phi_{\epsilon}$ induces the coordinate transformation
as
\begin{eqnarray}
  \label{eq:induced-coordinate-trans}
  y^{\mu}(q) := x^{\mu}(p) = \left((\Phi_{\epsilon}^{-1})^{*}x^{\mu}\right)(q)
\end{eqnarray}
in the passive point of
view~\cite{K.Nakamura-2003,M.Bruni-S.Matarrese-S.Mollerach-S.Sonego-1997},
where $p$, $q$ $\in$ $\scrM$ are identified to the same point
$``p\mbox{"}$ $\in$ $\scrM_{\epsilon}$ by the gauge choices
$\scrX_{\epsilon}$ and $\scrY_{\epsilon}$, respectively.
If we represent this coordinate transformation in terms of the Taylor
expansion (\ref{eq:Bruni-46-one}), we have the coordinate transformation
\begin{eqnarray}
  \label{eq:infinitesimal-coordinate-trans-explicit}
  y^{\mu}(q) = x^{\mu}(q) - \epsilon \xi^{\mu}_{(1)}(q) + O(\epsilon^{2}).
\end{eqnarray}
We should emphasize that the coordinate transformation
(\ref{eq:infinitesimal-coordinate-trans-explicit}) is not the starting
point of the gauge transformation but a result of the above framework.


On the other hand, we may consider the point replacement by the
one-parameter diffeomorphism $s=\Psi_{\lambda}(r)$, where
$s,r\in\scrM_{\epsilon}$ and $\lambda$ is an infinitesimal parameter
satisfying $\Psi_{\lambda=0}(r)=r$.
The pull-back $\Psi_{\lambda}^{*}$ of any tensor field $Q$ on
$\scrM_{\epsilon}$ is given by
\begin{eqnarray}
  Q(s)
  &=&
      Q(\Psi_{\lambda}(r)) = (\Psi_{\lambda}^{*})Q(r)
      \nonumber\\
  &=&
      Q(r) + \lambda \left.{\pounds}_{\zeta}Q\right|_{\lambda=0}(r)
      + O(\lambda^{2})
      ,
  \label{eq:Taylor-expansion-formula}
\end{eqnarray}
where $\zeta^{a}$ is the generator of the pull-back
$\Psi_{\lambda}^{*}$.
Equation (\ref{eq:Taylor-expansion-formula}) is just the formula of
the Taylor expansion on the manifold $\scrM_{\epsilon}$ without using
any coordinate system.
At the same time, Equation (\ref{eq:Taylor-expansion-formula}) is the
definitions of the Lie derivative ${\pounds}_{\zeta}$ and its generator
$\zeta^{a}$.
In many literatures, this formula is derived from the coordinate
transformation
\begin{eqnarray}
  \label{eq:infinitesimal-coordinate-trans-explicit-active}
  y^{\mu}(s) := x^{\mu}(r) + \lambda \zeta^{\mu}(r) + O(\lambda^{2})
\end{eqnarray}
as explained in the Part I paper~\cite{K.Nakamura-2021-PartI}.
The formula (\ref{eq:Taylor-expansion-formula}) of the Taylor
expansion defined without any coordinate system and the second
term in the right-hand side of Eq.~(\ref{eq:Taylor-expansion-formula})
represents the actual difference of the tensor fields $Q(s)$ and
$Q(r)$ in the different point $s\neq r$ and its physical meaning.


If we consider a formal metric decomposition
$\bar{g}_{ab}={}^{(0)}\!g_{ab}+\lambda h_{ab}+O(\lambda^{2})$ within
$\scrM_{\epsilon}$ as an example of the tensor field $Q$ in
Eq.~(\ref{eq:Taylor-expansion-formula}), the formula
(\ref{eq:Taylor-expansion-formula}) is given by
\begin{eqnarray}
  \label{eq:Taylor-expansion-of-formal-decomposition}
  {}^{(0)}g_{ab}(s) + \lambda h_{ab}(s)
  =
  {}^{(0)}g_{ab}(r) + \lambda \left(
  h_{ab}(r) + \left.{\pounds}_{\zeta}{}^{(0)}g_{ab}\right|_{r}
  \right)
  + O(\lambda^{2}).
\end{eqnarray}
Since the formula of the Taylor expansion
(\ref{eq:Taylor-expansion-formula}) is derived from the infinitesimal
coordinate transformation rule
(\ref{eq:infinitesimal-coordinate-trans-explicit-active}) in many
literatures, the term $\left.{\pounds}_{\zeta}{}^{(0)}g_{ab}\right|_{r}$
in the right-hand side of
Eq.~(\ref{eq:Taylor-expansion-of-formal-decomposition}) is often
called ``the degree of freedom of the coordinate transformation'' and
``unphysical degree of freedom''.
However, since the formula
(\ref{eq:Taylor-expansion-of-formal-decomposition}) is just the
re-expression of the Taylor expansion
(\ref{eq:Taylor-expansion-formula}) defined without any coordinate
system and the term $\left.{\pounds}_{\zeta}{}^{(0)}g_{ab}\right|_{r}$
in Eq.~(\ref{eq:Taylor-expansion-of-formal-decomposition}) should have
its physical meaning.
If we insist that the term
$\left.{\pounds}_{\zeta}{}^{(0)}g_{ab}\right|_{r}$ which appears due
to the point replacement on the single manifold is ``unphysical'',
this directly leads the statement that ``the famous arguments of the
Killing vector and the symmetry of the spacetime are physically
meaningless.''
For this reason, we have to emphasize that we {\it cannot} regard the
second term of the right-hand side of
Eq.~(\ref{eq:Taylor-expansion-formula}) as an ``unphysical degree of
freedom.''


In our series of
papers~\cite{K.Nakamura-2021-CQG,K.Nakamura-2021-LHEP,K.Nakamura-2021-PartI,K.Nakamura-2021-PartII,K.Nakamura-2021-PartIII}
and in this paper, we classify the term
$\left.{\pounds}_{\zeta}{}^{(0)}g_{ab}\right|_{r}$ in
Eq.~(\ref{eq:Taylor-expansion-of-formal-decomposition}) as one of
{\it gauge-degree of freedom of the first kind}, since this term can
be eliminate the infinitesimal coordinate transformation
(\ref{eq:infinitesimal-coordinate-trans-explicit-active}) which is the
coordinate transformation within the single manifold $\scrM_{\epsilon}$.
Furthermore, Equation
(\ref{eq:Taylor-expansion-of-formal-decomposition}) does not mean
$h_{ab}(s)$ $=$ $h_{ab}(r)$ $+$
$\left.{\pounds}_{\zeta}{}^{(0)}g_{ab}\right|_{r}$ but it just means
${}^{(0)}g_{ab}(s)$ $=$ ${}^{(0)}g_{ab}(r)$ $+$
$\lambda \left.{\pounds}_{\zeta}{}^{(0)}g_{ab}\right|_{r}$
$+$ $O(\lambda^{2})$.
Moreover, we also have to emphasize that the infinitesimal
coordinate transformation
(\ref{eq:infinitesimal-coordinate-trans-explicit-active}) is
essentially different from the infinitesimal coordinate transformation
(\ref{eq:infinitesimal-coordinate-trans-explicit}).
Actually, the infinitesimal coordinate transformation
(\ref{eq:infinitesimal-coordinate-trans-explicit-active}) is the
replacement of the point within the single manifold
$\scrM_{\epsilon}$.
On the other hand, the coordinate transformation
(\ref{eq:infinitesimal-coordinate-trans-explicit}) is just the change
of the coordinate label at the same point in the background spacetime
$\scrM$.


\subsection{Linear perturbations on spherically symmetric background spacetime}
\label{sec:spherical_background_case}


Here, we consider the 2+2 formulation of the perturbation of a
spherically symmetric background spacetime, which originally proposed
by Gerlach and Sengupta~\cite{Gerlach_Sengupta-1979a,Gerlach_Sengupta-1979b,Gerlach_Sengupta-1979c,Gerlach_Sengupta-1980}.
Spherically symmetric spacetimes are characterized by the direct
product $\scrM=\scrM_{1}\times S^{2}$ and their metric is
\begin{eqnarray}
  \label{eq:background-metric-2+2}
  g_{ab}
  &=&
  y_{ab} + r^{2}\gamma_{ab}
  , \\
  y_{ab} &=& y_{AB} (dx^{A})_{a}(dx^{B})_{b}, \quad
             \gamma_{ab} = \gamma_{pq} (dx^{p})_{a} (dx^{q})_{b},
\end{eqnarray}
where $x^{A} = (t,r)$, $x^{p}=(\theta,\phi)$, and $\gamma_{pq}$ is the
metric on the unit sphere.
In the Schwarzschild spacetime, the metric
(\ref{eq:background-metric-2+2}) is given by
\begin{eqnarray}
  \label{eq:background-metric-2+2-y-comp-Schwarzschild}
  y_{ab}
  &=&
      - f (dt)_{a}(dt)_{b}
      +
      f^{-1} (dr)_{a}(dr)_{b}
      ,
      \quad
      f = 1 - \frac{2M}{r}
  ,\\
  \label{eq:background-metric-2+2-gamma-comp-Schwarzschild}
  \gamma_{ab}
  &=&
      (d\theta)_{a}(d\theta)_{b}
      +
      \sin^{2}\theta(d\phi)_{a}(d\phi)_{b}
      =
      \theta_{a}\theta_{b} + \phi_{a}\phi_{b}
      ,
  \\
  \label{eq:S2-unit-basis-def}
  \theta_{a}
  &=&
      (d\theta)_{a}, \quad
      \phi_{a}
      =
      \sin\theta (d\phi)_{a}
      .
\end{eqnarray}


On this background spacetime $(\scrM,g_{ab})$, the components of
the metric perturbation is given by
\begin{eqnarray}
  \label{eq:metric-perturbation-components}
  h_{ab}
  =
  h_{AB} (dx^{A})_{a}(dx^{B})_{b}
  +
  2 h_{Ap} (dx^{A})_{(a}(dx^{p})_{b)}
  +
  h_{pq} (dx^{p})_{a}(dx^{q})_{b}
  .
\end{eqnarray}
Here, we note that the components $h_{AB}$, $h_{Ap}$, and
$h_{pq}$ are regarded as components of scalar, vector, and
tensor on $S^{2}$, respectively.
In the Part I paper~\cite{K.Nakamura-2021-PartI}, we showed the
linear independence of the set of harmonic functions
\begin{eqnarray}
  \label{eq:harmonic-fucntions-set}
  \left\{
  S_{\delta},
  \hat{D}_{p}S_{\delta},
  \epsilon_{pq}\hat{D}^{q}S_{\delta},
  \frac{1}{2}\gamma_{pq}S_{\delta},
  \left(\hat{D}_{p}\hat{D}_{q}-\frac{1}{2}\gamma_{pq}\right)S_{\delta},
  2\epsilon_{r(p}\hat{D}_{q)}\hat{D}^{r}S_{\delta}
  \right\}
  ,
\end{eqnarray}
where $\hat{D}_{p}$ is the covariant derivative associated with
the metric $\gamma_{pq}$ on $S^{2}$,
$\hat{D}^{p}=\gamma^{pq}\hat{D}_{q}$,
$\epsilon_{pq}=\epsilon_{[pq]}=2\theta_{[p}\phi_{q]}$ is the totally
antisymmetric tensor on $S^{2}$.
In the set of harmonic function (\ref{eq:harmonic-fucntions-set}), the
scalar harmonic function $S_{\delta}$ is given by
\begin{eqnarray}
  \label{eq:harmonics-extended-choice-sum}
  S_{\delta}
  =
  \left\{
  \begin{array}{ccccc}
    Y_{lm} & \quad & \mbox{for} & \quad & l\geq 2; \\
    k_{(\hat{\Delta}+2)m} & \quad & \mbox{for} & \quad & l=1; \\
    k_{(\hat{\Delta})} & \quad & \mbox{for} & \quad & l=0.
  \end{array}
  \right.
\end{eqnarray}
Here, functions $k_{(\hat{\Delta})}$ and $k_{(\hat{\Delta}+2)m}$ are
the kernel modes of the derivative operator $\hat{\Delta}$ and
$[\hat{\Delta}+2]$, respectively, and we employ the explicit form of
these functions as
\begin{eqnarray}
  k_{(\hat{\Delta})}
  &=&
      1 + \delta \ln\left(\frac{1-\cos\theta}{1+\cos\theta}\right)^{1/2},
      \quad \delta \in\RF
      ,
      \label{eq:l=0-general-mode-func-specific}
  \\
  k_{(\hat{\Delta}+2,m=0)}
  &=&
      \cos\theta
      +
      \delta \left(\frac{1}{2} \cos\theta \ln\frac{1+\cos\theta}{1-\cos\theta} -1\right)
      ,
     \quad \delta \in \RF
     ,
     \label{eq:l=1-m=0-mode-func-explicit}
     \\
  k_{(\hat{\Delta}+2,m=\pm1)}
  &=&
      \left[
      \sin\theta
      +
      \delta \left(
      + \frac{1}{2} \sin\theta \ln\frac{1+\cos\theta}{1-\cos\theta}
      + \cot\theta
      \right)
      \right]
      e^{\pm i\phi}
      .
      \label{eq:l=1-m=pm1-mode-func-explicit}
\end{eqnarray}


Then, we consider the mode decomposition of the components
$\{h_{AB},h_{Ap},h_{pq}\}$ as follows:
\begin{eqnarray}
  \label{eq:hAB-fourier}
  \!\!\!\!\!\!
  h_{AB}
  \!\!\!\!\!&=&\!\!\!\!\!
      \sum_{l,m} \tilde{h}_{AB} S_{\delta}
      ,
  \\
  \label{eq:hAp-fourier}
  \!\!\!\!\!\!
  h_{Ap}
  \!\!\!\!\!&=&\!\!\!\!\!
      r \sum_{l,m} \left[
      \tilde{h}_{(e1)A} \hat{D}_{p}S_{\delta}
      +
      \tilde{h}_{(o1)A} \epsilon_{pq} \hat{D}^{q}S_{\delta}
      \right]
      ,
  \\
  \label{eq:hpq-fourier}
  \!\!\!\!\!\!
  h_{pq}
  \!\!\!\!\!&=&\!\!\!\!\!
      r^{2} \sum_{l,m} \left[
      \frac{1}{2} \gamma_{pq} \tilde{h}_{(e0)} S_{\delta}
      +
      \tilde{h}_{(e2)} \left(
      \hat{D}_{p}\hat{D}_{q} - \frac{1}{2} \gamma_{pq} \hat{D}^{r}\hat{D}_{r}
      \right) S_{\delta}
      +
      2 \tilde{h}_{(o2)} \epsilon_{r(p} \hat{D}_{q)}\hat{D}^{r} S_{\delta}
      \right]
      .
\end{eqnarray}
Since the linear independence of each element of the set of harmonic
function (\ref{eq:harmonic-fucntions-set}) is guaranteed, the
one-to-one correspondence between the components $\{h_{AB},$ $h_{Ap},$
$h_{pq}\}$ and the mode coefficients $\{\tilde{h}_{AB},$
$\tilde{h}_{(e1)A},$ $\tilde{h}_{(o1)A},$ $\tilde{h}_{(e0)},$
$\tilde{h}_{(e2)},$ $\tilde{h}_{(o2)}\}$ with the decomposition
formulae (\ref{eq:hAB-fourier})-(\ref{eq:hpq-fourier}) is guaranteed
including $l=0,1$ mode if $\delta\neq 0$.
Then, the mode-by-mode analysis including $l=0,1$ is possible when
$\delta\neq 0$.
However, the mode functions
(\ref{eq:l=0-general-mode-func-specific})--(\ref{eq:l=1-m=pm1-mode-func-explicit})
are singular if $\delta\neq 0$.
When $\delta=0$, we have $k_{(\hat{\Delta})}\propto Y_{00}$ and
$k_{(\hat{\Delta}+2)m}\propto Y_{1m}$, but the linear dependence of
the set of harmonics (\ref{eq:harmonic-fucntions-set}) is lost in this
case.
Because of this situation, we proposed the following strategy:
\begin{proposal}
  \label{proposal:treatment-proposal-on-pert-on-spherical-BG}
  We decompose the metric perturbation $h_{ab}$ on the background
  spacetime with the metric
  (\ref{eq:background-metric-2+2})--(\ref{eq:background-metric-2+2-gamma-comp-Schwarzschild})
  through Eqs.~(\ref{eq:hAB-fourier})--(\ref{eq:hpq-fourier}) with the
  harmonic function $S_{\delta}$ given by
  Eq.~(\ref{eq:harmonics-extended-choice-sum}).
  Then, Eqs.~(\ref{eq:hAB-fourier})--(\ref{eq:hpq-fourier}) become
  invertible including $l=0,1$ modes.
  After deriving the mode-by-mode field equations such as linearized
  Einstein equations by using the harmonic functions $S_{\delta}$, we
  choose $\delta=0$ as the regular boundary condition for solutions
  when we solve these field equations.
\end{proposal}


As shown in the Part I paper~\cite{K.Nakamura-2021-PartI}, once we accept
Proposal~\ref{proposal:treatment-proposal-on-pert-on-spherical-BG},
the Conjecture~\ref{conjecture:decomposition-conjecture} becomes the
following statement:
\begin{theorem}
  \label{theorem:decomposition-theorem-with-spherical-symmetry}
  If the gauge-transformation rule for a perturbative pulled-back
  tensor field $h_{ab}$ to the background spacetime $\scrM$ is
  given by ${}_{\scrY}\!h_{ab}$ $-$ ${}_{\scrX}\!h_{ab}$ $=$
  ${\pounds}_{\xi_{(1)}}g_{ab}$ with the background metric $g_{ab}$
  with spherically symmetry, there then exist a tensor field
  $\scrF_{ab}$ and a vector  field $Y^{a}$ such that $h_{ab}$ is
  decomposed as $h_{ab}$ $=:$ $\scrF_{ab}$ $+$
  ${\pounds}_{Y}g_{ab}$, where $\scrF_{ab}$ and $Y^{a}$ are
  transformed into ${}_{\scrY}\!\scrF_{ab}$ $-$
  ${}_{\scrX}\!\scrF_{ab}$ $=$ $0$ and ${}_{\scrY}\!Y^{a}$
  $-$ ${}_{\scrX}\!Y^{a}$ $=$ $\xi^{a}_{(1)}$ under the gauge
  transformation, respectively.
\end{theorem}
Actually, the gauge-dependent variable $Y_{a}$ is given by
\begin{eqnarray}
  Y_{a}
  &:=&
       \sum_{l,m} \tilde{Y}_{A} S_{\delta} (dx^{A})_{a}
       +
       \sum_{l,m} \left(
       \tilde{Y}_{(e1)} \hat{D}_{p}S_{\delta}
       +
       \tilde{Y}_{(o1)} \epsilon_{pq}\hat{D}^{q}S_{\delta}
       \right)
       (dx^{p})_{a}
       ,
  \label{eq:2+2-Ya-def}
\end{eqnarray}
where
\begin{eqnarray}
  \tilde{Y}_{A}
  &:=&
       r \tilde{h}_{(e1)A}
       - \frac{r^{2}}{2} \bar{D}_{A}\tilde{h}_{(e2)}
       \label{eq:2+2-gauge-trans-tildeYA-def-sum}
       ,
  \\
  \label{eq:tildeYe-def}
  \tilde{Y}_{(e1)}
  &:=&
       \frac{r^{2}}{2} \tilde{h}_{(e2)}
       ,
  \\
  \label{eq:tildeYo-def}
  \tilde{Y}_{(o1)}
  &:=&
       - r^{2} \tilde{h}_{(o2)}
       .
\end{eqnarray}
Furthermore, including $l=0,1$ modes, the components of the
gauge-invariant part $\scrF_{ab}$ of the metric perturbation
$h_{ab}$ is given by
\begin{eqnarray}
  \label{eq:2+2-gauge-invariant-variables-calFAB}
  \scrF_{AB}
  &=&
      \sum_{l,m} \tilde{F}_{AB} S_{\delta}
      ,
  \\
  \label{eq:2+2-gauge-invariant-variables-calFAp}
  \scrF_{Ap}
  &=&
      r \sum_{l,m} \tilde{F}_{A} \epsilon_{pq}
      \hat{D}^{q}S_{\delta}, \quad
      \hat{D}^{p}\scrF_{Ap} = 0
      ,
  \\
  \label{eq:2+2-gauge-invariant-variables-calFpq}
  \scrF_{pq}
  &=&
      \frac{1}{2} \gamma_{pq} r^{2} \sum_{l,m} \tilde{F} S_{\delta}
      ,
\end{eqnarray}
where $\tilde{F}_{AB}$, $\tilde{F}_{A}$, and $\tilde{F}$ are given by
\begin{eqnarray}
  \label{eq:gauge-inv-tildeFAB-def-sum}
  \tilde{F}_{AB}
  &:=&
       \tilde{h}_{AB}
       - 2 \bar{D}_{(A}\tilde{Y}_{B)}
       ,
  \\
  \label{eq:2+2-gauge-inv-def-tildeFA-sum}
  \tilde{F}_{A}
  &:=&
       \tilde{h}_{(o1)A}
       + r \bar{D}_{A}\tilde{h}_{(o2)}
       ,
  \\
  \label{eq:2+2-gauge-inv-tildeF-def-sum}
  \tilde{F}
  &:=&
       \tilde{h}_{(e0)}
       - \frac{4}{r} \tilde{Y}_{A} \bar{D}^{A}r
       + \tilde{h}_{(e2)} l(l+1)
  .
\end{eqnarray}
Thus, we have constructed gauge-invariant metric perturbations on
the Schwarzschild background spacetime including $l=0,1$ modes.


Furthermore, from
Eqs.~(\ref{eq:2+2-gauge-invariant-variables-calFAB})--(\ref{eq:2+2-gauge-inv-tildeF-def-sum})
for the gauge-invariant variables $\scrF_{AB}$, $\scrF_{Ap}$, and
$\scrF_{pq}$ and gauge-dependent variables $Y_{a}$ defined by
Eqs.~(\ref{eq:2+2-Ya-def})--(\ref{eq:tildeYo-def}), the original
components $\{h_{AB},h_{Ap},h_{pq}\}$ of the metric perturbation
(\ref{eq:hAB-fourier})--(\ref{eq:hpq-fourier}) are given by
\begin{eqnarray}
  \label{eq:hab-gauge-decomp-final}
  h_{ab} &=& \scrF_{ab} + {\pounds}_{Y}g_{ab} .
\end{eqnarray}
This is the assertion of
Theorem~\ref{theorem:decomposition-theorem-with-spherical-symmetry}.


The aim of our proposal is to add a greater degree of freedom of the
metric perturbation so that the decomposition
(\ref{eq:hab-gauge-decomp-final}) is guaranteed even in the case of
$l=0,1$ modes.
Therefore, it is impossible to reach the final expression
(\ref{eq:hab-gauge-decomp-final}) if we treat the metric perturbation
by using $S_{\delta=0}=Y_{lm}$ in the decomposition formulae
(\ref{eq:hAB-fourier})--(\ref{eq:hpq-fourier}) from the starting point.


To see this more specifically, we show an explicit example of the
$l=1$ odd mode perturbation.
If we apply our proposal to the decomposition formulae
(\ref{eq:hAB-fourier})--(\ref{eq:hpq-fourier}), $l=1$ odd-mode
perturbation is given by
\begin{eqnarray}
  \label{eq:l=1-hab-fourier-deltaneq0}
  h_{ab}
  =
  2 r \tilde{h}_{(o1)A}^{(l=1)} \epsilon_{pq} \hat{D}^{q}S_{\delta}
  (dx^{A})_{(a} (dx^{p})_{b)}
  +
  2 r^{2} \tilde{h}_{(o2)}^{(l=1)} \epsilon_{r(p} \hat{D}_{q)}\hat{D}^{r} S_{\delta}
  (dx^{p})_{(a} (dx^{q})_{b)}
  .
\end{eqnarray}
If we use  $S_{\delta=0}=Y_{lm}$ from the starting point,
$2\epsilon_{r(p} \hat{D}_{q)}\hat{D}^{r} S_{\delta}=0$.
So, the mode coefficient $\tilde{h}_{(o2)}^{(l=1)}$ does not appear.
When $\delta\neq 0$, the gauge-transformation rule of the variables
$\tilde{h}_{(o1)A}^{(l=1)}$ and $\tilde{h}_{(o2)}^{(l=1)}$ are given
by
\begin{eqnarray}
  {}_{\scrY}\!\tilde{h}_{(o1)A}^{(l=1)}
  -
  {}_{\scrX}\!\tilde{h}_{(o1)A}^{(l=1)}
  =
  r \bar{D}_{A}\left(\frac{1}{r}\zeta_{(o)} \right)
  ,
  \quad
  {}_{\scrY}\!\tilde{h}_{(o2)}^{(l=1)}
  -
  {}_{\scrX}\!\tilde{h}_{(o2)}^{(l=1)}
  =
  - \frac{1}{r} \zeta_{(o)}
  ,
\end{eqnarray}
where the generator of the gauge transformation is given by
$\xi_{A}=0$ and $\xi_{p}=r \zeta_{(o)} \epsilon_{pq}
\hat{D}^{q}S_{\delta}$.
If we use  $S_{\delta=0}=Y_{lm}$ from the starting point, the
gauge-transformation rule for the variable $\tilde{h}_{(o2)}^{(l=1)}$
does not appear, either.
However, in our case, this gauge transformation appears due to the fact
$\epsilon_{r(p} \hat{D}_{q)}\hat{D}^{r} S_{\delta}\neq 0$.
Inspecting these gauge-transformation rules, we rewrite
Eq.~(\ref{eq:l=1-hab-fourier-deltaneq0}) as
\begin{eqnarray}
  h_{ab}
  &=&
      2 r \left[
      \left(
      \tilde{h}_{(o1)A}^{(l=1)}
      + r \bar{D}_{A}\tilde{h}_{(o2)}
      \right)
      - r \bar{D}_{A}\tilde{h}_{(o2)}
      \right] \epsilon_{pq} \hat{D}^{q}S_{\delta}
      (dx^{A})_{(a} (dx^{p})_{b)}
      \nonumber\\
  &&
      +
      2 r^{2} \tilde{h}_{(o2)}^{(l=1)} \epsilon_{r(p} \hat{D}_{q)}\hat{D}^{r} S_{\delta}
      (dx^{p})_{(a} (dx^{q})_{b)}
      \nonumber\\
  &=&
      2 r \left[
      \tilde{F}_{A}
      - r \bar{D}_{A}\left(\frac{1}{r^{2}}\tilde{Y}_{(o1)}\right)
      \right] \epsilon_{pq} \hat{D}^{q}S_{\delta}
      (dx^{A})_{(a} (dx^{p})_{b)}
      \nonumber\\
  &&
      +
      2 r^{2} \left(\frac{1}{r^{2}}\tilde{Y}_{(o1)}\right) \epsilon_{r(p} \hat{D}_{q)}\hat{D}^{r} S_{\delta}
      (dx^{p})_{(a} (dx^{q})_{b)}
      .
      \label{eq:l=1-hab-fourier-deltaneq0-rewrite}
\end{eqnarray}
In Eq.~(\ref{eq:l=1-hab-fourier-deltaneq0-rewrite}), $\tilde{F}_{A}$
is the gauge-invariant variable for the metric perturbation defined by
Eq.~(\ref{eq:2+2-gauge-inv-def-tildeFA-sum}) and $\tilde{Y}_{(o1)}$ is
the gauge-dependent part of the metric perturbation defined by
Eq.~(\ref{eq:tildeYo-def}).
The terms of $\tilde{Y}_{(o1)}$ in
Eq.~(\ref{eq:l=1-hab-fourier-deltaneq0-rewrite}) can be written in the
form of the Lie derivative of the background metric as in
Eq.~(\ref{eq:hab-gauge-decomp-final}).
When $\delta=0$, Eq.~(\ref{eq:l=1-hab-fourier-deltaneq0-rewrite}) is
given by
\begin{eqnarray}
  h_{ab}
  &=&
      2 r \left[
      \tilde{F}_{A}
      - r \bar{D}_{A}\left(\frac{1}{r^{2}}\tilde{Y}_{(o1)}\right)
      \right] \epsilon_{pq} \hat{D}^{q}S_{\delta}
      (dx^{A})_{(a} (dx^{p})_{b)}
      .
      \label{eq:l=1-hab-fourier-delta=0-rewrite}
\end{eqnarray}
This is also given in the form (\ref{eq:hab-gauge-decomp-final}).


When we construct the gauge-invariant variable $\tilde{F}_{A}$ in
Eqs.~(\ref{eq:l=1-hab-fourier-deltaneq0-rewrite}) and
(\ref{eq:l=1-hab-fourier-delta=0-rewrite}), the existence of the
coefficient $\tilde{h}_{(o2)}^{(l=1)}=\tilde{Y}_{(o1)}/r^{2}$ is
essential which does not appear in the treatment using the spherical
harmonics $Y_{lm}$ from the starting point.
Actually, we cannot define the gauge-invariant variable for $l=1$
odd-mode perturbation in the same manner as $l\geq 2$-mode case.
Our
proposal~\ref{proposal:treatment-proposal-on-pert-on-spherical-BG},
make the degree of freedom of the metric perturbations increase.
We have to emphasize that we can reach the
decomposition (\ref{eq:hab-gauge-decomp-final}) owing to this
additional degree of freedom.


To discuss the linearized Einstein equation
(\ref{eq:einstein-equation-gauge-inv}) and the linear perturbation of
the continuity equation
\begin{eqnarray}
  \label{eq:continuity-linearized-energy-momentum-general}
  \nabla_{a}{}^{(1)}\!\scrT_{b}^{\;\;a} = 0
\end{eqnarray}
of the gauge-invariant energy-momentum tensor
${}^{(1)}\!\scrT_{b}^{\;\;a}:=g^{ac}{}^{(1)}\!\scrT_{bc}$ on a
vacuum background spacetime, we consider the mode-decomposition of the
gauge-invariant part ${}^{(1)}\!\scrT_{bc}$ of the linear perturbation
of the energy-momentum tensor through the set
(\ref{eq:harmonic-fucntions-set}) of the harmonics as follows:
\begin{eqnarray}
  {}^{(1)}\!\scrT_{ab}
  &=&
      \sum_{l,m}
      \tilde{T}_{AB}
      S_{\delta}
      (dx^{A})_{a}(dx^{B})_{b}
      +
      r
      \sum_{l,m} \left\{
      \tilde{T}_{(e1)A} \hat{D}_{p}S_{\delta}
      +
      \tilde{T}_{(o1)A} \epsilon_{pr}\hat{D}^{r}S_{\delta}
      \right\}
      2 (dx^{A})_{(a}(dx^{p})_{b)}
      \nonumber\\
  &&
     +
     \sum_{l,m} \left\{
     \tilde{T}_{(e0)} \frac{1}{2} \gamma_{pq} S_{\delta}
     +
     \tilde{T}_{(e2)} \left(
     \hat{D}_{p}\hat{D}_{q}S_{\delta}
     -
     \frac{1}{2} \gamma_{pq} \hat{D}_{r}\hat{D}^{r}S_{\delta}
     \right)
     \right.
     \nonumber\\
  && \quad\quad\quad
     \left.
     +
     \tilde{T}_{(o2)}
     2 \epsilon_{r(p}\hat{D}_{q)}\hat{D}^{r}S_{\delta}
     \right\}
     (dx^{p})_{a}(dx^{q})_{b}
     .
     \label{eq:1st-pert-calTab-dd-decomp}
\end{eqnarray}


In the Part I paper~\cite{K.Nakamura-2021-PartI}, we derived the
linearized Einstein equations, discussed the odd-mode perturbation
$\tilde{F}_{Ap}$ in
Eq.~(\ref{eq:2+2-gauge-invariant-variables-calFAp}), and derived the
$l=1$ odd-mode solutions to these equations.
The Einstein equation for even mode $\tilde{F}_{AB}$ and $\tilde{F}$ in
Eqs.~(\ref{eq:2+2-gauge-invariant-variables-calFAB}) and
(\ref{eq:2+2-gauge-invariant-variables-calFpq}) also derived in the
Part I paper~\cite{K.Nakamura-2021-PartI}, and $l=0,1$
even-mode solutions are derived in the Part II
paper~\cite{K.Nakamura-2021-PartII}.
Since these solutions include the Kerr parameter perturbation and the
Schwarzschild mass parameter perturbation of the linear order in the
vacuum case, these are physically reasonable.
Then, we conclude that our proposal is also physically reasonable.
Furthermore, we also checked that our derived solutions include the
linearized LTB solution and non-rotating C-metric with the
Schwarzschild background in the Part III
paper~\cite{K.Nakamura-2021-PartIII}.


The purpose of this paper is to compare our proposed gauge-invariant
treatments for $l=0,1$-mode perturbations on the Schwarzschild
background spacetime with conventional ``complete gauge-fixing
treatments'' in which we use the spherical harmonics $Y_{lm}$ from the
starting point.
For this purpose, the linearized Einstein equations for the $l=0,1$-modes
and their solutions based on our proposal are necessary.
Therefore, we review them below.


\subsection{$l=1$ odd-mode linearized Einstein equations and solutions}
\label{sec:l=1-odd_Einstein_equations_solutions}


As derived in the Part I paper~\cite{K.Nakamura-2021-PartI}, the
$l=1$ odd-mode part in the linearized Einstein equations are
simplified as the constraint equation
\begin{eqnarray}
  \bar{D}_{D}(r\tilde{F}^{D})
  =
  0
  ,
  \label{eq:1st-p-Ein-non-vac-pq-traceless-odd-l=1}
\end{eqnarray}
and the evolution equation
\begin{eqnarray}
  &&
     - \left[ \bar{D}^{D}\bar{D}_{D} - \frac{2}{r^{2}} \right] (r\tilde{F}_{A})
     - \frac{2}{r^{2}} (\bar{D}^{D}r) (\bar{D}_{A}r) (r\tilde{F}_{D})
     + \frac{2}{r} (\bar{D}^{D}r) \bar{D}_{A}(r\tilde{F}_{D})
     \nonumber\\
  &=&
  16 \pi r \tilde{T}_{(o1)A}
  ,
  \label{eq:1st-p-Ein-non-vac-Aq-odd-sum-2-red-l=1}
\end{eqnarray}
where we choose $\tilde{T}_{(o2)}=0$ by hand.
Furthermore, we have the continuity equation
\begin{eqnarray}
  \bar{D}^{C}\tilde{T}_{(o1)C}
  + \frac{3}{r} (\bar{D}^{D}r) \tilde{T}_{(o1)D}
  =
  0
  \label{eq:div-barTab-linear-p-odd-l=1}
\end{eqnarray}
for the $l=1$ odd-mode matter perturbation which is derived from the
divergence of the first-order perturbation of the energy-momentum
tensor.


We also note that the constraint
(\ref{eq:1st-p-Ein-non-vac-pq-traceless-odd-l=1}) comes from the
traceless part of the $(q,p)$-component of the Einstein equation
(\ref{eq:einstein-equation-gauge-inv}), which is the coefficient of
the mode function $2\epsilon_{r(p}\hat{D}_{q)}\hat{D}^{p}S_{\delta}$.
If $\delta=0$, i.e., the scalar harmonics $S_{\delta}$ is the
spherical harmonics $Y_{lm}$,
$2\epsilon_{r(p}\hat{D}_{q)}\hat{D}^{p}S_{\delta}=0$ for $l=1$.
Therefore, the constraint
(\ref{eq:1st-p-Ein-non-vac-pq-traceless-odd-l=1}) does not appear when
we use the spherical harmonics $Y_{lm}$ from the starting point.


The explicit strategy to solve these odd-mode perturbations and
$l=0,1$ mode solutions was discussed in the Part I
paper~\cite{K.Nakamura-2021-PartI}.
As the result, we obtained the $l=1$ odd-mode ``formal'' solution as
follows:
\begin{eqnarray}
  \label{eq:l=1-odd-mode-propagating-sol-ver2-2}
  &&
     2 \scrF_{Ap} (dx^{A})_{(a}(dx^{p})_{b)}
     =
     6 M r^{2} \left[\int dr \frac{a_{1}(t,r)}{r^{4}}\right]
     \sin^{2}\theta
     (dt)_{(a} (d\phi)_{b)}
     +
     {\pounds}_{V}g_{ab}
     ,
  \\
  \label{eq:l=1-odd-mode-propagating-sol-ver2-Va-def-2}
  &&
     V_{a}
     =
     \left(\beta_{1}t + \beta_{0} + W_{(o)}(t,r)\right) r^{2} \sin^{2}\theta (d\phi)_{a}
     ,
\end{eqnarray}
where
\begin{eqnarray}
  a_{1}(t,r)
  &=&
      - \frac{16 \pi}{3M} r^{3} f \int dt \tilde{T}_{(o1)r} + a_{10}
      \nonumber\\
  &=&
      - \frac{16 \pi}{3M} \int dr r^{3} \frac{1}{f} \tilde{T}_{(o1)t} + a_{10}
      .
      \label{eq:Psio-Zo-relation-l=1-sum-6}
\end{eqnarray}
The constant $a_{10}$ in Eq.~(\ref{eq:Psio-Zo-relation-l=1-sum-6})
corresponds to the Kerr parameter.
Furthermore, the function $W_{(o)}(t,r)$ in
Eq.~(\ref{eq:l=1-odd-mode-propagating-sol-ver2-Va-def-2}) is related
to the solution to the $l=1$-mode Regge-Wheeler equation
\begin{eqnarray}
  \partial_{t}^{2}Z_{(o)}
  -  f \partial_{r}( f \partial_{r}Z_{(o)})
  + \frac{1}{r^{2}} f \left[ l(l+1) - 3 (1-f) \right] Z_{(o)}
  =
  16 \pi f^{2} \tilde{T}_{(o1)r}
  ,
  \label{eq:odd-master-equation-Regge-Wheeler}
\end{eqnarray}
where $Z_{(o)} = r f \partial_{r}W_{(o)}$.
We have to solve Eq.~(\ref{eq:odd-master-equation-Regge-Wheeler}) to
obtain the function $W_{(o)}(t,r)$.
In this sense, the solution
(\ref{eq:l=1-odd-mode-propagating-sol-ver2-Va-def-2})  should be
regarded as the ``formal'' one.


\subsection{$l=0,1$ even-mode linearized Einstein equations and solutions}
\label{sec:l=01-even_Einstein_equations_solutions}


The $l=0,1$ even-mode part of the linearized Einstein equation
(\ref{eq:einstein-equation-gauge-inv}) is summarized as follows:
\begin{eqnarray}
  \label{eq:linearized-Einstein-pq-traceless-even}
  \tilde{F}_{D}^{\;\;\;D}
  &=&
      0
      ,
  \\
  \bar{D}^{D}\tilde{\FF}_{AD}
  - \frac{1}{2} \bar{D}_{A}\tilde{F}
  &=&
      16 \pi r \tilde{T}_{(e1)A}
      ,
      \label{eq:even-FAB-divergence-3}
\end{eqnarray}
where the variable $\tilde{\FF}_{AB}$ is the traceless part of the
variable $\tilde{F}_{AB}$ defined by
\begin{eqnarray}
  \label{eq:FF-def}
  \tilde{\FF}_{AB} := \tilde{F}_{AB} - \frac{1}{2} y_{AB}
  \tilde{F}_{C}^{\;\;C}
\end{eqnarray}
and we choose the component of the energy-momentum tensor so that
$\tilde{T}_{(e2)}=0$ by hand.
Owing to the same choice $\tilde{T}_{(e2)}=0$, we also have the
evolution equations
\begin{eqnarray}
  &&
     \left(
     \bar{D}_{D}\bar{D}^{D}
     + \frac{2}{r} (\bar{D}^{D}r) \bar{D}_{D}
     -  \frac{(l-1)(l+2)}{r^{2}}
     \right) \tilde{F}
     - \frac{4}{r^{2}} (\bar{D}_{C}r) (\bar{D}_{D}r) \tilde{\FF}^{CD}
     \!\!=\!\!
     16 \pi S_{(F)}
     ,
     \label{eq:even-mode-tildeF-master-eq-mod-3}
  \\
  &&
  S_{(F)}
  :=
       \tilde{T}_{C}^{\;\;\;C}
       + 4 (\bar{D}_{D}r) \tilde{T}_{(e1)}^{D}
     ,
     \label{eq:sourcee0-def}
  \\
  &&
     \left[
     -  \bar{D}_{D}\bar{D}^{D}
     -  \frac{2}{r} (\bar{D}_{D}r) \bar{D}^{D}
     + \frac{4}{r} (\bar{D}^{D}\bar{D}_{D}r)
     + \frac{l(l+1)}{r^{2}}
     \right]
     \tilde{\FF}_{AB}
     \nonumber\\
  &&
     + \frac{4}{r} (\bar{D}^{D}r) \bar{D}_{(A}\tilde{\FF}_{B)D}
     -  \frac{2}{r} (\bar{D}_{(A}r) \bar{D}_{B)}\tilde{F}
     \nonumber\\
  && =
      16 \pi S_{(\FF)AB}
      ,
      \label{eq:1st-pert-Einstein-non-vac-AB-traceless-final-3}
\end{eqnarray}
\begin{eqnarray}
  S_{(\FF)AB}
  &:=&
      \tilde{T}_{AB} - \frac{1}{2} y_{AB} \tilde{T}_{C}^{\;\;\;C}
      - 2 \left( \bar{D}_{(A}(r \tilde{T}_{(e1)B)}) - \frac{1}{2} y_{AB} \bar{D}^{D}(r \tilde{T}_{(e1)D}) \right)
      \nonumber\\
  &&
      + 2 y_{AB} (\bar{D}^{C}r) \tilde{T}_{(e1)C}
     ,
  \label{eq:souce(FF)-def}
\end{eqnarray}
for the variable $\tilde{F}$ and the traceless variable
$\tilde{\FF}_{AB}$ with $l=0,1$.
We also have to take into account the even-mode part of the
continuity equation as follows:
\begin{eqnarray}
  &&
     \bar{D}^{C}\tilde{T}_{C}^{\;\;B}
     + \frac{2}{r} (\bar{D}^{D}r)\tilde{T}_{D}^{\;\;\;B}
     -  \frac{1}{r} l(l+1) \tilde{T}_{(e1)}^{B}
     -  \frac{1}{r} (\bar{D}^{B}r) \tilde{T}_{(e0)}
     =
     0
     ,
     \label{eq:div-barTab-linear-A}
  \\
  &&
     \bar{D}^{C}\tilde{T}_{(e1)C}
     + \frac{3}{r} (\bar{D}^{C}r) \tilde{T}_{(e1)C}
     + \frac{1}{2r} \tilde{T}_{(e0)}
     =
     0
     ,
     \label{eq:div-barTab-linear-p-even}
\end{eqnarray}
for $l=0,1$.


We also note that the constraint
(\ref{eq:linearized-Einstein-pq-traceless-even}) comes from the
traceless part of the $(q,p)$-component of the Einstein tensor
(\ref{eq:einstein-equation-gauge-inv}), which is the coefficient of
the mode function
$\left(\hat{D}_{p}\hat{D}_{q}-\frac{1}{2}\gamma_{pq}\hat{D}^{r}\hat{D}_{r}\right)S_{\delta}$.
If $\delta =0$, i.e., the scalar harmonics $S_{\delta}$ is the
spherical harmonics $Y_{lm}$,
$\left(\hat{D}_{p}\hat{D}_{q}-\frac{1}{2}\gamma_{pq}\hat{D}^{r}\hat{D}_{r}\right)S_{\delta}=0$
for $l=1$.
Therefore, the constraint
(\ref{eq:linearized-Einstein-pq-traceless-even}) does not appear when
we use the spherical harmonics $Y_{lm}$ from the starting point.


\subsubsection{$l=1$ even-mode solution}
\label{sec:l=1-even_solutions}


In the Part II paper~\cite{K.Nakamura-2021-PartII}, we derived the
$l=1$ solution to
Eqs.~(\ref{eq:linearized-Einstein-pq-traceless-even}),
(\ref{eq:even-FAB-divergence-3}),
(\ref{eq:even-mode-tildeF-master-eq-mod-3}),
(\ref{eq:1st-pert-Einstein-non-vac-AB-traceless-final-3}),
(\ref{eq:div-barTab-linear-A}), and
(\ref{eq:div-barTab-linear-p-even}) with $l=1$.
For $m=0$ mode, in the Part II paper~\cite{K.Nakamura-2021-PartII}, we
derived the following ``formal'' solution to the linearized Einstein equation
\begin{eqnarray}
  \scrF_{ab}
  &=&
      {\pounds}_{V}g_{ab}
      - \frac{16 \pi r^{2}}{3(1-f)}\left[
      f^{2} \left\{
      \frac{1+f}{2} \tilde{T}_{rr}
      + r f \partial_{r}\tilde{T}_{rr}
      -  \tilde{T}_{(e0)}
      -  4 \tilde{T}_{(e1)r}
      \right\}  (dt)_{a}(dt)_{b}
      \right.
      \nonumber\\
  && \quad\quad\quad\quad\quad\quad\quad\quad\quad
     \left.
      + \frac{2r}{f} \left\{
      \partial_{t}\tilde{T}_{tt}
      - \frac{3f(1-f)}{2r} \tilde{T}_{tr}
      \right\} (dt)_{(a}(dr)_{b)}
     \right.
      \nonumber\\
  && \quad\quad\quad\quad\quad\quad\quad\quad\quad
     \left.
      + \frac{r}{f}
      \left\{
      \partial_{r}\tilde{T}_{tt}
      - \frac{3(1-3f)}{2rf} \tilde{T}_{tt}
      \right\} (dr)_{a}(dr)_{b}
     \right.
      \nonumber\\
  && \quad\quad\quad\quad\quad\quad\quad\quad\quad
     \left.
      + r^{2} \tilde{T}_{tt} \gamma_{ab}
     \right] \cos\theta
      ,
      \label{eq:tildeFab-l=1-m=0-nonvacsum-2-cov}
\end{eqnarray}
where the vector field $V_{a}$ is given by
\begin{eqnarray}
  V_{a}
  &:=&
       -  r \partial_{t}\Phi_{(e)} \cos\theta (dt)_{a}
       + \left( \Phi_{(e)} - r \partial_{r}\Phi_{(e)} \right) \cos\theta (dr)_{a}
       -  r \Phi_{(e)} \sin\theta (d\theta)_{a}
       .
       \label{eq:generator-covariant-vacuum-l=1-m=0-result-2}
\end{eqnarray}
Here, the variable $\Phi_{(e)}$ is a solution to the $l=1$
Zerilli equation
\begin{eqnarray}
  &&
     -  \partial_{t}^{2}\Phi_{(e)}
     + f \partial_{r}\left[ f \partial_{r}\Phi_{(e)} \right]
     -
     \frac{f(1-f)}{r^{2}}
     \Phi_{(e)}
     \nonumber\\
  &=&
     \frac{4\pi r}{3(1-f)} \left[
     3(1-3f) \tilde{T}_{tt}
     + (1+f)f^{2} \tilde{T}_{rr}
     -  2 r f \partial_{r}\tilde{T}_{tt}
      \right.
      \nonumber\\
  && \quad\quad\quad\quad\quad
  \left.
     + 2 r f^{3} \partial_{r}\tilde{T}_{rr}
     -  2 f^{2} \tilde{T}_{(e0)}
     -  8 f^{2} \tilde{T}_{(e1)r}
     \right]
     .
     \label{eq:Zerilli-Moncrief-eq-final-l=1}
\end{eqnarray}
If we obtain the solution to the $l=1$ Zerilli equation
(\ref{eq:Zerilli-Moncrief-eq-final-l=1}), we can write the explicit
form of the solution (\ref{eq:tildeFab-l=1-m=0-nonvacsum-2-cov})
through the generator $V_{a}$ defined by
Eq.~(\ref{eq:generator-covariant-vacuum-l=1-m=0-result-2}).
In this sense, we have to regard the solution
(\ref{eq:tildeFab-l=1-m=0-nonvacsum-2-cov}) is a ``formal'' one.


\subsubsection{$l=0$ even-mode solution}
\label{sec:l=0-even_solutions}


On the other hand, for the $l=0$-mode, we may choose
$\tilde{T}_{(e1)A}=0$ in
Eqs.~(\ref{eq:linearized-Einstein-pq-traceless-even}),
(\ref{eq:even-FAB-divergence-3}),
(\ref{eq:even-mode-tildeF-master-eq-mod-3})--(\ref{eq:div-barTab-linear-p-even})
with $l=0$.
Then, we derived the $l=0$ mode solution
\begin{eqnarray}
  \label{eq:calFab+poundsVg-l=0-non-vac-final}
  \scrF_{ab}
  &=&
      \frac{2}{r} \left(M_{1}+4\pi \int dr \frac{r^{2}}{f} T_{tt}\right)
      \left((dt)_{a}(dt)_{a}+ \frac{1}{f^{2}} (dr)_{a}(dr)_{a}\right)
      \nonumber\\
  &&
      + 2 \left[4 \pi r \int dt \left(\frac{1}{f} \tilde{T}_{tt} + f \tilde{T}_{rr} \right)\right] (dt)_{(a}(dr)_{b)}
      + {\pounds}_{V}g_{ab}
      ,
\end{eqnarray}
where
\begin{eqnarray}
  \label{eq:Va-result-non-vac-final}
  V_{a}
  =
  \left(
  \frac{f}{4} \Upsilon
  + \frac{rf}{4} \partial_{r}\Upsilon
  -  \frac{r \Xi(r)}{(1-3f)}
  + f \int dr \frac{2 \Xi(r)}{f(1-3f)^{2}}
  \right) (dt)_{a}
  +
  \frac{1}{4f} r \partial_{t}\Upsilon (dr)_{a}
  .
\end{eqnarray}
Here, the variable $\tilde{F}=:\partial_{t}\Upsilon$ must satisfy the
equation
\begin{eqnarray}
  &&
     - \frac{1}{f} \partial_{t}^{2} \Upsilon
     + \partial_{r}(f \partial_{r}\Upsilon)
     + \frac{3(1-f)}{r^{2}}\Upsilon
     - \frac{8}{r^{3}} \int dt m_{1}(t,r)
     - \frac{4}{1-3f} \partial_{r}\Xi(r)
     \nonumber\\
  &=&
      16 \pi \int dt \left( - \frac{1}{f} \tilde{T}_{tt} + f \tilde{T}_{rr} \right)
      ,
      \label{eq:Upsilon-equation-1}
\end{eqnarray}
where $m_{1}(t,r)$ is given by
\begin{eqnarray}
  \label{eq:m1tr-def}
  m_{1}(t,r) = 4 \pi \int dr \frac{r^{2}}{f} \tilde{T}_{tt} + M_{1}
  = 4 \pi \int dt r^{2} f \tilde{T}_{tr} + M_{1},
\end{eqnarray}
$M_{1}$ is the constant corresponding to the Schwarzschild mass
perturbation, and $\Xi(r)$ is an arbitrary function of $r$.


\section{Rule of comparison and gauge-transformation rules in a
  conventional gauge-fixing}
\label{sec:Rule_of_comparison}


The main purpose of this paper is to check the following statement.
``{\it Gauge invariant formulations of perturbations are equivalent to
  complete gauge fixing approaches.}''
However, this statement is too ambiguous to check if we have the
background knowledge explained in
Sec.~\ref{sec:Brief_rev_gauge-inv_l=01modes}.
For example, there is no explanation of the terminology ``{\it
  gauge}'' in this statement.
Therefore, we have to clarify this statement as follows:


\begin{description}
\item[{\bf Our rule for comparison $\quad$}]
  First of all, we assume the terminology ``{\it gauge}'' in the
  statement ``{\it Gauge invariant formulations of perturbations are
    equivalent to complete gauge fixing approaches}'' is the second-kind
  gauge which is explained in
  Sec.~\ref{sec:general-framework-GI-perturbation-theroy}.
  Therefore, the gauge-transformation rule for this statement is given
  by (\ref{eq:Bruni-47-one}).
  Furthermore, the degree of freedom that changes under this
  gauge-transformation rule is regarded as ``{\it unphysical.}''
  The ``{\it gauge fixing}'' in the above statement is a specification
  of some perturbative variables through the degree of freedom of the
  generator $\xi_{(1)}^{a}$.
  Furthermore, ``{\it complete gauge fixing}'' in the above statement is
  a specification of some perturbative variables through the ``{\it
    entire}'' degree of freedom of the generator $\xi_{(1)}^{a}$.
\end{description}


These are all rules of our game to compare our gauge-invariant
formulation and a ``conventional complete gauge-fixing approach'' in
which we use the spherical harmonics $Y_{lm}$ from the starting point.


\subsection{Metric perturbations}
\label{sec:Metric_perturbations}


Based on the above conceptual premise, we consider the metric
perturbation $h_{ab}$ on the background Schwarzschild spacetime
$(\scrM,g_{ab})$ whose metric $g_{ab}$ is given by Eqs.~(\ref{eq:background-metric-2+2})--(\ref{eq:S2-unit-basis-def}).
We consider the components of the metric perturbation $h_{ab}$ as
Eq.~(\ref{eq:metric-perturbation-components}) and the decomposition of
these components $\{h_{AB}, h_{Ap}, h_{pq}\}$ as
Eqs.~(\ref{eq:hAB-fourier})--(\ref{eq:hpq-fourier}) but we concentrate
on the case $S_{\delta}=Y_{lm}=:S$, where $Y_{lm}$ is the conventional
spherical harmonics.


Since we choose $S=Y_{lm}$, we have $\hat{D}_{p}S$ $=$
$\epsilon_{pq}\hat{D}^{q}S$ $=$ $0$,
$\left(\hat{D}_{p}\hat{D}_{q}-\frac{1}{2}\gamma_{pq}\hat{D}^{r}\hat{D}_{r}\right)S$
$=$ $2 \epsilon_{r(p}\hat{D}_{q)}\hat{D}^{r}S$ $=$ $0$ for $l=0$
modes.
For $l=1$ modes,
$\left(\hat{D}_{p}\hat{D}_{q}-\frac{1}{2}\gamma_{pq}\hat{D}^{r}\hat{D}_{r}\right)S$
$=$ $2 \epsilon_{r(p}\hat{D}_{q)}\hat{D}^{r}S$ $=$ $0$.
Due to these facts, we cannot construct gauge-invariant variables for
$l=0,1$ modes in a similar manner to the derivation of
Eqs.~(\ref{eq:gauge-inv-tildeFAB-def-sum})--(\ref{eq:2+2-gauge-inv-tildeF-def-sum})
for $l\geq 2$ modes.
Then, we cannot use the gauge-invariant formulation reviewed in
Sec.~\ref{sec:Brief_rev_gauge-inv_l=01modes}.
Instead, we have to fix the second-kind gauge degree of freedom to
exclude the ``unphysical'' degree of freedom.


\subsection{Conventional gauge-transformation rules for $l=0,1$ metric perturbations}
\label{sec:Conventional-gauge-transformation_rules_for_metric_pert}


Here, we consider the gauge-transformation rule
\begin{eqnarray}
  \label{eq:gauge-transformation-rule-metric}
  {}_{\scrY}\!h_{ab}
  -
  {}_{\scrX}\!h_{ab}
  =
  {\pounds}_{\xi}g_{ab}
  =
  2 \nabla_{(a}\xi_{b)}
  .
\end{eqnarray}
From this gauge-transformation rule, we can derive the
gauge-transformation rule for the components
$\{h_{AB},h_{Ap},h_{qp}\}$ defined by
Eq.~(\ref{eq:metric-perturbation-components}) is given by
\begin{eqnarray}
  \label{eq:gauge-transformation-rule-AB}
  {}_{{\scrY}}\!h_{AB}
  -
  {}_{{\scrX}}\!h_{AB}
  &=&
      {\pounds}_{\xi}g_{AB}
      =
      \nabla_{A}\xi_{B}
      +
      \nabla_{B}\xi_{A}
      =
      \bar{D}_{A}\xi_{B}
      +
      \bar{D}_{B}\xi_{A}
      ,
  \\
  \label{eq:gauge-transformation-rule-Ap}
  {}_{{\scrY}}\!h_{Ap}
  -
  {}_{{\scrX}}\!h_{Ap}
  &=&
      {\pounds}_{\xi}g_{Ap}
      =
      \nabla_{A}\xi_{p}
      +
      \nabla_{p}\xi_{A}
      =
      \bar{D}_{A}\xi_{p}
      +
      \hat{D}_{p}\xi_{A}
      -
      \frac{2}{r} (\bar{D}_{A}r) \xi_{p}
      ,
  \\
  \label{eq:gauge-transformation-rule-pq}
  {}_{{\scrY}}\!h_{pq}
  -
  {}_{{\scrX}}\!h_{pq}
  &=&
      {\pounds}_{\xi}g_{pq}
      =
      \nabla_{p}\xi_{q}
      +
      \nabla_{q}\xi_{p}
      =
      \hat{D}_{p}\xi_{q}
      +
      \hat{D}_{q}\xi_{p}
      +
      2 r (\bar{D}^{A}r) \gamma_{pq} \xi_{A}
\end{eqnarray}
from the formulae summarized in Appendix B of the Part I
paper~\cite{K.Nakamura-2021-PartI}.
Here, we consider the Fourier transformation of
$\xi_{a}=:\xi_{A}(dx^{A})_{a}+\xi_{p}(dx^{p})_{a}$ as
\begin{eqnarray}
  \xi_{A} = \sum_{l,m} \zeta_{A} S, \quad
  \xi_{p} = r \sum_{l,m} \left(
  \zeta_{(e)} \hat{D}_{p}S
  +
  \zeta_{(o)} \epsilon_{pq}\hat{D}^{p}S
  \right)
  .
\end{eqnarray}
From these Fourier transformation rules,
Eqs.~(\ref{eq:gauge-transformation-rule-AB})--(\ref{eq:gauge-transformation-rule-pq})
are given by
\begin{eqnarray}
  \label{eq:gauge-transformation-rule-AB-Fourier}
  {}_{{\scrY}}\!h_{AB}
  -
  {}_{{\scrX}}\!h_{AB}
  &=&
      \bar{D}_{A}\xi_{B}
      +
      \bar{D}_{B}\xi_{A}
      =
      \sum_{l,m} 2 \bar{D}_{(A}\zeta_{B)} S
      ,
  \\
  \label{eq:gauge-transformation-rule-Ap-Fourier}
  {}_{{\scrY}}\!h_{Ap}
  -
  {}_{{\scrX}}\!h_{Ap}
  &=&
      \bar{D}_{A}\xi_{p}
      +
      \hat{D}_{p}\xi_{A}
      -
      \frac{2}{r} (\bar{D}_{A}r) \xi_{p}
      \nonumber\\
  &=&
      r
      \sum_{l,m} \left[
      \left(
      \frac{1}{r} \zeta_{A}
      + \bar{D}_{A}\zeta_{(e)}
      - \frac{1}{r} (\bar{D}_{A}r) \zeta_{(e)}
      \right) \hat{D}_{p}S
      \right.
      \nonumber\\
  && \quad\quad\quad\quad
     \left.
     +
     \left(
     \bar{D}_{A}\zeta_{(o)}
     - \frac{1}{r} (\bar{D}_{A}r) \zeta_{(o)}
     \right) \epsilon_{pq}\hat{D}^{p}S
     \right]
     ,
  \\
  \label{eq:gauge-transformation-rule-pq-Fourier}
  {}_{{\scrY}}\!h_{pq}
  -
  {}_{{\scrX}}\!h_{pq}
  &=&
      \hat{D}_{p}\xi_{q}
      +
      \hat{D}_{q}\xi_{p}
      +
      2 r (\bar{D}^{A}r) \gamma_{pq} \xi_{A}
      \nonumber\\
  &=&
      r^{2}
      \sum_{l,m} \left[
      \frac{4}{r}
      \left(
      -  \frac{l (l+1)}{2} \zeta_{(e)}
      + (\bar{D}^{A}r) \zeta_{A}
      \right)
      \frac{1}{2} \gamma_{pq} S
      \right.
      \nonumber\\
  && \quad\quad\quad\quad
     \left.
     +
     \frac{2}{r} \zeta_{(e)} \left(
     \hat{D}_{p}\hat{D}_{q}S
     -
     \frac{1}{2} \gamma_{pq} \hat{D}^{r}\hat{D}_{r}S
     \right)
      \right.
      \nonumber\\
  && \quad\quad\quad\quad
     \left.
     -
     \frac{1}{r} \zeta_{(o)} 2\epsilon_{r(q}\hat{D}_{p)}\hat{D}^{r}S
     \right]
     .
\end{eqnarray}
Here, we used the property of the mode function $S$ is an eigen
function of the Laplacian
$\hat{D}_{r}\hat{D}^{r}:=\gamma^{rs}\hat{D}_{r}\hat{D}_{s}$ as 
\begin{eqnarray}
  \hat{D}_{r}\hat{D}^{r}S = - l(l+1) S, \quad S=Y_{lm}.
\end{eqnarray}


\subsubsection{$l\geq 2$ perturbations}
\label{sec:Conventional-gauge-transformation_rules_lgeq2}


For $l\geq 2$ modes, the set of mode functions
\begin{eqnarray}
  \label{eq:set-of-mode-functions}
  \left\{
  S,
  \hat{D}_{p}S,
  \epsilon_{pq}\hat{D}^{q}S,
  \frac{1}{2}\gamma_{pq} S,
  \left(
  \hat{D}_{p}\hat{D}_{q} - \frac{1}{2} \gamma_{pq} \hat{D}_{r}\hat{D}^{r}
  \right) S,
  2 \epsilon_{r(p}\hat{D}^{r}\hat{D}_{q)} S
  \right\}
\end{eqnarray}
is a linear-independent set of the tensor field of the rank 0, 1, and
2, even if $S=Y_{lm}$.
Then, we may compare
Eqs.~(\ref{eq:hAB-fourier})--(\ref{eq:hpq-fourier}) and
Eqs.~(\ref{eq:gauge-transformation-rule-AB-Fourier})--(\ref{eq:gauge-transformation-rule-pq-Fourier}).
As the result of this comparison, we obtain the gauge-transformation
rules for the mode functions
$\{\tilde{h}_{AB},\tilde{h}_{(e1)A},\tilde{h}_{(o1)A},\tilde{h}_{(e0)},\tilde{h}_{(e2)},\tilde{h}_{(o2)}\}$
as
\begin{eqnarray}
  \label{eq:gauge-trans-tildehAB-lgeq2}
  {}_{\scrY}\tilde{h}_{AB}
  -
  {}_{\scrX}\tilde{h}_{AB}
  &=&
      2 \bar{D}_{(A}\zeta_{B)}
      ,
  \\
  \label{eq:gauge-trans-tildehe1A-lgeq2}
  {}_{\scrY}\tilde{h}_{(e1)A}
  -
  {}_{\scrX}\tilde{h}_{(e1)A}
  &=&
      \frac{1}{r} \zeta_{A}
      + \bar{D}_{A}\zeta_{(e)}
      - \frac{1}{r} (\bar{D}_{A}r) \zeta_{(e)}
      ,
  \\
  \label{eq:gauge-trans-tildeho1A-lgeq2}
  {}_{\scrY}\tilde{h}_{(o1)A}
  -
  {}_{\scrX}\tilde{h}_{(o1)A}
  &=&
      \bar{D}_{A}\zeta_{(o)}
      - \frac{1}{r} (\bar{D}_{A}r) \zeta_{(o)}
      ,
  \\
  \label{eq:gauge-trans-tildehe0-lgeq2}
  {}_{\scrY}\tilde{h}_{(e0)}
  -
  {}_{\scrX}\tilde{h}_{(e0)}
  &=&
      \frac{4}{r}
      \left(
      -  \frac{l (l+1)}{2} \zeta_{(e)}
      + (\bar{D}^{A}r) \zeta_{A}
      \right)
      ,
  \\
  \label{eq:gauge-trans-tildehe2-lgeq2}
  {}_{\scrY}\tilde{h}_{(e2)}
  -
  {}_{\scrX}\tilde{h}_{(e2)}
  &=&
      \frac{2}{r} \zeta_{(e)}
      ,
  \\
  \label{eq:gauge-trans-tildeho2-lgeq2}
  {}_{\scrY}\tilde{h}_{(o2)}
  -
  {}_{\scrX}\tilde{h}_{(o2)}
  &=&
     -
     \frac{1}{r} \zeta_{(o)}
      .
\end{eqnarray}
From these gauge-transformation rules, it is well-known that we can
construct gauge-invariant variables and the gauge-dependent variables
and Conjecture~\ref{conjecture:decomposition-conjecture} is valid for
$l\geq 2$ modes.


\subsubsection{$l=1$ perturbations}
\label{sec:Conventional-gauge-transformation_rules_l=1}


As shown in the Appendix A in Ref.~\cite{K.Nakamura-2021-PartI}, for
$l=1$ modes,
\begin{eqnarray}
  \label{eq:identically-nonvanishi-mode-function-l=1}
  \hat{D}_{p}Y_{1m}\neq 0\neq \epsilon_{pq}\hat{D}^{q}Y_{1m}
\end{eqnarray}
but
\begin{eqnarray}
  \label{eq:identically-vanishi-mode-function-l=1}
  \left(
  \hat{D}_{p}\hat{D}_{q} - \frac{1}{2} \hat{D}_{r}\hat{D}^{r}
  \right) Y_{1m}
  =
  0
  =
  2 \epsilon_{r(p}\hat{D}_{q)}\hat{D}^{r}Y_{1m}
  .
\end{eqnarray}
In this case, the Fourier transformation
(\ref{eq:hAB-fourier})--(\ref{eq:hpq-fourier})
of the metric perturbation with $S_{\delta}=S=Y_{lm}$ is given by
\begin{eqnarray}
  \label{eq:K.Nakamura-2021-PartI-3.7-l=1}
  h_{AB}
  &=&
      \sum_{m} \tilde{h}_{AB} S
      ,
  \\
  \label{eq:K.Nakamura-2021-PartI-3.8-l=1}
  h_{Ap}
  &=&
      \sum_{m} r \tilde{h}_{(e1)A} \hat{D}_{p}S
      +
      \sum_{m} r \tilde{h}_{(o1)A} \epsilon_{pq}\hat{D}^{q}S
      ,
  \\
  \label{eq:K.Nakamura-2021-PartI-3.9-l=1}
  h_{pq}
  &=&
      \sum_{m} \frac{r^{2}}{2} \gamma_{pq} \tilde{h}_{(e0)} S
      .
\end{eqnarray}
The gauge-transformation rules
(\ref{eq:gauge-transformation-rule-AB-Fourier})--(\ref{eq:gauge-transformation-rule-pq-Fourier})
for each $m$ mode are given by
\begin{eqnarray}
  \label{eq:gauge-transformation-rule-AB-Fourier-l=1}
  {}_{{\scrY}}\!\tilde{h}_{AB}
  -
  {}_{{\scrX}}\!\tilde{h}_{AB}
  &=&
      2 \bar{D}_{(A}\zeta_{B)}
      ,
  \\
  \label{eq:gauge-transformation-rule-e1A-Fourier-l=1}
  {}_{{\scrY}}\!\tilde{h}_{(e1)A}
  -
  {}_{{\scrX}}\!\tilde{h}_{(e1)A}
  &=&
      \frac{1}{r} \zeta_{A}
      + \bar{D}_{A}\zeta_{(e)}
      - \frac{1}{r} (\bar{D}_{A}r) \zeta_{(e)}
      ,
  \\
  \label{eq:gauge-transformation-rule-o1A-Fourier-l=1}
  {}_{{\scrY}}\!\tilde{h}_{(o1)A}
  -
  {}_{{\scrX}}\!\tilde{h}_{(o1)A}
  &=&
      \bar{D}_{A}\zeta_{(o)}
      - \frac{1}{r} (\bar{D}_{A}r) \zeta_{(o)}
      ,
  \\
  \label{eq:gauge-transformation-rule-e0-Fourier-l=1}
  {}_{{\scrY}}\!\tilde{h}_{(e0)}
  -
  {}_{{\scrX}}\!\tilde{h}_{(e0)}
  &=&
      -  \frac{4}{r} \zeta_{(e)}
      + \frac{4}{r} (\bar{D}^{A}r) \zeta_{A}
      .
\end{eqnarray}
Comparing the gauge-transformation rules
(\ref{eq:gauge-trans-tildehAB-lgeq2})--(\ref{eq:gauge-trans-tildeho2-lgeq2})
for $l\geq 2$ with the gauge-transformation rules
(\ref{eq:gauge-transformation-rule-AB-Fourier-l=1})--(\ref{eq:gauge-transformation-rule-e0-Fourier-l=1}),
it is easy to find the gauge-transformation rule
(\ref{eq:gauge-trans-tildehe2-lgeq2}) and
(\ref{eq:gauge-trans-tildeho2-lgeq2}) for $l\geq 2$ do not appear in
the $l=1$ mode gauge transformation.
This is due to the fact
Eqs.~(\ref{eq:identically-vanishi-mode-function-l=1}), and the reason
why we cannot construct gauge-invariant variables for $l=1$ mode
perturbations in a similar method to the $l\geq 2$ case.


\subsubsection{$l=0$ perturbations}
\label{sec:Conventional-gauge-transformation_rules_l=0}


The spherical harmonic function $S=Y_{lm}$ is constant when $l=0$.
In this case, the Fourier transformation
(\ref{eq:hAB-fourier})--(\ref{eq:hpq-fourier})
of the metric perturbation with $S_{\delta}=S=Y_{lm}$ is given by
\begin{eqnarray}
  \label{eq:K.Nakamura-2021-PartI-3.7-l=0}
  h_{AB}
  &=&
      \tilde{h}_{AB} S
      ,
  \\
  \label{eq:K.Nakamura-2021-PartI-3.8-l=0}
  h_{Ap}
  &=&
      0
      ,
  \\
  \label{eq:K.Nakamura-2021-PartI-3.9-l=0}
  h_{pq}
  &=&
      \frac{r^{2}}{2} \gamma_{pq} \tilde{h}_{(e0)} S
      .
\end{eqnarray}
The gauge-transformation rules
(\ref{eq:gauge-transformation-rule-AB-Fourier})--(\ref{eq:gauge-transformation-rule-pq-Fourier})
are given by
\begin{eqnarray}
  \label{eq:gauge-transformation-rule-AB-Fourier-l=0}
  {}_{{\scrY}}\!\tilde{h}_{AB}
  -
  {}_{{\scrX}}\!\tilde{h}_{AB}
  &=&
      2 \bar{D}_{(A}\zeta_{B)}
      =
      \bar{D}_{A}\zeta_{B}
      +
      \bar{D}_{B}\zeta_{A}
      ,
  \\
  \label{eq:gauge-transformation-rule-pq-Fourier-l=0}
  {}_{\scrY}\!\tilde{h}_{(e0)}
  -
  {}_{\scrX}\!\tilde{h}_{(e0)}
  &=&
      \frac{4}{r} (\bar{D}^{A}r) \zeta_{A}
      .
\end{eqnarray}
It is easy to see that many gauge-transformation rules in
(\ref{eq:gauge-trans-tildehAB-lgeq2})--(\ref{eq:gauge-trans-tildeho2-lgeq2})
are missing.
This is due to the fact
\begin{eqnarray}
  \hat{D}_{p}S = \epsilon_{pq}\hat{D}^{q}S = 0, \quad
  \left(\hat{D}_{q}\hat{D}_{p}- \frac{1}{2} \gamma_{pq}
  \hat{D}_{r}\hat{D}^{r}\right)S
  =
  2 \epsilon_{r(p} \hat{D}_{q)} \hat{D}^{r} S = 0
  \label{eq:l=0-harmonics-derivative-zero}
\end{eqnarray}
This is the reason why we cannot construct gauge-invariant variables
for $l=0$ mode perturbations in a similar method to the $l\geq 2$ case.


\section{l=1 odd-mode perturbation in the conventional approach}
\label{sec:l=1-odd-ConventionalGaugeFixing}


As discussed in
Sec.~\ref{sec:Conventional-gauge-transformation_rules_l=1}, $l=1$ mode
perturbations are given by
Eqs.~(\ref{eq:K.Nakamura-2021-PartI-3.7-l=1})--(\ref{eq:K.Nakamura-2021-PartI-3.9-l=1}).
In particular, among these expressions of the $l=1$ mode perturbation, 
odd-mode perturbation is given by
\begin{eqnarray}
  \label{eq:K.Nakamura-2021-PartI-3.7-3.8-3.9-l=1-odd-2}
  h_{AB}
  &=&
      0
      , \quad
  h_{Ap}
  =
      r \tilde{h}_{(o1)A} \epsilon_{pq}\hat{D}^{q}S
      , \quad
  h_{pq}
  =
      0
      .
\end{eqnarray}
As the property of the $l=1$ spherical harmonics, we obtain the
condition (\ref{eq:identically-vanishi-mode-function-l=1}).
For odd-mode perturbation we have
\begin{eqnarray}
  \label{eq:identically-vanishi-mode-function-l=1-odd}
  2 \epsilon_{r(p}\hat{D}_{q)}\hat{D}^{r}Y_{1m}
  =
  0
  .
\end{eqnarray}


Here, we apply the notation which is introduced by
Eqs.~(\ref{eq:barh-defs})--(\ref{eq:barDhatD-defs}) and we obtain
\begin{eqnarray}
  \bar{h}_{AB} = 0, \quad
  \bar{h}_{A}^{\;\;D} = 0, \quad
  \bar{h}^{pD} := \gamma^{pq} y^{DE} r \tilde{h}_{(o1)E} \epsilon_{qr}\hat{D}^{r}S
, \quad
  \bar{h}_{p}^{\;\;q} = 0, \quad
  \bar{h}^{pq} = 0, \quad
\end{eqnarray}
for $l=1$ odd mode perturbations.
Through this notation of the metric perturbation, the linear
perturbations
(\ref{eq:linearized-Einstein-tensor-AB-result-sum})--(\ref{eq:linearized-Einstein-tensor-pq-result-sum})
of the Einstein tensor are given by
\begin{eqnarray}
  {}^{(1)}\!G_{A}^{\;\;B}
  \!\!\!\!&=&\!\!\!\!
      0
      ,
     \label{eq:linearized-Einstein-tensor-AB-result-l=1-odd}
\end{eqnarray}
\begin{eqnarray}
  {}^{(1)}\!G_{A}^{\;\;q}
  \!\!\!\!&=&\!\!\!\!
      \frac{1}{2r}
      \left[
      \bar{D}_{A}\bar{D}^{C}\tilde{h}_{(o1)C}
      -  \bar{D}_{C}\bar{D}^{C}\tilde{h}_{(o1)A}
      -  \frac{2}{r} (\bar{D}^{C}r) \bar{D}_{C}\tilde{h}_{(o1)A}
      \right.
      \nonumber\\
  && \quad
     \left.
      + \frac{3}{r} (\bar{D}^{C}r) \bar{D}_{A}\tilde{h}_{(o1)C}
      -  \frac{1}{r} (\bar{D}_{A}r) \bar{D}^{C}\tilde{h}_{(o1)C}
      + \frac{1}{2r^{2}} (\bar{D}_{C}r) (\bar{D}^{C}r) \tilde{h}_{(o1)A}
      \right.
      \nonumber\\
  && \quad
     \left.
      -  \frac{2}{r^{2}} (\bar{D}_{A}r)  (\bar{D}^{C}r) \tilde{h}_{(o1)C}
      + \frac{3}{2r^{2}} \tilde{h}_{(o1)A}
      \right]
      \epsilon^{qr}\hat{D}_{r}S
      \label{eq:linearized-Einstein-tensor-Aq-result-l=1-odd}
     ,
\end{eqnarray}
\begin{eqnarray}
  {}^{(1)}\!G_{p}^{\;\;q}
  \!\!\!\!&=&\!\!\!\!
      - \frac{1}{2r^{2}} \bar{D}^{C}\left( r \tilde{h}_{(o1)C} \right)
      \gamma^{qr}
      2
      \epsilon_{s(p}\hat{D}_{r)}\hat{D}^{s}S
  =
      0
      .
      \label{eq:linearized-Einstein-tensor-pq-result-l=1-odd}
\end{eqnarray}
The final equality in
Eq.~(\ref{eq:linearized-Einstein-tensor-pq-result-l=1-odd}) is due to
the property of the spherical harmonics $Y_{lm}$
(\ref{eq:identically-vanishi-mode-function-l=1-odd}).
Furthermore, the expression
(\ref{eq:linearized-Einstein-tensor-Aq-result-l=1-odd}) is gauge
invariant under the gauge transformation rule
(\ref{eq:gauge-transformation-rule-o1A-Fourier-l=1}).
On the other hand, the $\bar{D}^{C}\left( r \tilde{h}_{(o1)C} \right)$
in Eq.~(\ref{eq:linearized-Einstein-tensor-pq-result-l=1-odd}) is not
gauge invariant as shown in below.
However, the gauge-invariance of the linear-order Einstein tensor
${}^{(1)}\!G_{a}^{\;\;b}$ is guaranteed by the identity
$2\epsilon_{s(p}\hat{D}_{r)}\hat{D}^{s}S=0$ for the $l=1$ odd-mode
perturbations.


For the $l=1$ odd-mode perturbations, the components of the linearized
energy-momentum tensor are summarized as
\begin{eqnarray}
  {}^{(1)}\!{\cal T}_{A}^{\;\;\;B}
  =
      0
      ,
  \quad
  {}^{(1)}\!{\cal T}_{A}^{\;\;q}
  =
      \frac{1}{r}
      \tilde{T}_{(o1)A} \epsilon^{qr}\hat{D}_{r}S
      ,
  \quad
  {}^{(1)}\!{\cal T}_{p}^{\;\;q}
  =
      0
     .
     \label{eq:1st-pert-calTab-du-decomp-2-pq-l=1-odd}
\end{eqnarray}
Then, the linearized Einstein equations for the odd-mode perturbations
are given by
\begin{eqnarray}
  &&
     \bar{D}_{A}\bar{D}^{C}\tilde{h}_{(o1)C}
     -  \bar{D}_{C}\bar{D}^{C}\tilde{h}_{(o1)A}
      \nonumber\\
  && 
      -  \frac{2}{r} (\bar{D}^{C}r) \bar{D}_{C}\tilde{h}_{(o1)A}
      + \frac{3}{r} (\bar{D}^{C}r) \bar{D}_{A}\tilde{h}_{(o1)C}
      -  \frac{1}{r} (\bar{D}_{A}r) \bar{D}^{C}\tilde{h}_{(o1)C}
      \nonumber\\
  && 
     + \frac{1}{2r^{2}} (\bar{D}_{C}r) (\bar{D}^{C}r) \tilde{h}_{(o1)A}
     -  \frac{2}{r^{2}} (\bar{D}_{A}r)  (\bar{D}^{C}r) \tilde{h}_{(o1)C}
     + \frac{3}{2r^{2}} \tilde{h}_{(o1)A}
     \nonumber\\
  &=&
      16\pi \tilde{T}_{(o1)A}
      ,
      \label{eq:linearized-Einstein-eq-Aq-result-l=1-odd-2}
\end{eqnarray}


Although the equation
\begin{eqnarray}
  \label{eq:linearized-Einstein-tensor-pq-result-l=1-odd-mod}
 \bar{D}^{C}\left( r \tilde{h}_{(o1)C} \right) = 0
\end{eqnarray}
does not appear from
Eq.~(\ref{eq:linearized-Einstein-tensor-pq-result-l=1-odd}) due to the
fact that $2\epsilon_{s(p}\hat{D}_{r)}\hat{D}^{s}Y_{1m}=0$,
we may use
Eq.~(\ref{eq:linearized-Einstein-tensor-pq-result-l=1-odd-mod}) as a
gauge condition.
As noted in
Sec.~\ref{sec:Conventional-gauge-transformation_rules_l=1}, the
gauge-transformation rule for the variable $\tilde{h}_{(o1)C}$ is
given by Eq.~(\ref{eq:gauge-transformation-rule-o1A-Fourier-l=1}).
From the gauge transformation rule
(\ref{eq:gauge-transformation-rule-o1A-Fourier-l=1}), we consider the
gauge-transformation rule for the left-hand side of
Eq.~(\ref{eq:linearized-Einstein-tensor-pq-result-l=1-odd-mod}) as
\begin{eqnarray}
  \bar{D}^{A} \left(r {}_{{\scrY}}\!\tilde{h}_{(o1)A}\right)
  -
  \bar{D}^{A} \left(r {}_{{\scrX}}\!\tilde{h}_{(o1)A}\right)
  &=&
      r \bar{D}^{A}\bar{D}_{A}\zeta_{(o)}
      - (\bar{D}^{A}\bar{D}_{A}r) \zeta_{(o)}
      .
\end{eqnarray}
Using the background Einstein equation (Eq.~(B67) of Appendix B in
Ref.~\cite{K.Nakamura-2021-PartI}), we consider the equation
\begin{eqnarray}
  \label{eq:gauge-fix-1A-Fourier-l=1-div-r-eq}
  r \bar{D}^{C}\bar{D}_{C}\zeta_{(o)}
  - \frac{1}{r} \left(
  1 - (\bar{D}^{C}r)(\bar{D}_{C}r)
  \right) \zeta_{(o)}
  =
  - \bar{D}^{C} \left(r {}_{{\scrX}}\!\tilde{h}_{(o1)C}\right)
\end{eqnarray}
More explicitly, Eq.~(\ref{eq:gauge-fix-1A-Fourier-l=1-div-r-eq})
\begin{eqnarray}
  \label{eq:gauge-fix-1A-Fourier-l=1-div-r-eq-exp}
  -    \partial_{t}^{2}\zeta_{(o)}
  + f \partial_{r}\left(f\partial_{r}\zeta_{(o)}\right)
  - \frac{f}{r^{2}} \left(
  1 - f
  \right) \zeta_{(o)}
  =
  - \frac{f}{r} \bar{D}^{C} \left(r {}_{{\scrX}}\!\tilde{h}_{(o1)C}\right)
\end{eqnarray}
If we choose $\zeta_{(o)}$ as a special solution to
Eq.~(\ref{eq:gauge-fix-1A-Fourier-l=1-div-r-eq-exp}) or equivalently
Eq.~(\ref{eq:gauge-fix-1A-Fourier-l=1-div-r-eq}),
Eq.~(\ref{eq:linearized-Einstein-tensor-pq-result-l=1-odd-mod}) is
regarded as a gauge condition in the $\scrY$-gauge, i.e.,
\begin{eqnarray}
  \bar{D}^{C} \left(r {}_{{\scrY}}\!\tilde{h}_{(o1)C}\right)
  &=&
      0
      .
  \label{eq:gauge-fix-condition-o1A-Fourier-l=1}
\end{eqnarray}


We have to emphasize that this is not a complete gauge fixing, since
there is a room of the choice of $\zeta_{(o)}$ which satisfy the
homogeneous equation of
Eq.~(\ref{eq:gauge-fix-1A-Fourier-l=1-div-r-eq}):
\begin{eqnarray}
  \label{eq:gauge-fix-1A-Fourier-l=1-div-r-eq-hom}
  r \bar{D}^{C}\bar{D}_{C}\zeta_{(o)}
  - \frac{1}{r} \left(
  1 - (\bar{D}^{C}r)(\bar{D}_{C}r)
  \right) \zeta_{(o)}
  =
  0
  ,
\end{eqnarray}
i.e.,
\begin{eqnarray}
  \label{eq:gauge-fix-1A-Fourier-l=1-div-r-eq-hom-exp}
  -    \partial_{t}^{2}\zeta_{(o)}
  + f \partial_{r}\left(f\partial_{r}\zeta_{(o)}\right)
  - \frac{f}{r^{2}} \left(
  1 - f
  \right) \zeta_{(o)}
  =
  0
  .
\end{eqnarray}


Under the gauge choice (\ref{eq:gauge-fix-condition-o1A-Fourier-l=1}),
Eq.~(\ref{eq:linearized-Einstein-eq-Aq-result-l=1-odd-2}) is given by
\begin{eqnarray}
&&
  \left[
  \bar{D}_{C}\bar{D}^{C}
  - \frac{2}{r^{2}}
  \right] \left(r \tilde{h}_{(o1)A}\right)
  -  \frac{2}{r^{2}} (\bar{D}_{A}r) (\bar{D}^{C}r) \left(r \tilde{h}_{(o1)C}\right)
  + \frac{2}{r} (\bar{D}^{C}r) \bar{D}_{A}\left(r \tilde{h}_{(o1)C}\right)
   \nonumber\\
  &=&
  16\pi r \tilde{T}_{(o1)A}
  .
  \label{eq:linearized-Einstein-eq-Aq-redued-wgauefix-l=1-odd}
\end{eqnarray}
Although the variable $r\tilde{h}_{(o1)C}$ is a gauge-dependent
variable which is different from the gauge-invariant variable
$r\tilde{F}_{A}$, we can obtain the equation
(\ref{eq:linearized-Einstein-eq-Aq-redued-wgauefix-l=1-odd}) if we
replace the variable $r\tilde{F}_{A}$ in
Eq.~(\ref{eq:1st-p-Ein-non-vac-Aq-odd-sum-2-red-l=1}) with the
variable $r\tilde{h}_{(o1)C}$.


We also note that the gauge condition
(\ref{eq:gauge-fix-condition-o1A-Fourier-l=1}) coincides with
Eq.~(\ref{eq:1st-p-Ein-non-vac-pq-traceless-odd-l=1}), though we have
to replace the gauge-invariant variable $r \tilde{F}^{D}$ with the
gauge-dependent variable $r \tilde{h}_{(o1)C}$ in
Eq.~(\ref{eq:1st-p-Ein-non-vac-pq-traceless-odd-l=1}) to confirm this
coincidence.
Furthermore, we also take into account the continuity equation
(\ref{eq:div-barTab-linear-p-odd-l=1}) of the $l=1$ odd-mode
perturbation of the energy-momentum tensor.
Thus, the equations to be solved for the $l=1$ odd-mode perturbation
are the gauge condition
(\ref{eq:linearized-Einstein-tensor-pq-result-l=1-odd-mod}), the
evolution equation
(\ref{eq:linearized-Einstein-eq-Aq-redued-wgauefix-l=1-odd}), and the
continuity equation (\ref{eq:div-barTab-linear-p-odd-l=1}).
These equations coincide with
Eqs.~(\ref{eq:1st-p-Ein-non-vac-pq-traceless-odd-l=1})--(\ref{eq:div-barTab-linear-p-odd-l=1})
except for the fact that the variable to be obtained is not the
gauge-invariant $r\tilde{F}^{D}$ variable but the gauge-dependent
variable $r \tilde{h}_{(o1)C}$.
Then, through the same logic in Ref.~\cite{K.Nakamura-2021-PartI}, we
obtain the solution to
the gauge condition (\ref{eq:gauge-fix-condition-o1A-Fourier-l=1}),
the evolution equation
(\ref{eq:linearized-Einstein-eq-Aq-redued-wgauefix-l=1-odd}), and the
continuity equation (\ref{eq:div-barTab-linear-p-odd-l=1}) as follows:
\begin{eqnarray}
  &&
     2 r \tilde{h}_{(o1)C} \sin^{2}\theta (dx^{C})_{(a} (d\phi)_{b)}
     =
     6 M r^{2} \left[\int dr\frac{a_{1}(t,r)}{r^{4}}\right]
     \sin^{2}\theta (dt)_{(a}(d\phi)_{b)}
     \nonumber\\
  && \quad\quad\quad\quad\quad\quad\quad\quad\quad\quad\quad\quad\quad\quad
     + {\pounds}_{V}g_{ab},
     \label{eq:K.Nakamura-2021-PartI-6.45-gaue-fix}
     , \\
  &&
     V_{a} = (\beta_{1}t + \beta_{0} + W_{(o)}(t,r)) r^{2} \sin^{2}\theta (d\phi)_{a},
     \label{eq:K.Nakamura-2021-PartI-6.46-gaue-fix}
\end{eqnarray}
Here, we concentrate only on the $m=0$ solution, $\beta_{1}$ and
$\beta_{0}$ are constant, and $Z_{(o)}=rf\partial_{r}W_{(o)}(t,r)$ is
an arbitrary function which satisfies the equation
\begin{eqnarray}
  \label{eq:Regge-Wheeler-Zo-eq-gaue-fix}
  \partial_{t}^{2}Z_{(o)}
  -
  f\partial_{r}(f\partial_{r}Z_{(o)})
  +
  \frac{1}{r^{2}} f \left[
  2 - 3 (1-f)
  \right]
  Z_{(o)}
  =
  16\pi f^{2} \tilde{T}_{(o1)r}
  .
\end{eqnarray}
$a_{1}(t,r)$ in Eq.~(\ref{eq:K.Nakamura-2021-PartI-6.45-gaue-fix}) is
given by Eq.~(6.44) in Ref.~\cite{K.Nakamura-2021-PartI}, i.e.,
\begin{eqnarray}
  \label{eq:K.Nakamura-2021-PartI-6.44-gaue-fix}
  a_{1}(t,r)
  &=&
      - \frac{16\pi}{3M} r^{3} f \int dt \tilde{T}_{(o1)r} + a_{10}
      \nonumber\\
  &=&
      - \frac{16\pi}{3M} \int dr r^{3} \frac{1}{f} \tilde{T}_{(o1)t} + a_{10}
      ,
\end{eqnarray}
where $a_{10}$ is the perturbative Kerr parameter.


Note that the formal solution described by
Eqs.~(\ref{eq:K.Nakamura-2021-PartI-6.45-gaue-fix})--(\ref{eq:K.Nakamura-2021-PartI-6.44-gaue-fix})
has the same form as the formal solution described by
Eqs.~(\ref{eq:l=1-odd-mode-propagating-sol-ver2-2})--(\ref{eq:odd-master-equation-Regge-Wheeler}).
However, we have to emphasize that the variable $\scrF_{Ap}$ in
Eq.~(\ref{eq:l=1-odd-mode-propagating-sol-ver2-2}) is gauge invariant
as noted by Eq.~(\ref{eq:l=1-hab-fourier-delta=0-rewrite}) and the
vector field $V_{a}$ in
Eq.~(\ref{eq:l=1-odd-mode-propagating-sol-ver2-2}) is also gauge
invariant.
On the other hand, the variable $\tilde{h}_{(o1)C}$ is still
gauge-dependent.
Actually, there are remaining gauge degrees of freedom of the
generator $\zeta_{(o)}$ which satisfy
Eq.~(\ref{eq:gauge-fix-1A-Fourier-l=1-div-r-eq-hom-exp}) as noted
above.
This is clear from the fact that the gauge transformation rule
(\ref{eq:gauge-transformation-rule-o1A-Fourier-l=1}) with
Eq.~(\ref{eq:gauge-fix-1A-Fourier-l=1-div-r-eq-hom-exp}) gives a still
non-trivial transformation rule.
This remaining gauge degree of freedom is so-called ``residual gauge''.
For this reason, there is a possibility that $V_{a}$ in
Eq.~(\ref{eq:K.Nakamura-2021-PartI-6.45-gaue-fix}) includes this
residual gauge.


To clarify whether $V_{a}$ in
Eq.~(\ref{eq:K.Nakamura-2021-PartI-6.45-gaue-fix}) includes the
``residual gauge'', we have to confirm
Eq.~(\ref{eq:gauge-fix-1A-Fourier-l=1-div-r-eq-hom-exp}).
Within our rules to compare our gauge-invariant formulation with a
conventional gauge-fixed approach, we regard the terms in $V_{a}$ in
Eq.~(\ref{eq:K.Nakamura-2021-PartI-6.45-gaue-fix}) which
satisfies Eq.~(\ref{eq:gauge-fix-1A-Fourier-l=1-div-r-eq-hom-exp}) is
the second-kind gauge degree of freedom and we regard these degrees of
freedom as ``unphysical degree of freedom.''
To clarify this ``unphysical degree of freedom,'' we introduce the
indicator function $\FrakR_{(o)}[*]$ as
\begin{eqnarray}
  \label{eq:SecondKindIndicator-l=1-odd}
  \FrakR_{(o)}[\zeta_{(o)}]
  :=
  -    \partial_{t}^{2}\zeta_{(o)}
  + f \partial_{r}\left(f\partial_{r}\zeta_{(o)}\right)
  - \frac{f}{r^{2}} \left(
  1 - f
  \right) \zeta_{(o)}
  .
\end{eqnarray}
If $\FrakR_{(o)}[\zeta_{(o)}]=0$, we should regard $\zeta_{(o)}$ is
the second-kind gauge and we regard $\zeta_{(o)}$ as ``unphysical
degree of freedom.''


We can easily confirm that
\begin{eqnarray}
  \label{eq:SecondKindInjigator-l=1-odd-rigid-rotation}
  \FrakR_{(o)}[(\beta_{1}t+\beta_{0})r] = 0
  .
\end{eqnarray}
Then, according to our above rule of comparison, we have to conclude
that the rigid rotating term $(\beta_{1}t+\beta_{0})r$ in $V_{a}$ in
Eq.~(\ref{eq:K.Nakamura-2021-PartI-6.46-gaue-fix}) should be regarded
as the second-kind gauge degree of freedom which is ``unphysical degree
of freedom.''


Next, we consider the term $W_{(o)}(t,r)r$ in
Eq.~(\ref{eq:K.Nakamura-2021-PartI-6.46-gaue-fix}).
In this case, the direct calculation of the indicator
$\FrakR_{(o)}[W_{(o)}(t,r)r]$ is useless.
However, it is useful to consider the $r$-derivative of
$\FrakR_{(o)}[W_{(o)}(t,r)r]$ and we can show that
\begin{eqnarray}
  r f \partial_{r}\left(
  \frac{1}{r} \FrakR_{(o)}[W_{(o)}(t,r)r]
  \right)
  \!\!\!\!&=&\!\!\!\!
      -  \partial_{t}^{2}Z_{(o)}
      + f \partial_{r}(f \partial_{r}Z_{(o)})
      -  \frac{f}{r^{2}} \left( 2 - 3 (1-f) \right) Z_{(o)}
      ,
      \label{eq:rfderivativerSecondKindInjigator-l=1-odd}
\end{eqnarray}
where we used $Z_{(o)}=rf\partial_{r}W_{(o)}(t,r)$.
Since the left-hand side of
Eq.~(\ref{eq:rfderivativerSecondKindInjigator-l=1-odd}) coincides with
the right-hand side of Eq.~(\ref{eq:Regge-Wheeler-Zo-eq-gaue-fix}).
Through Eq.~(\ref{eq:Regge-Wheeler-Zo-eq-gaue-fix}), we obtain
\begin{eqnarray}
  r f \partial_{r}\left(
  \frac{1}{r} \FrakR_{(o)}[W_{(o)}(t,r)r]
  \right)
  &=&
      - 16 \pi f^{2} \tilde{T}_{(o1)r}
      .
      \label{eq:rfderivativerSecondKindInjigator-l=1-odd-withRWheeler-eq}
\end{eqnarray}
or, equivalently,
\begin{eqnarray}
  \label{eq:rfpartialr1overrcalE-To1r-int}
  \FrakR_{(o)}[W_{(o)}(t,r)r]
  &=&
      - 16 \pi r \int dr \frac{f}{r} \tilde{T}_{(o1)r}
     .
\end{eqnarray}
Then, if we consider the case where $\tilde{T}_{(o1)r} \neq 0$, we
have nonvanishing $\FrakR_{(o)}[W_{(o)}(t,r)r]$.
This means that the $W_{(o)}(t,r)$ term does not belong to the
second-kind gauge, and we have to regard the degree of freedom
$W_{(o)}(t,r)$ in Eq.~(\ref{eq:K.Nakamura-2021-PartI-6.46-gaue-fix})
is a ``physical one''.


On the other hand, even in the case where $\tilde{T}_{(o1)r}=0$,
Eq.~(\ref{eq:rfderivativerSecondKindInjigator-l=1-odd}) is valid.
If $W_{(o)}(t,r)$ belongs to the second-kind gauge, i.e.,
$\FrakR_{(o)}[W_{(o)}(t,r)r]=0$,
Eq.~(\ref{eq:rfderivativerSecondKindInjigator-l=1-odd}) yields
\begin{eqnarray}
  \label{eq:RW-homogeneous-equation}
  -  \partial_{t}^{2}Z_{(o)}
  + f \partial_{r}(f \partial_{r}Z_{(o)})
  -  \frac{f}{r^{2}} \left( 2 - 3 (1-f) \right) Z_{(o)}
  =
  0
  .
\end{eqnarray}
However, even if $Z_{(o)}$ is a solution to
Eq.~(\ref{eq:RW-homogeneous-equation}), we cannot directly yield
$\FrakR_{(o)}[W_{(o)}(t,r)r]=0$.
Therefore, considering the following two sets of function
$W_{(o)}(t,r)r$ as
\begin{eqnarray}
  \label{eq:WorIsGauge}
  \FrakG_{(o)}
  \!\!\!\!&:=&\!\!\!\!
       \left\{
       \left.
       r W_{(o)}(t,r)
       \right|
       \FrakR_{(o)}[W_{(o)}(t,r)r] = 0
       \right\}
       ,
  \\
  \label{eq:RW-homogeneous-sol}
  \FrakH_{(o)}
  \!\!\!\!&:=&\!\!\!\!
       \left\{
       \left.
       r W_{(o)}(t,r)
       \right|
       Z_{(o)} = rf \partial_{r}W_{(o)}(t,r),
       \right.
       \nonumber\\
  && \quad\quad\quad\quad\quad\quad
       \left.
       - \partial_{t}^{2}Z_{(o)}
       + f \partial_{r}(f \partial_{r}Z_{(o)})
       -
       \frac{f}{r^{2}} \left( 2 - 3(1-f) \right) Z_{(o)}
       = 0
       .
       \right\}
       ,
\end{eqnarray}
we obtain the relation
\begin{eqnarray}
  \label{eq:subset-Zo-homogeneous-calE=0}
  \FrakG_{(o)} \subset \FrakH_{(o)}.
\end{eqnarray}
This indicates that a part of solutions to
Eq.~(\ref{eq:RW-homogeneous-equation}) should be regarded
as a gauge degree of freedom of the second kind which is the
``unphysical degree of freedom.''


Furthermore, to obtain the explicit solution $W_{(o)}(t,r)$, we have
to solve Eq.~(\ref{eq:Regge-Wheeler-Zo-eq-gaue-fix}) with
appropriate boundary conditions.
Equation~(\ref{eq:Regge-Wheeler-Zo-eq-gaue-fix}) is an inhomogeneous
second-order linear differential equation for $Z_{(o)}$ and its boundary conditions
are adjusted by the homogeneous solutions to
Eq.~(\ref{eq:Regge-Wheeler-Zo-eq-gaue-fix}), i.e., the element of the
set of function $\FrakH_{(o)}$.
However, a part of this homogeneous solution to
Eq.~(\ref{eq:Regge-Wheeler-Zo-eq-gaue-fix}) should be regarded as
an unphysical degree of freedom in the ``complete gauge fixing approach''
as mentioned above.
Therefore, in the conventional ``complete gauge fixing approach'', the
boundary conditions for Eq.~(\ref{eq:Regge-Wheeler-Zo-eq-gaue-fix}) is
restricted.
In this sense, a conventional ``complete gauge-fixing approach''
includes a stronger restriction than our proposed gauge-invariant
formulation.


\section{l=1 even-mode perturbation in the conventional approach}
\label{sec:l=1-even-ConventionalGaugeFixing}


As discussed in
Sec.~\ref{sec:Conventional-gauge-transformation_rules_l=1}, $l=1$ mode
perturbations are given by
Eqs.~(\ref{eq:K.Nakamura-2021-PartI-3.7-l=1})--(\ref{eq:K.Nakamura-2021-PartI-3.9-l=1}).
In particular, among these expressions of the $l=1$ mode
perturbations, even-mode perturbation is given by
\begin{eqnarray}
  \label{eq:K.Nakamura-2021-PartI-3.7-l=1-even-2}
  h_{AB}
  =
      \tilde{h}_{AB} S
      ,
  \quad
  h_{Ap}
  =
      r \tilde{h}_{(e1)A} \hat{D}_{p}S
      ,
  \quad
  h_{pq}
  =
      \frac{r^{2}}{2} \gamma_{pq} \tilde{h}_{(e0)} S
      .
\end{eqnarray}
As the property of the $l=1$ spherical harmonics, we obtain the
condition (\ref{eq:identically-vanishi-mode-function-l=1}).
For even-mode perturbation, we have
\begin{eqnarray}
  \label{eq:identically-vanishi-mode-function-l=1-even}
  \left(
  \hat{D}_{p}\hat{D}_{q} - \frac{1}{2} \gamma_{pq} \hat{D}_{r}\hat{D}^{r}
  \right) Y_{1m}
  =
  0
  .
\end{eqnarray}


The gauge-transformation rules for the $l=1$ even-mode perturbations
are given by Eqs.~(\ref{eq:gauge-transformation-rule-AB-Fourier-l=1}),
(\ref{eq:gauge-transformation-rule-e1A-Fourier-l=1}), and
(\ref{eq:gauge-transformation-rule-e0-Fourier-l=1}).
Inspecting the gauge-transformation rules
(\ref{eq:gauge-transformation-rule-e1A-Fourier-l=1}) and
(\ref{eq:gauge-transformation-rule-e0-Fourier-l=1}), we define the
variable $\tilde{H}$ by
\begin{eqnarray}
  \tilde{H}
  &:=&
       \tilde{h}_{(e0)} - 4 (\bar{D}^{A}r) \tilde{h}_{(e1)A}
      .
  \label{eq:partial-gauge-inv-tildeHe0-def-l=1-even-sum}
\end{eqnarray}
The gauge-transformation rule of $\tilde{H}$ is given by
\begin{eqnarray}
  {}_{{\scrY}}\!\tilde{H}
  -
  {}_{{\scrX}}\!\tilde{H}
  &=&
      - 4 (\bar{D}^{A}r) \bar{D}_{A}\zeta_{(e)}
      -  \frac{4}{r} \left( 1 - (\bar{D}^{A}r) (\bar{D}_{A}r) \right) \zeta_{(e)}
      .
  \label{eq:gauge-trans-tildeHe0-l=1-even-sum}
\end{eqnarray}
Furthermore, inspecting gauge-transformation rules
(\ref{eq:gauge-transformation-rule-AB-Fourier-l=1}) and
(\ref{eq:gauge-transformation-rule-e1A-Fourier-l=1}) we also define
the variable $\tilde{H}_{AB}$ by
\begin{eqnarray}
  \label{eq:partial-gauge-inv-tildeHAB-def-l=1-even-sum}
  \tilde{H}_{AB}
  &:=&
       \tilde{h}_{AB}
       -
       \bar{D}_{A} \left( r \tilde{h}_{(e1)B} \right)
       -
       \bar{D}_{B} \left( r \tilde{h}_{(e1)A} \right)
       .
\end{eqnarray}
The gauge-transformation of $\tilde{H}_{AB}$ is given by
\begin{eqnarray}
  {}_{{\scrY}}\!\tilde{H}_{AB}
  -
  {}_{{\scrX}}\!\tilde{H}_{AB}
  &=&
      - 2 r \bar{D}_{A}\bar{D}_{B}\zeta_{(e)}
      + 2 (\bar{D}_{A}\bar{D}_{B}r) \zeta_{(e)}
      .
      \label{eq:gauge-trans-tildeHAB-l=1-even-sum}
\end{eqnarray}


The definitions (\ref{eq:partial-gauge-inv-tildeHe0-def-l=1-even-sum})
and (\ref{eq:partial-gauge-inv-tildeHAB-def-l=1-even-sum}) of the
variables $\tilde{H}$ and $\tilde{H}_{AB}$, respectively, are
analogous to the gauge-invariant variable $\tilde{F}$ and
$\tilde{F}_{AB}$ define by Eqs.~(\ref{eq:gauge-inv-tildeFAB-def-sum})
and (\ref{eq:2+2-gauge-inv-tildeF-def-sum}) for $l\geq 2$ modes,
respectively.
However, the variables $\tilde{H}$ and $\tilde{H}_{AB}$ are not gauge
invariant.
The employment of the variables $\tilde{H}$ and $\tilde{H}_{AB}$
corresponds to the gauge fixing of the gauge degree of freedom of the
generator $\zeta_{A}$ and the specification of the component
$\tilde{h}_{(e1)A}$ of the metric perturbation.
On the other hand, the gauge-transformation rules
(\ref{eq:gauge-transformation-rule-AB-Fourier-l=1}),
(\ref{eq:gauge-transformation-rule-e1A-Fourier-l=1}), and
(\ref{eq:gauge-transformation-rule-e0-Fourier-l=1}) includes the gauge
degree of freedom of the generator $\zeta_{(e)}$.
This gauge degree of freedom appears in the gauge-transformation rules
(\ref{eq:gauge-trans-tildeHe0-l=1-even-sum}) and
(\ref{eq:gauge-trans-tildeHAB-l=1-even-sum}).


The linearized Einstein tensor for $l=1$ even modes in terms of
$\tilde{H}_{AB}$ and $\tilde{H}$ are given as
\begin{eqnarray}
  y_{CB}{}^{(1)}\!G_{A}^{\;\;C}/S
  &=&
      \left[
      -  \frac{1}{2} \bar{D}_{C}\bar{D}^{C}
      + \frac{3}{r^{2}}
      -  \frac{1}{r} (\bar{D}^{C}r) \bar{D}_{C}
      \right] \tilde{H}_{AB}
      + \bar{D}_{(A}\bar{D}^{C}\tilde{H}_{B)C}
      -  \frac{1}{2} \bar{D}_{A}\bar{D}_{B}\tilde{H}_{C}^{\;\;C}
     \nonumber\\
  &&
     +  \frac{2}{r} (\bar{D}^{C}r) \left[
     \bar{D}_{(A}\tilde{H}_{B)C}
     -  \frac{1}{r} (\bar{D}_{C}r) \tilde{H}_{AB}
     \right]
     -  \frac{1}{2} \bar{D}_{A}\bar{D}_{B}\tilde{H}
     -  \frac{1}{r} (\bar{D}_{(A}r) \bar{D}_{B)}\tilde{H}
     \nonumber\\
  &&
     + \frac{1}{2} y_{AB} \left[
     \left[
     \bar{D}_{C}\bar{D}^{C}
     + \frac{3}{r} (\bar{D}^{C}r) \bar{D}_{C}
     \right] \tilde{H}
     +
     \left\{
     \bar{D}_{D}\bar{D}^{D}
     -  \frac{5}{r^{2}}
     \right\} \tilde{H}_{C}^{\;\;C}
     \right.
     \nonumber\\
  && \quad\quad\quad\quad
     \left.
     + \frac{1}{r} (\bar{D}^{C}r) \left\{
     2 \bar{D}_{C}\tilde{H}_{D}^{\;\;D}
     + \frac{3}{r} (\bar{D}_{C}r) \tilde{H}_{D}^{\;\;D}
     \right\}
     -  \bar{D}^{C}\bar{D}^{D}\tilde{H}_{CD}
     \right.
     \nonumber\\
  && \quad\quad\quad\quad
     \left.
     - \frac{2}{r} (\bar{D}^{D}r) \left\{
     2 \bar{D}^{C}\tilde{H}_{CD}
     + \frac{1}{r} (\bar{D}^{C}r) \tilde{H}_{CD}
     \right\}
     \right]
     ,
     \label{eq:linearized-Einstein-tensor-AB-l=1-partial-GI-result-sum}
\end{eqnarray}
\begin{eqnarray}
  {}^{(1)}\!G_{A}^{\;\;q}
  &=&
      \left[
      \frac{1}{2r^{2}} \bar{D}^{C}\tilde{H}_{AC}
      -  \frac{1}{2r^{2}} \bar{D}_{A}\tilde{H}_{C}^{\;\;C}
      + \frac{1}{2r^{3}} (\bar{D}_{A}r) \tilde{H}_{C}^{\;\;C}
     -  \frac{1}{4r^{2}} \bar{D}_{A}\tilde{H}
     \right] \hat{D}^{q}S
     ,
     \label{eq:linearized-Einstein-tensor-Aq-l=1-partial-GI-result-sum}
\end{eqnarray}
\begin{eqnarray}
  \gamma_{qr}{}^{(1)}\!G_{p}^{\;\;r}
  &=&
      \left[
      + \frac{1}{4} \bar{D}_{C}\bar{D}^{C}\tilde{H}
      + \frac{1}{2r} (\bar{D}^{C}r) \bar{D}_{C}\tilde{H}
     + \frac{1}{2} \left\{
     + \bar{D}_{C}\bar{D}^{C}
     + \frac{1}{r} (\bar{D}^{C}r) \bar{D}_{C}
     -  \frac{1}{r^{2}}
     \right\} \tilde{H}_{D}^{\;\;D}
     \right.
     \nonumber\\
  && \quad
     \left.
     - \frac{1}{2} \left\{
     \bar{D}_{C}
     + \frac{2}{r} (\bar{D}_{C}r)
     \right\} \bar{D}_{D}\tilde{H}^{CD}
      \right] \gamma_{pq} S
      \nonumber\\
  &&
     -  \frac{1}{2r^{2}} \tilde{H}_{C}^{\;\;C}
     \left(
     \hat{D}_{p}\hat{D}_{q}S
     - \frac{1}{2} \gamma_{pq} \hat{D}^{r}\hat{D}_{r} S
     \right)
     .
      \label{eq:linearized-Einstein-tensor-pq-l=1-partial-GI-result-sum}
\end{eqnarray}
Since there is the identity
(\ref{eq:identically-vanishi-mode-function-l=1-even}) for $l=1$ mode
perturbations, the last term of
Eq.~(\ref{eq:linearized-Einstein-tensor-pq-l=1-partial-GI-result-sum})
does not appear for $l=1$ even modes.
Furthermore, the expressions
(\ref{eq:linearized-Einstein-tensor-AB-l=1-partial-GI-result-sum}) and
(\ref{eq:linearized-Einstein-tensor-Aq-l=1-partial-GI-result-sum}) are
gauge invariant under the gauge-transformation rules
(\ref{eq:gauge-trans-tildeHe0-l=1-even-sum}) and
(\ref{eq:gauge-trans-tildeHAB-l=1-even-sum}).
Even in
Eq.~(\ref{eq:linearized-Einstein-tensor-pq-l=1-partial-GI-result-sum})
the component $\gamma_{qr}{}^{(1)}\!G_{p}^{\;\;r}$ is gauge invariant
except for the last term in
Eq.~(\ref{eq:linearized-Einstein-tensor-pq-l=1-partial-GI-result-sum}).
Actually, the variable $\tilde{H}_{C}^{\;\;C}$ is gauge dependent as
shown below.
However, the gauge-invariance $\gamma_{qr}{}^{(1)}\!G_{p}^{\;\;r}$ is
guaranteed by the identity
(\ref{eq:identically-vanishi-mode-function-l=1-even}).


On the other hand, the $l=1$ even components of them are given by
\begin{eqnarray}
  {}^{(1)}\!{\cal T}_{A}^{\;\;\;B}
  &=&
      S
      \tilde{T}_{A}^{\;\;\;B}
      ,
      \label{eq:1st-pert-calTab-du-decomp-2-AB-l=1-even}
\end{eqnarray}
\begin{eqnarray}
  \gamma_{qr} {}^{(1)}\!{\cal T}_{A}^{\;\;r}
  &=&
      \frac{1}{r}
      \tilde{T}_{(e1)A} \hat{D}_{q}S
      ,
      \label{eq:1st-pert-calTab-du-decomp-2-Aq-l=1-even}
\end{eqnarray}
\begin{eqnarray}
  \gamma_{qr}{}^{(1)}\!{\cal T}_{p}^{\;\;r}
  &=&
     \tilde{T}_{(e0)} \frac{1}{2} \gamma_{pq} S
     +
     \tilde{T}_{(e2)} \left(
     \hat{D}_{p}\hat{D}_{q}S
     -
     \frac{1}{2} \gamma_{pq} \hat{D}_{r}\hat{D}^{r}S
     \right)
     \label{eq:1st-pert-calTab-du-decomp-2-pq-l=1-even}
\end{eqnarray}
Here again, we note that
Eq.~(\ref{eq:identically-vanishi-mode-function-l=1-even}) satisfies as
a mathematical identity for $l=1$ mode perturbations.
Therefore, the last term of
Eq.~(\ref{eq:1st-pert-calTab-du-decomp-2-pq-l=1-even}) does not appear
for $l=1$ even modes.


If equation~(\ref{eq:identically-vanishi-mode-function-l=1-even}) is
not satisfied, we have the equation
\begin{eqnarray}
     \left[
     -  \frac{1}{2r^{2}} \tilde{H}_{C}^{\;\;C}
     \right]
     \left(
     \hat{D}_{p}\hat{D}_{q}S
     - \frac{1}{2} \gamma_{pq} \hat{D}^{r}\hat{D}_{r} S
     \right)
  =
    8\pi
     \tilde{T}_{(e2)} \left(
     \hat{D}_{p}\hat{D}_{q}S
     -
     \frac{1}{2} \gamma_{pq} \hat{D}_{r}\hat{D}^{r}S
     \right)
\end{eqnarray}
as one of the components of the linearized Einstein equation.
However, Eq.~(\ref{eq:identically-vanishi-mode-function-l=1-even})
implies that this equation is identically satisfied.
Therefore, this equation does not restrict $\tilde{H}_{C}^{\;\;C}$ nor
$\tilde{T}_{(e2)}$ and there is no other restriction of them.
Here, we note that the tensor field $\tilde{H}_{AB}$ is not
gauge-invariant as shown in
Eq.~(\ref{eq:gauge-trans-tildeHAB-l=1-even-sum}).
Since we may freely choose $\tilde{H}_{C}^{\;\;C}$ and
$\tilde{T}_{(e2)}$, we choose $\tilde{T}_{(e2)}=0$ by hand and choose
\begin{eqnarray}
  \label{eq:HCC-vanish-gauge-condition}
  H_{C}^{\;\;C} = 0.
\end{eqnarray}
as a gauge condition.


Actually, from the gauge-transformation rule
(\ref{eq:gauge-trans-tildeHAB-l=1-even-sum}), we obtain
\begin{eqnarray}
  y^{AB}{}_{{\scrY}}\!\tilde{H}_{AB}
  -
  y^{AB}{}_{{\scrX}}\!\tilde{H}_{AB}
  &=&
      - 2 r \left[
      -  \frac{1}{f} \partial_{t}^{2}\zeta_{(e)}
      + \partial_{r}( f \partial_{r}\zeta_{(e)})
      - \frac{1-f}{r^{2}} \zeta_{(e)}
      \right]
      .
      \label{eq:gauge-trans-tildeHAB-l=1-even-sum-trace}
\end{eqnarray}
Then, if we choose $\zeta_{(e)}$ as a special solution to the equation
\begin{eqnarray}
  \frac{1}{2r} y^{AB}{}_{{\scrX}}\!\tilde{H}_{AB}
  &=&
      -  \frac{1}{f} \partial_{t}^{2}\zeta_{(e)}
      + \partial_{r}( f \partial_{r}\zeta_{(e)})
      - \frac{1-f}{r^{2}} \zeta_{(e)}
      ,
      \label{eq:gauge-trans-tildeHAB-l=1-even-trace-gauge-fix}
\end{eqnarray}
we may regard that ${}_{{\scrY}}\!\tilde{H}_{C}^{C}=0$ in the
$\scrY$-gauge.
We have to note that this gauge fixing
(\ref{eq:gauge-trans-tildeHAB-l=1-even-sum-trace}) is not complete
gauge fixing.
There is a remaining degree of freedom in the choice of $\zeta_{(e)}$
as the homogeneous solution $\zeta_{(e)h}$ to the wave equation
\begin{eqnarray}
  -  \frac{1}{f} \partial_{t}^{2}\zeta_{(e)h}
  + \partial_{r}( f \partial_{r}\zeta_{(e)h})
  - \frac{1-f}{r^{2}} \zeta_{(e)h}
  =
  0
  .
  \label{eq:gauge-trans-tildeHAB-l=1-even-residual-gauge}
\end{eqnarray}


Due to the gauge condition (\ref{eq:HCC-vanish-gauge-condition}),
the components
(\ref{eq:linearized-Einstein-tensor-AB-l=1-partial-GI-result-sum})--(\ref{eq:linearized-Einstein-tensor-pq-l=1-partial-GI-result-sum})
of the linearized Einstein tensor and components
(\ref{eq:1st-pert-calTab-du-decomp-2-AB-l=1-even})--(\ref{eq:1st-pert-calTab-du-decomp-2-pq-l=1-even})
of the linear-order energy-momentum tensor yield the linearized
Einstein equations.
The $(A,p)$-components (equivalently, $(p,B)$-components) of the
linearized Einstein equations are given by
\begin{eqnarray}
  \bar{D}^{C}\tilde{H}_{AC}
  -  \frac{1}{2} \bar{D}_{A}\tilde{H}
  =
  16 \pi r
  \tilde{T}_{(e1)A}
  .
  \label{eq:linearized-Einstein-Eq-Aq-l=1-partial-GI-traceGF-2}
\end{eqnarray}
Since the tensor $\tilde{H}_{AC}$ is traceless due to the gauge
condition (\ref{eq:HCC-vanish-gauge-condition}), equation
(\ref{eq:linearized-Einstein-Eq-Aq-l=1-partial-GI-traceGF-2})
coincides with Eq.~(\ref{eq:even-FAB-divergence-3}).


From the trace part of the $(p,q)$-component of the linearized
Einstein equation and
Eq.~(\ref{eq:linearized-Einstein-Eq-Aq-l=1-partial-GI-traceGF-2})
yields the component (\ref{eq:div-barTab-linear-p-even}) of the
continuity equation of the linearized energy-momentum tensor.


The trace part of $(A,B)$-component of the linearized Einstein
equation with
Eq.~(\ref{eq:linearized-Einstein-Eq-Aq-l=1-partial-GI-traceGF-2})
gives
\begin{eqnarray}
     \bar{D}_{C}\bar{D}^{C}\tilde{H}
     + \frac{2}{r} (\bar{D}_{C}r) \bar{D}^{C}\tilde{H}
     -  \frac{4}{r^{2}} (\bar{D}^{C}r) (\bar{D}^{D}r) \tilde{H}_{CD}
  =
      16 \pi
      \left(
      \tilde{T}_{C}^{\;\;C}
      +  4 (\bar{D}^{D}r) \tilde{T}_{(e1)D}
      \right)
      .
      \label{eq:linearized-Einstein-Eq-AB-l=1-partial-GI-traceGF-trace-mod}
\end{eqnarray}
Since $\tilde{H}_{CD}$ is a traceless tensor,
Eq.~(\ref{eq:linearized-Einstein-Eq-AB-l=1-partial-GI-traceGF-trace-mod})
coincides with the $l=1$ version of
Eq.~(\ref{eq:even-mode-tildeF-master-eq-mod-3}) with the source term
(\ref{eq:sourcee0-def}).
However, we have to emphasize that $\tilde{H}$ nor $\tilde{H}_{CD}$
are not gauge invariant as shown above.


Finally, the traceless part of $(A,B)$-component of the linearized
Einstein equation with
Eqs.~(\ref{eq:linearized-Einstein-Eq-Aq-l=1-partial-GI-traceGF-2}) and
(\ref{eq:linearized-Einstein-Eq-AB-l=1-partial-GI-traceGF-trace-mod})
yields
\begin{eqnarray}
  &&
     \left[
      -  \bar{D}_{C}\bar{D}^{C}
      -  \frac{2}{r} (\bar{D}^{C}r) \bar{D}_{C}
      + \frac{4}{r} (\bar{D}^{C}\bar{D}_{C}r)
      + \frac{2}{r^{2}}
     \right] \tilde{H}_{AB}
     + \frac{4}{r} (\bar{D}^{C}r) \bar{D}_{(A}\tilde{H}_{B)C}
     \nonumber\\
  &&
     -  \frac{2}{r} (\bar{D}_{(A}r) \bar{D}_{B)}\tilde{H}
     \nonumber\\
  &=&
      16 \pi
      \left[
      \tilde{T}_{AB}
      -
      \frac{1}{2} y_{AB} \tilde{T}_{C}^{\;\;C}
     -  2 \left(
      \bar{D}_{(A}(r \tilde{T}_{(e1)B)})
      - \frac{1}{2} y_{AB} \bar{D}^{C}(r \tilde{T}_{(e1)C})
      \right)
      \right.
      \nonumber\\
  && \quad\quad\quad
     \left.
     + 2 y_{AB} (\bar{D}^{D}r) \tilde{T}_{(e1)D}
      \right]
      ,
      \label{eq:linearized-Einstein-Eq-AB-l=1-partial-GI-traceGF-traceless}
\end{eqnarray}
Since the variable $H_{AB}$ is traceless due to the gauge condition
(\ref{eq:HCC-vanish-gauge-condition}),
Eq.~(\ref{eq:linearized-Einstein-Eq-AB-l=1-partial-GI-traceGF-traceless})
coincides with the $l=1$ version of
Eq.~(\ref{eq:1st-pert-Einstein-non-vac-AB-traceless-final-3}) with the
source term (\ref{eq:souce(FF)-def}).


In addition to these linearized Einstein equations, the following
linearized perturbations of the continuity equation should be
satisfied
\begin{eqnarray}
  &&
     \bar{D}^{C}\tilde{T}_{C}^{\;\;B}
     + \frac{2}{r} (\bar{D}^{D}r)\tilde{T}_{D}^{\;\;\;B}
     -  \frac{2}{r} \tilde{T}_{(e1)}^{B}
     -  \frac{1}{r} (\bar{D}^{B}r) \tilde{T}_{(e0)}
     =
     0
     ,
     \label{eq:div-barTab-linear-vac-back-u-A-mode-dec-sum-l=1}
  \\
  &&
     \bar{D}^{C}\tilde{T}_{(e1)C}
     + \frac{3}{r} (\bar{D}^{C}r) \tilde{T}_{(e1)C}
     + \frac{1}{2r} \tilde{T}_{(e0)}
     =
     0
     .
     \label{eq:div-barTab-linear-vac-back-u-p-mode-dec-even-sum-l=1}
\end{eqnarray}


Together with the gauge condition
(\ref{eq:HCC-vanish-gauge-condition}),
Eqs.~(\ref{eq:linearized-Einstein-Eq-Aq-l=1-partial-GI-traceGF-2})--(\ref{eq:div-barTab-linear-vac-back-u-p-mode-dec-even-sum-l=1})
coincide with
Eqs.~(\ref{eq:linearized-Einstein-pq-traceless-even})--(\ref{eq:div-barTab-linear-p-even})
with $l=1$ in our gauge-invariant formulation developed in Refs.~\cite{K.Nakamura-2021-CQG,K.Nakamura-2021-LHEP,K.Nakamura-2021-PartI,K.Nakamura-2021-PartII,K.Nakamura-2021-PartIII}.
In Ref.~\cite{K.Nakamura-2021-PartII}, we derived the formal solution to
Eqs.~(\ref{eq:linearized-Einstein-pq-traceless-even})--(\ref{eq:div-barTab-linear-p-even})
with $l=1$ as Eq.~(\ref{eq:tildeFab-l=1-m=0-nonvacsum-2-cov}) with
Eqs.~(\ref{eq:generator-covariant-vacuum-l=1-m=0-result-2}),
(\ref{eq:Zerilli-Moncrief-eq-final-l=1}), and $m=0$.
Due to the coincidence of the set of equations, this formal solution
should be the $m=0$ formal solution to
Eqs.~(\ref{eq:HCC-vanish-gauge-condition}), and Eqs.~(\ref{eq:linearized-Einstein-Eq-Aq-l=1-partial-GI-traceGF-2})--(\ref{eq:div-barTab-linear-vac-back-u-p-mode-dec-even-sum-l=1}).
Then, as the $m=0$ solution, we obtain
\begin{eqnarray}
  \scrH_{ab}
  &:=&
       H_{AB} \cos\theta (dx^{A})_{(a}(dx^{B})_{b)} + \frac{r^{2}}{2} H \cos\theta  \gamma_{ab}
       \nonumber\\
  &=&
      {\pounds}_{V}g_{ab}
      - \frac{16\pi r^{2}}{3(1-f)} \left[
      f^{2} \left\{
      \frac{1+f}{2} \tilde{T}_{rr} + r f \partial_{r}\tilde{T}_{rr} -
      \tilde{T}_{(e0)} - 4 \tilde{T}_{(e1)r}
      \right\} (dt)_{a} (dt)_{b}
      \right.
      \nonumber\\
  && \quad\quad\quad\quad\quad\quad\quad\quad\quad
     \left.
      + \frac{2r}{f} \left\{
      \partial_{t}\tilde{T}_{tt} - \frac{3f(1-f)}{2r} \tilde{T}_{tr}
      \right\} (dt)_{(a}(dr)_{b)}
      \right.
      \nonumber\\
  && \quad\quad\quad\quad\quad\quad\quad\quad\quad
     \left.
      + \frac{r}{f} \left\{
      \partial_{r}\tilde{T}_{tt} - \frac{3(1-3f)}{2rf} \tilde{T}_{tt}
      \right\} (dr)_{a} (dr)_{b})
      \right.
      \nonumber\\
  && \quad\quad\quad\quad\quad\quad\quad\quad\quad
     \left.
      + r^{2} \tilde{T}_{tt} \gamma_{ab}
      \right] \cos\theta
     ,
  \label{eq:l=1-general-solution-gauge-fix}
\end{eqnarray}
where $\gamma_{ab} = \gamma_{pq}(dx^{p})_{a}(dx^{q})_{b}$,
$V_{a}=g_{ab}V^{b}$ is the vector field defined by
\begin{eqnarray}
  \label{eq:l=1-general-solution-generator-gauge-fix}
  V_{a} := - r \partial_{t}\Phi_{(e)} \cos\theta (dt)_{a} +
  (\Phi_{(e)} - r \partial_{r}\Phi_{(e)}) \cos (dr)_{a} - r \Phi_{(e)}
  \sin\theta (d\theta)_{a}
\end{eqnarray}
and $\Phi_{(e)}$ is a solution to the equation
\begin{eqnarray}
  \label{eq:l=1-master-equation-general-gauge-fix}
  &&
  - \frac{1}{f} \partial_{t}^{2}\Phi_{(e)} + \partial_{r}\left[f \partial_{r}\Phi_{(e)}\right]
  - \frac{1-f}{r^{2}} \Phi_{(e)}
  =
  \frac{16\pi r}{3(1-f)} S_{(\Phi_{(e)})},
  \\
  &&
     S_{(\Phi_{(e)})} = \frac{1}{2} r \partial_{t} T_{tr} - \frac{1}{2}
     r \partial_{r}\tilde{T}_{tt} + \frac{1-4f}{2f} \tilde{T}_{tt}
     - \frac{f}{2} \tilde{T}_{rr} - f \tilde{T}_{(e1)r}
     .
     \label{eq:l=1-master-equation-general-source}
\end{eqnarray}


As in the case of the $l=1$ odd-mode perturbation, the tensor field
$\scrF_{ab}$ in Eq.~(\ref{eq:tildeFab-l=1-m=0-nonvacsum-2-cov}) is
gauge invariant.
Then, $V_{a}$ in Eq.~(\ref{eq:tildeFab-l=1-m=0-nonvacsum-2-cov}),
which defined by
Eq.~(\ref{eq:generator-covariant-vacuum-l=1-m=0-result-2}) is also
gauge invariant.
Therefore, we regarded $V_{a}$ in
Eq.~(\ref{eq:tildeFab-l=1-m=0-nonvacsum-2-cov}) as the first kind gauge.
On the other hand, the tensor field $\scrH_{ab}$ in
Eq.~(\ref{eq:l=1-general-solution-gauge-fix}) is gauge-dependent in
the sense of the second kind due to the gauge-transformation rules
(\ref{eq:gauge-trans-tildeHe0-l=1-even-sum}) and
(\ref{eq:gauge-trans-tildeHAB-l=1-even-sum}) for the components
$\tilde{H}$ and $\tilde{H}_{AB}$, respectively.
Actually, there is a remaining gauge degree of freedom of the generator
$\zeta_{(e)}$ which satisfies
Eq.~(\ref{eq:gauge-trans-tildeHAB-l=1-even-residual-gauge}) as noted
above.
This remaining gauge degree of freedom is so-called ``residual
gauge''.
For this reason, there is a possibility that $V_{a}$ in
Eq.~(\ref{eq:l=1-general-solution-gauge-fix}) includes this ``residual
gauge'', while $V_{a}$ in
Eq.~(\ref{eq:tildeFab-l=1-m=0-nonvacsum-2-cov}) is gauge invariant.


Since we already fixed the gauge-degree of freedom $\zeta_{(e)}$ so that
Eq.~(\ref{eq:HCC-vanish-gauge-condition}) is satisfied, the
remaining gauge degree of freedom $\zeta_{(e)h}$ must satisfy the
equation (\ref{eq:gauge-trans-tildeHAB-l=1-even-residual-gauge}).
Therefore, to clarify whether $V_{a}$ in
Eq.~(\ref{eq:l=1-general-solution-gauge-fix}) includes the ``residual
gauge,'' we have to confirm
Eq.~(\ref{eq:gauge-trans-tildeHAB-l=1-even-residual-gauge}).
Within our rules to compare our gauge-invariant formulation with a
conventional gauge-fixed approach, we regard the term in $V_{a}$ in
Eq.~(\ref{eq:l=1-general-solution-gauge-fix}) satisfies
Eq.~(\ref{eq:gauge-trans-tildeHAB-l=1-even-residual-gauge}) is the
second-kind gauge degree of freedom and we regard this degree of
freedom as an ``unphysical degree of freedom.''
As in the case of the $l=1$ odd-mode perturbations, we introduce the
indicator function $\FrakR_{(e)}[*]$ as
\begin{eqnarray}
  \FrakR_{(e)}[\zeta_{(e)h}]
  :=
  -  \frac{1}{f} \partial_{t}^{2}\zeta_{(e)h}
  + \partial_{r}( f \partial_{r}\zeta_{(e)h})
  - \frac{1-f}{r^{2}} \zeta_{(e)h}
  .
  \label{eq:SecondKindIndicator-l=1-even}
\end{eqnarray}
If $\FrakR_{(e)}[\zeta_{(e)}]=0$, we should regard $\zeta_{(e)}$ is
the second-kind gauge degree of freedom and we regard $\zeta_{(e)}$ as
an ``unphysical degree of freedom.''


To clarify the second-kind gauge degree of freedom, we consider the
gauge-transformation rule of the tensor field $\scrH_{ab}$ through the
gauge-transformation rules
(\ref{eq:gauge-trans-tildeHe0-l=1-even-sum}) and
(\ref{eq:gauge-trans-tildeHAB-l=1-even-sum}) as
\begin{eqnarray}
  &&
  {}_{{\scrY}}\scrH_{ab}
  -
  {}_{{\scrX}}\scrH_{ab}
     \nonumber\\
  &:=&
       \left(
       {}_{\scrY}\!H_{AB}
       -
       {}_{\scrX}\!H_{AB}
       \right) \cos\theta (dx^{A})_{(a}(dx^{B})_{b)}
       + \frac{r^{2}}{2} \left(
       {}_{\scrY}\!H
       -
       {}_{\scrX}\!H
       \right) \cos\theta \gamma_{ab}
       \nonumber
\end{eqnarray}
\begin{eqnarray}
  &=&
      - 2 r \left[
      \partial_{t}^{2}\zeta_{(e)h}
      -  \frac{f(1-f)}{2r} \partial_{r}\zeta_{(e)h}
      +  \frac{f(1-f)}{2r^{2}} \zeta_{(e)h}
      \right] \cos\theta
      (dt)_{a}(dt)_{b}
      \nonumber\\
  &&
      - 4 r \left[
      \partial_{t}\partial_{r}\zeta_{(e)h}
      -  \frac{1-f}{2rf} \partial_{t}\zeta_{(e)h}
      \right] \cos\theta
     (dt)_{(a}(dr)_{b)}
     \nonumber\\
  &&
      - 2 r \left[
      \partial_{r}^{2}\zeta_{(e)h}
      + \frac{1-f}{2rf} \partial_{r}\zeta_{(e)h}
      -  \frac{1-f}{2r^{2}f} \zeta_{(e)h}
      \right] \cos\theta
     (dr)_{a}(dr)_{b}
      \nonumber\\
  &&
     - 2 r^{2} \left(
     f \partial_{r}\zeta_{(e)h}
     + \frac{1}{r} (1-f) \zeta_{(e)h}
     \right) \cos\theta \gamma_{ab}
     \label{eq:l=1-gauge-trans-scrHab-explicit}
  \\
  &=&
      {\pounds}_{W}g_{ab}
      ,
     \label{eq:l=1-gauge-trans-scrHab-explicit-sum}
\end{eqnarray}
where
\begin{eqnarray}
  W_{a}
  &=&
      - r \partial_{t}\zeta_{(e)h} \cos\theta (dt)_{a}
      + \left( \zeta_{(e)h} - r \partial_{r}\zeta_{(e)h} \right) \cos\theta(dr)_{a}
      - r \zeta_{(e)h} \sin\theta (d\theta)_{a}
      .
      \label{eq:l=1-gauge-trans-scrHab-Lie-expression-zetaeh}
\end{eqnarray}
Comparing Eq.~(\ref{eq:l=1-gauge-trans-scrHab-explicit-sum}) with the
generator (\ref{eq:l=1-gauge-trans-scrHab-Lie-expression-zetaeh}) and
Eq.~(\ref{eq:l=1-general-solution-gauge-fix}) with the generator
(\ref{eq:l=1-general-solution-generator-gauge-fix}), there is the
possibility that the Lie derivative term ${\pounds}_{V}g_{ab}$ in the
solution (\ref{eq:l=1-general-solution-gauge-fix}) is ``residual gauge
degree of freedom'' with the identification
\begin{eqnarray}
  \label{eq:Phie-zetaeh-identification}
  \Phi_{(e)} = \zeta_{(e)h}.
\end{eqnarray}
For this reason, we check the indicator
(\ref{eq:SecondKindIndicator-l=1-even}).
From Eq.~(\ref{eq:l=1-master-equation-general-gauge-fix}), we obtain
\begin{eqnarray}
  \FrakR_{(e)}[\Phi_{(e)}]
  :=
  \frac{16\pi r}{3(1-f)} S_{(\Phi_{(e)})}
  .
  \label{eq:SecondKindIndicator-l=1-even-Phie}
\end{eqnarray}
Therefore, the variable $\Phi_{(e)}$ should be regarded as the
second-kind gauge degree of freedom and is the ``unphysical degree of
freedom'' if $S_{(\Phi_{(e)})}=0$, while in the non-vacuum case
$S_{\Phi_{(e)}}\neq 0$, the indicator
(\ref{eq:SecondKindIndicator-l=1-even-Phie}) yields
$\FrakR_{(e)}[\Phi_{(e)}]\neq 0$.
This means that $\Phi_{(e)}$ is not the second-kind gauge degree of
freedom but is the ``physical degree of freedom'' in the non-vacuum
case $S_{\Phi_{(e)}}\neq 0$.
Due to this existence of the source term, the identification
(\ref{eq:Phie-zetaeh-identification}) is impossible in the case of the
non-vacuum situation.
Rather, this term is gauge-invariant in the sense of the second kind,
and we regard this term as a gauge-degree of freedom of the first
kind in the non-vacuum case.


Although we have $\FrakR_{(e)}[\Phi_{(e)}]= 0$ in the case where
$S_{\Phi_{(e)}}=0$ and we should regard $\Phi_{(e)}$ is a ``gauge
degree of freedom of the second kind'',
Eq.~(\ref{eq:SecondKindIndicator-l=1-even-Phie}) indicates that
\begin{eqnarray}
  \FrakR_{(e)}[\Phi_{(e)}]
  =
  -  \frac{1}{f} \partial_{t}^{2}\Phi_{(e)}
  + \partial_{r}( f \partial_{r}\Phi_{(e)})
  - \frac{1-f}{r^{2}} \Phi_{(e)}
  =
  0
  .
  \label{eq:SecondKindIndicator-l=1-even-Phie-explicit}
\end{eqnarray}
This coincides with the left-hand side of
Eq.~(\ref{eq:l=1-master-equation-general-gauge-fix}).
Therefore, as in the case of the $l=1$ odd-mode perturbations, we
consider the following two sets of function $\Phi_{(e)}$ as
\begin{eqnarray}
  \label{eq:PhieGauge}
  \FrakG_{(e)}
  &:=&
       \left\{
       \left.
       \Phi_{(e)}
       \right|
       \FrakR_{(o)}[\Phi_{(e)}] = 0
       \right\}
       ,
  \\
  \label{eq:Zerilli-homogeneous-sol}
  \FrakH_{(e)}
  &:=&
       \left\{
       \left.
       \Phi_{(e)}
       \right|
       -  \frac{1}{f} \partial_{t}^{2}\Phi_{(e)}
       + \partial_{r}( f \partial_{r}\Phi_{(e)})
       - \frac{1-f}{r^{2}} \Phi_{(e)}
       =
       0
       \right\}
       .
\end{eqnarray}
From Eq.~(\ref{eq:SecondKindIndicator-l=1-even-Phie-explicit}), we
obtain the relation
\begin{eqnarray}
  \label{eq:equal-Phie-homogeneous-calE=0}
  \FrakG_{(e)} = \FrakH_{(e)}.
\end{eqnarray}
This indicates that any homogeneous solution (without source term) to
Eq.~(\ref{eq:l=1-master-equation-general-gauge-fix}) should be
regarded as a gauge degree of freedom of the second kind which is
the ``unphysical degree of freedom.''


On the other hand, to obtain the explicit solution $\Phi_{(e)}$ in the
case $S_{\Phi_{(e)}}\neq 0$, which is regarded as a ``physical degree
of freedom,'' we have to solve
Eq.~(\ref{eq:l=1-master-equation-general-gauge-fix}) with appropriate
boundary conditions.
Equation~(\ref{eq:l=1-master-equation-general-gauge-fix}) is an inhomogeneous
second-order linear differential equation for $\Phi_{(e)}$ and its boundary conditions
are adjusted by the homogeneous solutions to
Eq.~(\ref{eq:l=1-master-equation-general-gauge-fix}), i.e., the
element of the set of function $\FrakH_{(e)}$.
However, any homogeneous solution to
Eq.~(\ref{eq:l=1-master-equation-general-gauge-fix}) should be
regarded as an ``unphysical degree of freedom'' in the ``complete
gauge fixing approach'' as mentioned above.
Therefore, in the conventional ``complete gauge fixing approach'', we
have to impose the boundary conditions for
Eq.~(\ref{eq:Regge-Wheeler-Zo-eq-gaue-fix}) using a homogeneous
solution, which is regarded as an ``unphysical degree of freedom,'' to
obtain a ``physical solution'' to
Eq.~(\ref{eq:Regge-Wheeler-Zo-eq-gaue-fix}) with nonvanishing source
term $S_{\Phi_{(e)}}\neq 0$.
This situation is a dilemma.
In this sense, as in the case of $l=1$ odd-mode perturbations, a
conventional ``complete gauge-fixing approach'' includes a stronger
restriction than our proposed gauge-invariant formulation.


\section{l=0 mode perturbation in the conventional approach}
\label{sec:l=0ConventionalGaugeFixing}


\subsection{Einstein equations for $l=0$ mode perturbations}
\label{sec:l=0Einstein_equations}


Here, we consider the $l=0$ mode perturbations which are described by
Eqs.~(\ref{eq:K.Nakamura-2021-PartI-3.7-l=0})--(\ref{eq:K.Nakamura-2021-PartI-3.9-l=0}).
Substitution of
Eqs.~(\ref{eq:K.Nakamura-2021-PartI-3.7-l=0})--(\ref{eq:K.Nakamura-2021-PartI-3.9-l=0})
into
Eqs.~(\ref{eq:linearized-Einstein-tensor-AB-result-sum})--(\ref{eq:linearized-Einstein-tensor-pq-result-sum}),
we obtain the $l=0$ mode perturbations of the linearized Einstein
tensor.
We use the background Einstein equations (Eq.~(B67) and (B68) of
Appendix B in the Part I paper~\cite{K.Nakamura-2021-PartI}).
Then, we obtain the non-trivial components of the $l=0$ mode Einstein
equation ${}^{(1)}\!G_{a}^{\;\;b}=8\pi {}^{(1)}\!T_{a}^{\;\;b}$ are  given by
\begin{eqnarray}
  &&
      -  \frac{1}{2} \bar{D}_{C}\bar{D}^{C}\tilde{h}_{A}^{\;\;B}
      + \frac{1}{2} \bar{D}^{B}\bar{D}^{C}\tilde{h}_{AC}
      + \frac{1}{2} \bar{D}_{A}\bar{D}_{C}\tilde{h}^{BC}
      -  \frac{1}{2} \bar{D}_{A}\bar{D}^{B}\tilde{h}_{C}^{\;\;C}
      \nonumber\\
  &&
     -  \frac{1}{r} (\bar{D}^{C}r) \bar{D}_{C}\tilde{h}_{A}^{\;\;B}
     + \frac{1}{r} (\bar{D}^{C}r) \bar{D}^{B}\tilde{h}_{AC}
     +  \frac{1}{r} (\bar{D}_{C}r) \bar{D}_{A}\tilde{h}^{BC}
     -  \frac{2}{r^{2}} (\bar{D}^{C}r) (\bar{D}_{C}r) \tilde{h}_{A}^{\;\;B}
     \nonumber\\
  &&
     -  \frac{1}{2r} (\bar{D}^{B}r) \bar{D}_{A}\tilde{h}_{(e0)}
     -  \frac{1}{2r} (\bar{D}_{A}r) \bar{D}^{B}\tilde{h}_{(e0)}
     -  \frac{1}{2} \bar{D}_{A}\bar{D}^{B}\tilde{h}_{(e0)}
     + \frac{2}{r^{2}} \tilde{h}_{A}^{\;\;B}
     \nonumber\\
  &&
     + y_{A}^{\;\;B} \left(
      -  \frac{1}{2} \bar{D}_{C}\bar{D}_{D}\tilde{h}^{CD}
     + \frac{1}{2} \bar{D}_{C}\bar{D}^{C}\tilde{h}_{D}^{\;\;D}
     -  \frac{2}{r} (\bar{D}_{D}r) \bar{D}_{C}\tilde{h}^{CD}
     + \frac{1}{r} (\bar{D}^{C}r) \bar{D}_{C}\tilde{h}_{D}^{\;\;D}
     \right.
     \nonumber\\
  && \quad\quad\quad\quad
     \left.
     -  \frac{1}{r^{2}} (\bar{D}_{C}r) (\bar{D}_{D}r) \tilde{h}^{CD}
     + \frac{3}{2r^{2}} (\bar{D}^{D}r) (\bar{D}_{D}r) \tilde{h}_{C}^{\;\;C}
     -  \frac{3}{2r^{2}} \tilde{h}_{C}^{\;\;C}
     \right.
     \nonumber\\
  && \quad\quad\quad\quad
     \left.
     + \frac{1}{2r^{2}} \tilde{h}_{(e0)}
     + \frac{3}{2r} (\bar{D}^{C}r) \bar{D}_{C}\tilde{h}_{(e0)}
     + \frac{1}{2} \bar{D}_{C}\bar{D}^{C}\tilde{h}_{(e0)}
     \right)
     \nonumber\\
     &=&
         8 \pi \tilde{T}_{A}^{\;\;\;B}
         ,
     \label{eq:linearized-Einstein-eq-AB-l=0-sum}
  \\
  &&
     \bar{D}_{C}\bar{D}^{C}\tilde{h}_{D}^{\;\;D}
     -  \bar{D}_{C}\bar{D}_{D}\tilde{h}^{CD}
     -  \frac{2}{r} (\bar{D}_{C}r) \bar{D}_{D}\tilde{h}^{CD}
     + \frac{1}{r} (\bar{D}^{C}r) \bar{D}_{C}\tilde{h}_{D}^{\;\;D}
     \nonumber\\
  &&
     + \frac{1}{2} \bar{D}_{C}\bar{D}^{C}\tilde{h}_{(e0)}
     + \frac{1}{r} (\bar{D}^{C}r) \bar{D}_{C}\tilde{h}_{(e0)}
     \nonumber\\
  &=&
     8 \pi \tilde{T}_{(e0)}
     .
      \label{eq:linearized-Einstein-eq-pq-l=0-sum}
\end{eqnarray}


Here, we decomposition of the component $\tilde{h}_{AB}$ as
\begin{eqnarray}
  \label{eq:decomposition-trace-traceless-hAB}
  \tilde{h}_{AB} =: \tilde{\HF}_{AB} + \frac{1}{2} y_{AB} \tilde{h}_{C}^{\;\;C}.
\end{eqnarray}
In terms of the variables defined by
Eq.~(\ref{eq:decomposition-trace-traceless-hAB}), the linearization of
the Einstein equations for $l=0$ mode are summarized as follows.
The trace of ${}^{(1)}\!G_{A}^{\;\;B}=8\pi {}^{(1)}\!T_{A}^{\;\;B}$
for $l=0$ mode is given by
\begin{eqnarray}
  &&
     -  \frac{1}{r^{2}} \tilde{h}_{C}^{\;\;C}
     - \frac{2}{r} (\bar{D}_{C}r) \left(
     + \bar{D}_{D}\tilde{\HF}^{DC}
     + \frac{1}{r} (\bar{D}_{D}r)  \tilde{\HF}^{CD}
     \right)
     \nonumber\\
  &&
     + \frac{1}{2} \bar{D}_{C}\bar{D}^{C}\tilde{h}_{(e0)}
     + \frac{2}{r} (\bar{D}_{C}r) \bar{D}^{C}\tilde{h}_{(e0)}
     + \frac{1}{r^{2}} \tilde{h}_{(e0)}
     =
      8 \pi \tilde{T}_{C}^{\;\;\;C}
      .
      \label{eq:linearized-Einstein-eq-AB-l=0-trace-sum}
\end{eqnarray}
The traceless part of ${}^{(1)}\!G_{A}^{\;\;B}=8\pi
{}^{(1)}\!T_{A}^{\;\;B}$ for $l=0$ mode is given by
\begin{eqnarray}
  &&
     - \frac{1}{2} \left[
     \bar{D}_{C}\bar{D}^{C}
     + \frac{2}{r} (\bar{D}^{C}r) \bar{D}_{C}
     + \frac{4}{r^{2}} (\bar{D}^{C}r) (\bar{D}_{C}r)
     - \frac{4}{r^{2}}
     \right] \tilde{\HF}_{AB}
     \nonumber\\
  &&
     + \bar{D}_{(A}\bar{D}^{C}\tilde{\HF}_{B)C}
     -  \frac{1}{2} y_{AB} \bar{D}_{C}\bar{D}_{D}\tilde{\HF}^{CD}
     + \frac{2}{r} (\bar{D}^{C}r) \left(
     \bar{D}_{(A}\tilde{\HF}_{B)C}
     -  \frac{1}{2} y_{AB} \bar{D}^{D}\tilde{\HF}_{DC}
     \right)
     \nonumber\\
  &&
     + \frac{1}{r} \left(
     (\bar{D}_{(A}r) \bar{D}_{B)}\tilde{h}_{E}^{\;\;E}
     -  \frac{1}{2} y_{AB} (\bar{D}^{C}r) \bar{D}_{C}\tilde{h}_{D}^{\;\;D}
     \right)
     - \frac{1}{2} \left[
     \bar{D}_{A}\bar{D}_{B}\tilde{h}_{(e0)}
     - \frac{1}{2} y_{AB} \bar{D}_{C}\bar{D}^{C}\tilde{h}_{(e0)}
     \right]
     \nonumber\\
  &&
     - \frac{1}{r} \left[
     (\bar{D}_{(A}r) \bar{D}_{B)}\tilde{h}_{(e0)}
     -  \frac{1}{2} y_{AB} (\bar{D}^{C}r) \bar{D}_{C}\tilde{h}_{(e0)}
     \right]
     =
      8 \pi \left[
      \tilde{T}_{AB} - \frac{1}{2} y_{AB} \tilde{T}_{C}^{\;\;\;C}
      \right]
      .
      \label{eq:linearized-Einstein-eq-AB-l=0-traceless-sum}
\end{eqnarray}
Furthermore, ${}^{(1)}\!G_{p}^{\;\;q}=8\pi {}^{(1)}\!T_{p}^{\;\;q}$
for $l=0$ mode is given by
\begin{eqnarray}
  &&
      \frac{1}{2} \bar{D}_{C}\bar{D}^{C}\tilde{h}_{(e0)}
      + \frac{1}{r} (\bar{D}^{C}r) \bar{D}_{C}\tilde{h}_{(e0)}
     - \bar{D}_{C}\bar{D}_{D}\tilde{\HF}^{CD}
     -  \frac{2}{r} (\bar{D}_{C}r) \bar{D}_{D}\tilde{\HF}^{CD}
     + \frac{1}{2} \bar{D}_{C}\bar{D}^{C}\tilde{h}_{D}^{\;\;D}
     \nonumber\\
  &=&
      8 \pi \tilde{T}_{(e0)}
      .
      \label{eq:linearized-Einstein-eq-pq-l=0-trace-traceless-sum}
\end{eqnarray}


\subsection{Gauge-fixing for $l=0$ mode perturbations}
\label{sec:Gauge-fixing-for-l=0-node-perturbation}


We consider the gauge-transformation rules
(\ref{eq:gauge-transformation-rule-AB-Fourier-l=0}) and
(\ref{eq:gauge-transformation-rule-pq-Fourier-l=0}) of the mode
coefficient $\tilde{h}_{AB}$ and $\tilde{h}_{(e0)}$.
If these gauge-transformation rules are those of the second kind, we
should exclude these gauge degrees of freedom through some
gauge-fixing procedure, because the degree of freedom of the
second-kind gauge is the unphysical degree of freedom.


In the static chart in which the metric $y_{AB}$ is given by
Eq.~(\ref{eq:background-metric-2+2-y-comp-Schwarzschild}).
Through the static chart
(\ref{eq:background-metric-2+2-y-comp-Schwarzschild}), the
gauge-transformation rule
(\ref{eq:gauge-transformation-rule-pq-Fourier-l=0}) is given by
\begin{eqnarray}
  \label{eq:gauge-transformation-rule-pq-Fourier-l=0-static}
  {}_{\scrY}\!\tilde{h}_{(e0)}
  -
  {}_{\scrX}\!\tilde{h}_{(e0)}
  =
  \frac{4}{r} f  \zeta_{r}
  .
\end{eqnarray}
Then we may choose $\zeta_{r}$ so that
\begin{eqnarray}
  \label{eq:gauge-fixing-pq-Fourier-l=0-static}
  {}_{\scrY}\!\tilde{h}_{(e0)}
  &=&
      {}_{\scrX}\!\tilde{h}_{(e0)}
      + \frac{4}{r} f  \zeta_{r}
      =
      0
      ,
\end{eqnarray}
i.e.,
\begin{eqnarray}
  \label{eq:gauge-fixing-pq-Fourier-l=0-static-2}
  \zeta_{r}
  =
  - \frac{r}{4f} {}_{\scrX}\!\tilde{h}_{(e0)}
  .
\end{eqnarray}
Through this gauge-fixing, we may regard $\tilde{h}_{(e0)}=0$ in the
$\scrY$-gauge.


As the gauge-fixing for $\tilde{h}_{AB}$, we fix the gauge so that
$\tilde{h}_{AB}$ is traceless.
This gauge-fixing so that $\tilde{h}_{AB}$ is traceless makes easy to
compare with our gauge-invariant expression in Refs.~\cite{K.Nakamura-2021-CQG,K.Nakamura-2021-LHEP,K.Nakamura-2021-PartI,K.Nakamura-2021-PartII,K.Nakamura-2021-PartIII}.
The trace of the gauge transformation rue
Eq.~(\ref{eq:gauge-transformation-rule-AB-Fourier-l=0}), we obtain
\begin{eqnarray}
  \label{eq:gauge-transformation-rule-AB-Fourier-l=0-trace}
  y^{AB}{}_{{\scrY}}\!\tilde{h}_{AB}
  -
  y^{AB}{}_{{\scrX}}\!\tilde{h}_{AB}
  &=&
      {}_{{\scrY}}\!\tilde{h}_{C}^{\;\;C}
      -
      {}_{{\scrX}}\!\tilde{h}_{C}^{\;\;C}
      =
      2
      \bar{D}^{C}\zeta_{C}
      .
\end{eqnarray}
In the static chart
(\ref{eq:background-metric-2+2-y-comp-Schwarzschild}) of the
Schwarzschild spacetime, the gauge transformation rule
(\ref{eq:gauge-transformation-rule-AB-Fourier-l=0-trace}) is given by
\begin{eqnarray}
  {}_{{\scrY}}\!\tilde{h}_{C}^{\;\;C}
  -
  {}_{{\scrX}}\!\tilde{h}_{C}^{\;\;C}
  &=&
      - 2 f^{-1} \partial_{t}\zeta_{t}
      + 2 \partial_{r}(f\zeta_{r})
      .
      \label{eq:gauge-transformation-rule-AB-Fourier-l=0-trace-static}
\end{eqnarray}
From this gauge-transformation rule and the gauge fixing
(\ref{eq:gauge-fixing-pq-Fourier-l=0-static-2}), we may fix the gauge
degree of freedom $\zeta_{t}$ so that
\begin{eqnarray}
  0
  &=&
  {}_{{\scrY}}\!\tilde{h}_{C}^{\;\;C}
      \nonumber\\
  &=&
      {}_{{\scrX}}\!\tilde{h}_{C}^{\;\;C}
      - 2 f^{-1} \partial_{t}\zeta_{t}
      + 2 \partial_{r}\left(
      - \frac{r}{4} {}_{\scrX}\!\tilde{h}_{(e0)}
      \right)
      .
      \label{eq:gauge-transformation-rule-AB-Fourier-l=0-trace-static-2}
\end{eqnarray}
Through this gauge-fixing, we may regard $\tilde{h}_{C}^{\;\;C}=0$.
However, we have to emphasize that the degree of freedom of the choice
of the generator $\zeta_{t}$ remains the degree of freedom of an
arbitrary function of $r$.


Thus, through the gauge-fixing
(\ref{eq:gauge-fixing-pq-Fourier-l=0-static-2}) and
(\ref{eq:gauge-transformation-rule-AB-Fourier-l=0-trace-static-2}), we
may regard
\begin{eqnarray}
  \label{eq:tildehe0=tildehCC=0}
  \tilde{h}_{(e0)} = \tilde{h}_{C}^{\;\;C} = 0
\end{eqnarray}
Under the gauge-fixing condition
(\ref{eq:tildehe0=tildehCC=0}) linearization of the Einstein equations
for $l=0$ mode are summarized as
\begin{eqnarray}
  - (\bar{D}_{C}r) \bar{D}_{D}\tilde{\HF}^{DC}
  -  \frac{1}{r} (\bar{D}_{C}r) (\bar{D}_{D}r)  \tilde{\HF}^{CD}
  =
  4 \pi r \tilde{T}_{C}^{\;\;\;C}
  ,
  \label{eq:linearized-Einstein-eq-AB-l=0-trace-GF}
\end{eqnarray}
\begin{eqnarray}
  &&
     - \frac{1}{2} \left[
     \bar{D}_{C}\bar{D}^{C}
     + \frac{2}{r} (\bar{D}^{C}r) \bar{D}_{C}
     + \frac{4}{r^{2}} (\bar{D}^{C}r) (\bar{D}_{C}r)
     - \frac{4}{r^{2}}
     \right] \tilde{\HF}_{AB}
     + \bar{D}_{(A}\bar{D}^{C}\tilde{\HF}_{B)C}
     \nonumber\\
  &&
     -  \frac{1}{2} y_{AB} \bar{D}_{C}\bar{D}_{D}\tilde{\HF}^{CD}
     + \frac{2}{r} (\bar{D}^{C}r) \left(
     + \bar{D}_{(A}\tilde{\HF}_{B)C}
     -  \frac{1}{2} y_{AB} \bar{D}^{D}\tilde{\HF}_{DC}
     \right)
     \nonumber\\
  &=&
      8 \pi \left[
      \tilde{T}_{AB} - \frac{1}{2} y_{AB} \tilde{T}_{C}^{\;\;\;C}
      \right]
      ,
      \label{eq:linearized-Einstein-eq-AB-l=0-traceless-GF}
\end{eqnarray}
and
\begin{eqnarray}
  - \bar{D}_{C}\bar{D}_{D}\tilde{\HF}^{CD}
  - \frac{2}{r} (\bar{D}_{C}r) \bar{D}_{D}\tilde{\HF}^{CD}
  =
  8 \pi \tilde{T}_{(e0)}
  .
  \label{eq:linearized-Einstein-eq-pq-l=0-trace-traceless-GF}
\end{eqnarray}


\subsection{Component expression of the linearized $l=0$ gauge-fixed
  field equations}
\label{sec:component-expression-gauge-fixed-l=0}


Here, we consider the component representations of
Eqs.~(\ref{eq:linearized-Einstein-eq-AB-l=0-trace-GF})--(\ref{eq:linearized-Einstein-eq-pq-l=0-trace-traceless-GF}).
To do this, we consider the components of the traceless tensor
$\tilde{\HF}_{AB}$ as
\begin{eqnarray}
  \label{eq:l=0-componentHFAB}
  \tilde{\HF}_{AB}
  =:
  X_{(e)} \left[(dt)_{A} (dt)_{B} + f^{-2} (dr)_{A}(dr)_{B}\right]
  +
  2 Y_{(e)} (dt)_{(A} (dr)_{B)}
  .
\end{eqnarray}
Here, we use the background Einstein equation (B65) of Appendix B in
the Part I paper~\cite{K.Nakamura-2021-PartI}, i.e.,
\begin{eqnarray}
  \label{eq:background-Einstein-equation-for-f}
  \partial_{r}f = \frac{1-f}{r}.
\end{eqnarray}
Then, we obtain
\begin{eqnarray}
  \label{eq:background-Einstein-equation-for-f-2}
  \partial_{r}^{2}f
  =
  \partial_{r}\left(\frac{1-f}{r}\right)
  =
  - \frac{2(1-f)}{r^{2}}
\end{eqnarray}


Through Eqs.~(\ref{eq:background-Einstein-equation-for-f}) and
(\ref{eq:background-Einstein-equation-for-f-2}),
Eq.~(\ref{eq:linearized-Einstein-eq-AB-l=0-trace-GF}) is given by
\begin{eqnarray}
  r f \partial_{t}Y_{(e)}
  - r f \partial_{r}X_{(e)}
  -  f X_{(e)}
  =
  - 4 \pi r^{2} \left( \tilde{T}_{tt} - f^{2} \tilde{T}_{rr} \right)
  .
  \label{eq:linearized-Einstein-eq-AB-l=0-trace-GF-component-sum-f}
\end{eqnarray}
The $(A,B)=(t,t)$ and $(A,B)=(r,r)$ components of
Eq.~(\ref{eq:linearized-Einstein-eq-AB-l=0-traceless-GF}) yield the
same equation as
\begin{eqnarray}
  r f \partial_{t}Y_{(e)}
  =
  4 \pi r^{2} \left( \tilde{T}_{tt} + f^{2} \tilde{T}_{rr} \right)
  .
  \label{eq:linearized-Einstein-eq-AB-l=0-traceless-GF-tt-sum-f-2}
\end{eqnarray}
The $(A,B)=(t,r)$ component of
Eq.~(\ref{eq:linearized-Einstein-eq-AB-l=0-traceless-GF}) is given by
\begin{eqnarray}
  \partial_{t}X_{(e)}
  =
  8 \pi r f \tilde{T}_{tr}
  .
  \label{eq:linearized-Einstein-eq-AB-l=0-traceless-GF-tr-sum-f-2}
\end{eqnarray}
In terms of the components (\ref{eq:l=0-componentHFAB}),
Eq.~(\ref{eq:linearized-Einstein-eq-pq-l=0-trace-traceless-GF}) is
given by
\begin{eqnarray}
  -  \partial_{t}^{2}X_{(e)}
  -  f^{2} \partial_{r}^{2}X_{(e)}
  -  \frac{2}{r} f^{2} \partial_{r}X_{(e)}
  + 2 f^{2} \partial_{t}\partial_{r}Y_{(e)}
  + \frac{1}{r} f(1+f) \partial_{t}Y_{(e)}
  =
  8 \pi f^{2} \tilde{T}_{(e0)}
  .
  \label{eq:linearized-Einstein-eq-pq-l=0-trace-traceless-GF-comp-sum-f}
\end{eqnarray}


On the other hand, the even-mode perturbation of the divergence of the
energy-momentum tensor is given by Eqs.~(\ref{eq:div-barTab-linear-A})
and (\ref{eq:div-barTab-linear-A}).
However, Eq.~(\ref{eq:div-barTab-linear-A}) does not appear due to
$\hat{D}_{p}S=\epsilon_{pq}\hat{D}^{q}S=0$.
Then, the non-trivial $l=0$ components of the divergence of the
energy-momentum tensor are given by
\begin{eqnarray}
  \bar{D}^{C}\tilde{T}_{C}^{\;\;B}
  + \frac{2}{r} (\bar{D}^{D}r)\tilde{T}_{D}^{\;\;\;B}
  -  \frac{1}{r} (\bar{D}^{B}r) \tilde{T}_{(e0)}
  =
  0
  .
  \label{eq:div-barTab-linear-vac-back-u-A-mode-dec-l=0}
\end{eqnarray}
The $B=t$ component of
Eq.~(\ref{eq:div-barTab-linear-vac-back-u-A-mode-dec-l=0}) is given by
\begin{eqnarray}
  \partial_{t}\tilde{T}_{tt}
  -  f^{2} \partial_{r}\tilde{T}_{rt}
  - \frac{1}{r} f (1+f) \tilde{T}_{rt}
  =
  0
  .
  \label{eq:div-barTab-linear-vac-back-u-A-mode-dec-l=0-t-sum-f}
\end{eqnarray}
and $B=r$ component of
Eq.~(\ref{eq:div-barTab-linear-vac-back-u-A-mode-dec-l=0}) is given by
\begin{eqnarray}
  -  f \partial_{t}\tilde{T}_{tr}
  + f^{3} \partial_{r}\tilde{T}_{rr}
  + \frac{1}{2r} (1-f) \tilde{T}_{tt}
  + \frac{1}{2r} f^{2} (3+f) \tilde{T}_{rr}
  -  \frac{1}{r} f^{2} \tilde{T}_{(e0)}
  =
  0
  .
  \label{eq:div-barTab-linear-vac-back-u-A-mode-dec-l=0-r-sum-f}
\end{eqnarray}


Substituting
Eq.~(\ref{eq:linearized-Einstein-eq-AB-l=0-traceless-GF-tt-sum-f-2})
into
Eq.~(\ref{eq:linearized-Einstein-eq-AB-l=0-trace-GF-component-sum-f}),
we obtain
\begin{eqnarray}
  \partial_{r}\left(r X_{(e)}\right)
  =
  r \partial_{r}X_{(e)}
  + X_{(e)}
  =
  8 \pi \frac{r^{2}}{f} \tilde{T}_{tt}
  .
  \label{eq:linearized-Einstein-eq-AB-l=0-trace-GF-component-sum-f-2}
\end{eqnarray}
Substituting
Eqs.~(\ref{eq:linearized-Einstein-eq-AB-l=0-traceless-GF-tt-sum-f-2}),
(\ref{eq:linearized-Einstein-eq-AB-l=0-traceless-GF-tr-sum-f-2}), and (\ref{eq:linearized-Einstein-eq-AB-l=0-trace-GF-component-sum-f-2})
into
Eq.~(\ref{eq:linearized-Einstein-eq-pq-l=0-trace-traceless-GF-comp-sum-f}),
we obtain
\begin{eqnarray}
  0
  &=&
      -  f \partial_{t}\tilde{T}_{tr}
      + f^{3} \partial_{r}\tilde{T}_{rr}
      + \frac{1}{2r} (1-f) \tilde{T}_{tt}
      + \frac{1}{2r} f^{2} (3+f) \tilde{T}_{rr}
      -  \frac{1}{r} f^{2} \tilde{T}_{(e0)}
      \nonumber
      .
\end{eqnarray}
This coincides with
Eq.~(\ref{eq:div-barTab-linear-vac-back-u-A-mode-dec-l=0-r-sum-f})
which indicates that
Eq.~(\ref{eq:linearized-Einstein-eq-pq-l=0-trace-traceless-GF-comp-sum-f})
does not give any new information other than
Eq.~(\ref{eq:div-barTab-linear-vac-back-u-A-mode-dec-l=0-r-sum-f}).


Here, we consider the integrability of
Eq.~(\ref{eq:linearized-Einstein-eq-AB-l=0-traceless-GF-tr-sum-f-2})
and
(\ref{eq:linearized-Einstein-eq-AB-l=0-trace-GF-component-sum-f-2}) as follows:
\begin{eqnarray}
  \label{eq:Integrability-of-X_e}
  \partial_{r}(\partial_{t}(rX_{(e)}))
  -
  \partial_{t}(\partial_{r}(rX_{(e)}))
  &=&
      \partial_{r}\left(
      8 \pi r^{2} f \tilde{T}_{tr}
      \right)
      -
      \partial_{t}\left(
      8 \pi \frac{r^{2}}{f} \tilde{T}_{tt}
      \right)
      \nonumber\\
  &=&
      - 8 \pi r^{2} \left[
      + \partial_{t}\tilde{T}_{tt}
      - f^{2} \partial_{r}\tilde{T}_{tr}
      - \frac{1}{r} f (1+f) \tilde{T}_{tr}
      \right]
      \nonumber\\
  &=&
      0.
\end{eqnarray}
Here, we used
Eq.~(\ref{eq:div-barTab-linear-vac-back-u-A-mode-dec-l=0-t-sum-f}) in
the last equality.
This means that the $t$-component
(\ref{eq:div-barTab-linear-vac-back-u-A-mode-dec-l=0-t-sum-f}) of the
continuity equation guarantees the integrability of
Eqs.~(\ref{eq:linearized-Einstein-eq-AB-l=0-traceless-GF-tr-sum-f-2})
and
(\ref{eq:linearized-Einstein-eq-AB-l=0-trace-GF-component-sum-f-2}).
Then, from
Eq.~(\ref{eq:linearized-Einstein-eq-AB-l=0-trace-GF-component-sum-f-2}),
we may write the solution to
Eqs.~(\ref{eq:linearized-Einstein-eq-AB-l=0-traceless-GF-tr-sum-f-2})
and
(\ref{eq:linearized-Einstein-eq-AB-l=0-trace-GF-component-sum-f-2})
under the integrability condition
(\ref{eq:div-barTab-linear-vac-back-u-A-mode-dec-l=0-t-sum-f}) as
\begin{eqnarray}
  X_{(e)}
  =
  \frac{1}{r} \left[2M_{1} + 8\pi \int dr \frac{r^{2}}{f} \tilde{T}_{tt}\right]
  ,
  \label{eq:linearized-Einstein-eq-AB-l=0-GF-X_e-sol}
\end{eqnarray}
where $M_{1}$ is the constant of integration.
This $M_{1}$ corresponds to the perturbation of the Schwarzschild mass
parameter.


On the other hand, in the linearized Einstein equation, there is no
equation for $\partial_{r}Y_{(e)}$ which guarantees the integrability
condition for
Eq.~(\ref{eq:linearized-Einstein-eq-AB-l=0-traceless-GF-tt-sum-f-2}).
However, we may write the solution to the equation
(\ref{eq:linearized-Einstein-eq-AB-l=0-traceless-GF-tt-sum-f-2}) as
\begin{eqnarray}
  Y_{(e)}
  =
  \frac{4\pi r}{f} \int dt \left( \tilde{T}_{tt} + f^{2} \tilde{T}_{rr} \right)
  +
  Y_{(e)0}(r)
  .
  \label{eq:linearized-Einstein-eq-AB-l=0-traceless-GF-tt-sum-f-3}
\end{eqnarray}
where $Y_{(e)0}(r)$ is an arbitrary function of $r$.
There is no equation that determines the arbitrary function
$Y_{(e)0}(r)$ of $r$ within the linearized Einstein equations.


From the definition (\ref{eq:l=0-componentHFAB}) of the metric
perturbation $\tilde{\HF}_{AB}$ and the solutions
(\ref{eq:linearized-Einstein-eq-AB-l=0-GF-X_e-sol}) and
(\ref{eq:linearized-Einstein-eq-AB-l=0-traceless-GF-tt-sum-f-3}), we
obtain
\begin{eqnarray}
  \tilde{\HF}_{AB}
  &=&
      \frac{2}{r} \left(M_{1} + 4\pi \int dr \frac{r^{2}}{f} \tilde{T}_{tt}\right)
      \left((dt)_{A} (dt)_{B} + f^{-2} (dr)_{A}(dr)_{B}\right)
      \nonumber\\
  &&
      2 \left[
      4\pi r \int dt \left( \frac{1}{f} \tilde{T}_{tt} + f \tilde{T}_{rr} \right)
      +
      Y_{(e)0}(r)
      \right] (dt)_{(A} (dr)_{B)}
      .
     \label{eq:l=0-componentHFAB-solution}
\end{eqnarray}
Here, the components of the energy-momentum tensor satisfy the
continuity equations
(\ref{eq:div-barTab-linear-vac-back-u-A-mode-dec-l=0-t-sum-f}) and
(\ref{eq:div-barTab-linear-vac-back-u-A-mode-dec-l=0-r-sum-f}).


To interpret the term of $Y_{(e)0}(r)$ in
Eq.~(\ref{eq:l=0-componentHFAB-solution}), we consider the term
${\pounds}_{V}g_{ab}$ with the generator $V_{a}$ whose components are
given by
\begin{eqnarray}
  \label{eq:LieVgab-generator-components}
  V_{a} = V_{t}(r) (dt)_{a}.
\end{eqnarray}
Then, the nonvanishing components of ${\pounds}_{V}g_{ab}$ are given
by
\begin{eqnarray}
  \label{eq:LieVgab-components-tr-Vr=0-partialtVt=0}
  {\pounds}_{V}g_{tr}
  &=&
      f \partial_{r}\left( \frac{1}{f} V_{t} \right)
      .
\end{eqnarray}
Choosing $V_{t}$ so that
\begin{eqnarray}
  \label{eq:LieVgab-components-tr-Vr=0-partialtVt=0-Vt-choice}
  f \partial_{r}\left( \frac{1}{f} V_{t} \right)
  =
  Y_{(e)0}(r)
  , \quad
  V_{t}
  =
  f \int dr \frac{1}{f} Y_{(e)0}(r)
  ,
\end{eqnarray}
we obtain
\begin{eqnarray}
  \label{eq:LieVgab-components-tr-Vr=0-partialtVt=0-Ye0-choice}
  {\pounds}_{V}g_{tr}
  &=&
      Y_{(e)0}(r)
      .
\end{eqnarray}
Then, the solution (\ref{eq:l=0-componentHFAB-solution}) is given by
\begin{eqnarray}
  \tilde{\HF}_{ab}
  &=&
      \frac{2}{r} \left(M_{1} + 4\pi \int dr \frac{r^{2}}{f} \tilde{T}_{tt}\right)
      \left((dt)_{a} (dt)_{b} + f^{-2} (dr)_{a}(dr)_{b}\right)
      \nonumber\\
  &&
     +
     8\pi r \int dt \left( \frac{1}{f} \tilde{T}_{tt} + f \tilde{T}_{rr} \right)
     (dt)_{(a} (dr)_{b)}
     +
     {\pounds}_{V}g_{ab}
     ,
     \label{eq:l=0-componentHFAB-solution-2}
\end{eqnarray}
where
\begin{eqnarray}
  \label{eq:LieVgab-generator-components-Ye0r-choice}
  V_{a}
  =
  \left(
  f \int dr \frac{1}{f} Y_{(e)0}(r)
  \right) (dt)_{a}
  .
\end{eqnarray}
As noted just after
Eq.~(\ref{eq:gauge-transformation-rule-AB-Fourier-l=0-trace-static-2}),
there is still the remaining degree of freedom of gauge whose
generator is given by
\begin{eqnarray}
  \label{eq:remaining-degree-of-freedom}
  \zeta_{a} = \zeta_{t}(r) dt_{a},
\end{eqnarray}
where $\zeta_{t}(r)$ is an arbitrary function of $r$.
Therefore, we may regard the degree of freedom of $V_{a}$ given in
Eq.~(\ref{eq:LieVgab-generator-components-Ye0r-choice}) can be
eliminated as the second-kind gauge degree of freedom
(\ref{eq:remaining-degree-of-freedom}).
According to our rule of the comparison between our gauge-invariant
formulation and a ``conventional complete gauge-fixing approach,'' we
have to regard that the degree of freedom of $V_{a}$ given in
Eq.~(\ref{eq:LieVgab-generator-components-Ye0r-choice}) is
``unphysical.''


Thus, as the ``physical solution'' for $l=0$-mode linearized Einstein
equation based of a ``conventional complete gauge-fixing approach,'' we
obtain
\begin{eqnarray}
  h_{ab}
  &=&
      \frac{2}{r} \left(M_{1} + 4\pi \int dr \frac{r^{2}}{f} \tilde{T}_{tt}\right)
      \left((dt)_{a} (dt)_{b} + f^{-2} (dr)_{a}(dr)_{b}\right)
      \nonumber\\
  &&
     +
     8\pi r \int dt \left( \frac{1}{f} \tilde{T}_{tt} + f \tilde{T}_{rr} \right)
     (dt)_{(a} (dr)_{b)}
     .
     \label{eq:l=0-componentHFAB-solution-complete-gauga-fixing-final}
\end{eqnarray}
This coincides with the solution obtained in
Refs.~\cite{K.Nakamura-2021-CQG,K.Nakamura-2021-PartII} except for the
terms of the Lie derivative of the background metric $g_{ab}$.
Therefore, we conclude that the $l=0$ solution except for the
terms of the Lie derivative of the background metric $g_{ab}$
obtained in Refs.~\cite{K.Nakamura-2021-CQG,K.Nakamura-2021-PartII} can
be also obtained as
(\ref{eq:l=0-componentHFAB-solution-complete-gauga-fixing-final})
through a complete gauge-fixing approach.


The difference between our solution in
Refs.~\cite{K.Nakamura-2021-CQG,K.Nakamura-2021-PartII} is in the term
of the Lie derivative of the background metric.
As shown above, according to our rule of comparison between our
gauge-invariant formulation and a ``conventional complete gauge-fixing
approach,'' all terms of the Lie derivative of the background metric
should be regarded as the second-kind gauge degree of freedom,
and these are ``unphysical degree of freedom'' in the above solution
(\ref{eq:l=0-componentHFAB-solution-complete-gauga-fixing-final}).
On the other hand, in our gauge-invariant formulation developed in
Refs.~\cite{K.Nakamura-2021-CQG,K.Nakamura-2021-LHEP,K.Nakamura-2021-PartI,K.Nakamura-2021-PartII,K.Nakamura-2021-PartIII},
$l=0,1$-mode metric perturbations are given in a gauge-invariant form
and the terms of the Lie derivative of the background metric is
included in these gauge-invariant variables.
Since the second-kind gauge degree of freedom is completely excluded
in our gauge-invariant formulation, we cannot regard these terms of
the Lie derivative of the background metric as an ``unphysical degree
of freedom.''
Therefore, we regard the terms of the Lie derivative of the
background metric in these gauge-invariant variables as
first-kind gauges that have some physical meaning.
This point is the essential difference of the solutions in
Refs.~\cite{K.Nakamura-2021-CQG,K.Nakamura-2021-LHEP,K.Nakamura-2021-PartI,K.Nakamura-2021-PartII,K.Nakamura-2021-PartIII}
and the above solutions based on a ``conventional complete
gauge-fixing approach.''
This difference leads to the confusion of the interpretation of the
perturbative Tolman Bondi solution, as shown in the next subsection.


\subsection{Comparing with Lema\^itre-olman-Bondi solution}
\label{sec:Comparing with Tolman-Bondi solution}


\subsubsection{Perturbative expression of the LTB solution on
  Schwarzschild background spacetime}
\label{sec:Perturbative_expression_of_the_LTB_sol._on_Schwarzschild_BG_spacetime}


Here, we consider the Lema\^itre-Tolman-Bondi (LTB)
solution~\cite{L.Landau-E.Lifshitz-1962} which is an exact solution to
the Einstein equation with the matter field
\begin{eqnarray}
  \label{eq:dast-energy-momentum-tensor}
  T_{ab} = \rho u_{a}u_{b}, \quad u_{a} = - (d\tau)_{a},
\end{eqnarray}
and the metric
\begin{eqnarray}
  \label{eq:LTB-metric}
  g_{ab}
  &=&
      - (d\tau)_{a} (d\tau)_{b}
      + \frac{(\partial_{R}r)^{2}}{1+f(R)} (dR)_{a}(dR)_{b}
      + r^{2} \gamma_{ab}
      ,
      \quad
      r = r(\tau,R)
     .
\end{eqnarray}
This solution is a spherically symmetric solution to the Einstein equation.
The function $r=r(\tau,R)$ satisfies the differential equation
\begin{eqnarray}
  \label{eq:LTB-Hubble-equation}
  (\partial_{\tau}r)^{2}
  =
  \frac{F(R)}{r}
  +
  f(R)
  .
\end{eqnarray}
Here, we note that $F(R)$ is an arbitrary function of $R$ representing
the dust matter's initial distribution.
$f(R)$ is also an arbitrary function of $R$ that represents the
initial distribution of the energy of the dust field in the Newtonian
sense.
The solution to Eq.~(\ref{eq:LTB-Hubble-equation}) is given in the
three cases
\begin{description}
\item[(i) $f(R)>0$ : ]
  \begin{eqnarray}
    \label{eq:Morita-Nakamura-Kasai-1998-f-positive}
    r = \frac{F(R)}{2f(R)}(\cosh\eta - 1), \quad
    \tau_{0}(R) - \tau  = \frac{F(R)}{2f(R)^{3/2}} (\sinh\eta -\eta)
    ,
  \end{eqnarray}
\item[(ii) $f(R)<0$ : ]
  \begin{eqnarray}
    \label{eq:Morita-Nakamura-Kasai-1998-f-negative}
    r = \frac{F(R)}{-2f(R)}(1-\cos\eta), \quad
    \tau_{0}(R) - \tau = \frac{F(R)}{2(-f(R))^{3/2}} (\eta - \sin\eta)
    ,
  \end{eqnarray}
\item[(iii) $f(R)=0$ : ]
  \begin{eqnarray}
    \label{eq:Morita-Nakamura-Kasai-1998-f=0}
    r = \left(\frac{9F(R)}{4}\right)^{1/3} \left[\tau_{0}(R)-\tau\right]^{2/3}
    .
  \end{eqnarray}
\end{description}
The energy density $\rho$ is given by
\begin{eqnarray}
  \label{eq:LTB-energy-density}
  8\pi \rho = \frac{\partial_{R}F}{(\partial_{R}r)r^{2}}.
\end{eqnarray}
The LTB solution includes the three arbitrary functions $f(R)$,
$F(R)$, and $\tau_{0}(R)$.


Here, we consider the vacuum case $\rho=0$.
In this case, from Eq.~(\ref{eq:LTB-energy-density}), we have
\begin{eqnarray}
  \label{eq:LTB-energy-density-vacuum}
  \partial_{R}F = 0, \quad F = 2M
  ,
\end{eqnarray}
where $M$ is the Schwarzschild metric with the mass parameter $M$.
Furthermore, we consider the case $f(R)=0$.
Here, we chose $\tau_{0}=R$, i.e., $\partial_{R}\tau_{0}=1$.
In this case, Eq.~(\ref{eq:Morita-Nakamura-Kasai-1998-f=0}) yields
\begin{eqnarray}
  (dR)_{a}
  &=&
      (d\tau)_{a}
      +
      \left(\frac{2M}{r}\right)^{-1/2}
      (dr)_{a}
      .
      \label{eq:dRa-Morita-Nakamura-Kasai-1998-f=0}
\end{eqnarray}
Then, the metric (\ref{eq:LTB-metric}) is given by
\begin{eqnarray}
  g_{ab}
  &=&
      - (d\tau)_{a} (d\tau)_{b}
      + (\partial_{R}r)^{2} (dR)_{a}(dR)_{b}
      + r^{2} \gamma_{ab}
      \nonumber\\
  &=&
      -
      f
      (dt)_{a}(dt)_{b}
     +
     f^{-1}
     (dr)_{a}
     (dr)_{b}
     + r^{2} \gamma_{ab}
     ,
      \quad
      f = 1 - \frac{2M}{r}
      ,
      \label{eq:Schwarzschild-static-chart-background}
\end{eqnarray}
where $(dt)_{a}$ is defined by
\begin{eqnarray}
  (dt)_{a}
  :=
  (d\tau)_{a}
  -
  f^{-1}
  \left(1-f\right)^{1/2}
  (dr)_{a}
  .
  \label{eq:Killing-time-function-LTB-1998-f=0}
\end{eqnarray}
We also note that the degree of freedom of the choice of the
coordinate $R$ is completely fixed through the choice $\tau_{0}=R$,
though there remains a degree of freedom of the choice of
$R=R(\tilde{R})$ in the exact solution (\ref{eq:LTB-metric}).


Now, we consider the perturbation of the Schwarzschild spacetime which
is derived by the exact LTB solution (\ref{eq:LTB-metric}) so that
\begin{eqnarray}
  F(R)
  &=&
      2 \left[
      M
      +
      \epsilon
      m_{1}(R)
      \right]
      + O(\epsilon^{2})
      ,
      \label{eq:FR-perturbation-def}
  \\
  f(R)
  &=&
      0
      +
      \epsilon
      f_{1}(R)
      + O(\epsilon^{2})
      ,
      \label{eq:fR-perturbation-def}
  \\
  \tau_{0}(R)
  &=&
      R
      +
      \epsilon
      \tau_{1}(R)
      + O(\epsilon^{2})
      .
      \label{eq:tau0R-perturbation-def}
\end{eqnarray}
Through these perturbations
(\ref{eq:FR-perturbation-def})--(\ref{eq:tau0R-perturbation-def}), we
consider the perturbative expansion of the function $r$ which is
determined by Eq.~(\ref{eq:LTB-Hubble-equation}):
\begin{eqnarray}
  \label{eq:circum-ference-perturbations}
  r(\tau,R) = r_{s}(\tau,R) + \epsilon r_{1}(\tau,R)
  + O(\epsilon^{2})
  .
\end{eqnarray}
Here, the function $r_{s}(\tau,R)$ is given by
Eq.~(\ref{eq:Morita-Nakamura-Kasai-1998-f=0}), i.e.,
\begin{eqnarray}
  \label{eq:LTB-flat-vacuum-circumference}
  r_{s}(\tau,R)
  =
  r(\tau,R)
  =
  \left(\frac{9M}{2}\right)^{1/3}
  \left[R-\tau\right]^{2/3}
  .
\end{eqnarray}
In Eqs.~(\ref{eq:tau0R-perturbation-def}) and
(\ref{eq:LTB-flat-vacuum-circumference}), we chose the background
value of the function $\tau_{0}(R)$ to be $R$.


Through this perturbative expansion, we evaluate
$O(\epsilon^{1})$ perturbation of Eq.~(\ref{eq:LTB-Hubble-equation})
through Eq.~(\ref{eq:dRa-Morita-Nakamura-Kasai-1998-f=0}) as
\begin{eqnarray}
     (1-f)^{1/2}
     (\partial_{\tau}r_{1})
     +
     \frac{m_{1}(R)}{r}
     -
     \frac{M}{r^{2}}
     r_{1}
     +
     \frac{1}{2}
     f_{1}(R)
     =
     0
     ,
     \label{eq:LTB-Hubble-equation-linear-pert}
\end{eqnarray}
where we used Eq.~(\ref{eq:dRa-Morita-Nakamura-Kasai-1998-f=0})
and the replacement $r_{s}\rightarrow r$.
The solution to Eq.~(\ref{eq:LTB-Hubble-equation-linear-pert}) is
given by
\begin{eqnarray}
  r_{1}
  &=&
      \left(\frac{M}{6}\right)^{1/3}
      \frac{m_{1}(R)}{M}
      \left[R-\tau\right]^{2/3}
      -
      \frac{3}{20}
      \left(\frac{6}{M}\right)^{1/3}
      f_{1}(R)
      \left[R-\tau\right]^{+4/3}
      \nonumber\\
  &&
      +
      B(R)
      \left[R-\tau\right]^{-1/3}
      .
      \label{eq:LTB-Hubble-equation-linear-pert-sol.-sum}
\end{eqnarray}
From the comparison with Eq.~(\ref{eq:LTB-flat-vacuum-circumference}),
$B(R)$ is the perturbation of the $\tau_{1}(R)$ as
$\tau_{0}(R)=R+\tau_{1}(R)$ in the exact solution
(\ref{eq:Morita-Nakamura-Kasai-1998-f-positive})--(\ref{eq:Morita-Nakamura-Kasai-1998-f=0}).
Furthermore, the solution
(\ref{eq:LTB-Hubble-equation-linear-pert-sol.-sum}) can be also derived
from the exact solution
(\ref{eq:Morita-Nakamura-Kasai-1998-f-positive})--(\ref{eq:Morita-Nakamura-Kasai-1998-f=0}).
From Eq.~(\ref{eq:LTB-energy-density}), the perturbative dust energy
density given by
\begin{eqnarray}
  \label{eq:LTB-energy-density-perturbation}
  8\pi \rho = \frac{2\partial_{R}m_{1}(R)}{(\partial_{R}r)r^{2}}.
\end{eqnarray}


Through the perturbative solution
(\ref{eq:LTB-Hubble-equation-linear-pert-sol.-sum}), the metric
(\ref{eq:LTB-metric}) is given by
\begin{eqnarray}
  g_{ab}
  &=&
      - (d\tau)_{a} (d\tau)_{b}
      +
      (\partial_{R}r)^{2}
      (dR)_{a}(dR)_{b}
      +
      r^{2}
      \gamma_{ab}
      \nonumber\\
  &&
     +
     \epsilon
     \left[
     \left(
     2(\partial_{R}r_{1})
     -
     f_{1}
     (\partial_{R}r)
     \right)
     (\partial_{R}r)
     (dR)_{a}(dR)_{b}
     +
     2
     r
     r_{1}
     \gamma_{ab}
     \right]
     + O(\epsilon^{2})
     \nonumber\\
  &=:&
       g_{ab}^{(0)}
       +
       \epsilon
       {}_{\scrX}\!h_{ab}
       + O(\epsilon^{2})
       .
       \label{eq:LTB-metric-perturbative-expansion}
\end{eqnarray}
As shown in Eq.~(\ref{eq:Schwarzschild-static-chart-background}),
the background metric $g_{ab}^{(0)}$ is given by the Schwarzschild
metric in the static chart.
On the other hand, the linear order perturbation ${}_{\scrX}\!h_{ab}$
(in the gauge $\scrX_{\epsilon}$) is given by
\begin{eqnarray}
  {}_{\scrX}\!h_{ab}
  :=
  \left(
  2
  (\partial_{R}r_{1})
  -
  f_{1}(R)
  (\partial_{R}r)
  \right)
  (\partial_{R}r)
  (dR)_{a}(dR)_{b}
  +
  2
  r
  r_{1}
  \gamma_{ab}
  .
  \label{eq:LTB-linear-pert-with-Schwarzschild-BG}
\end{eqnarray}
Here, we fixed the second-kind gauge so that
\begin{eqnarray}
  \label{eq:second-gauge-fixed-LTB}
  \scrX_{\epsilon} :
  (\tau,R,\theta,\phi)\in\scrM_{ph} \mapsto
  (\tau,R,\theta,\phi)\in\scrM.
\end{eqnarray}
Through this second-kind gauge choice and the choice of the background
radial coordinate $R$ in
Eq.~(\ref{eq:Schwarzschild-static-chart-background}), the degree of
freedom of the choice of the radial coordinate $R=R(\tilde{R})$ is
also completely fixed even in the linearized version of the exact
solution (\ref{eq:LTB-linear-pert-with-Schwarzschild-BG}), which
though there remains a degree of freedom of the choice of
$R=R(\tilde{R})$ in the exact solution (\ref{eq:LTB-metric}).


Of course, if we employ the different gauge choice
$\scrY_{\epsilon}$ from the above gauge-choice
$\scrX_{\epsilon}$, we obtain the different expression of the
metric perturbation ${}_{\scrY}\!h_{ab}$.
Actually, we may choose $\scrY_{\epsilon}$ as the identification of
\begin{eqnarray}
  \label{eq:second-gauge-fixed-LTB-2}
  \scrY_{\epsilon} :
  (\tau+\epsilon\xi^{\tau}(\tau,R),R+\epsilon\xi^{R}(\tau,R),\theta,\phi)\in\scrM_{ph} \mapsto
  (\tau,R,\theta,\phi)\in\scrM.
\end{eqnarray}
In this identification, the metric on $\scrM_{ph}$ pulled back to
$\scrM$ is given by
\begin{eqnarray}
  {}_{\scrY_{\epsilon}}g_{ab}
  &=&
      {}_{\scrX_{\epsilon}}g_{ab}
      +
      \epsilon
      {\pounds}_{\xi}g_{ab}^{(0)}
      +
      O(\epsilon^{2})
      \nonumber\\
  &=&
      g_{ab}^{(0)}
      +
      \epsilon
      \left(
      {}_{\scrX}\!h_{ab}
      +
      {\pounds}_{\xi}g_{ab}^{(0)}
      \right)
      +
      O(\epsilon^{2})
       ,
       \label{eq:general-second-kind-gauge-trans-expression}
\end{eqnarray}
where
$\xi^{a}=\xi^{\tau}(\partial_{\tau})^{a}+\xi^{R}(\partial_{R})^{a}$ is
the generator of second-kind gauge transformation
$\scrX_{\epsilon}\rightarrow\scrY_{\epsilon}$.


\subsubsection{Expression of the perturbative LTB solution in static chart}
\label{sec:Schwarzschild_static_chart_LTB_solution}


Here, we consider the expression of the linear perturbation
${}_{\scrX}\!h_{ab}$ given by
Eq.~(\ref{eq:LTB-linear-pert-with-Schwarzschild-BG}).
From Eqs.~(\ref{eq:dRa-Morita-Nakamura-Kasai-1998-f=0}) and
(\ref{eq:Killing-time-function-LTB-1998-f=0}) with $F=2M$, we obtain
\begin{eqnarray}
  (dR)_{a}
  &=&
      (dt)_{a}
      +
      f^{-1}
      (1-f)^{-1/2}
      (dr)_{a}
      ,
      \quad
      f = 1 - \frac{2M}{r}
      ,
      \label{eq:dRa-f=0-tau0=R-with-Killing-time-sum-R}
  \\
  (d\tau)_{a}
  &=&
      (dt)_{a}
      +
      f^{-1}
      (1-f)^{1/2}
      (dr)_{a}
      .
      \label{eq:dRa-f=0-tau0=R-with-Killing-time-sum-tau}
\end{eqnarray}


First, we consider the perturbation of the energy-momentum tensor of
the matter field.
In the case of the LTB solution, the matter field is characterized by
the dust field whose energy-momentum tensor
(\ref{eq:dast-energy-momentum-tensor}) is given by
\begin{eqnarray}
  \label{eq:dust-energy-momentum-tesor-general}
  T_{ab} = \rho u_{a} u_{b}, \quad u_{a} = - (d\tau)_{a},
  \quad u^{a} = \left(\partial_{\tau}\right)^{a}.
\end{eqnarray}
In our case, the linearized Einstein equation gives
Eq.~(\ref{eq:LTB-energy-density-perturbation}), i.e.,
\begin{eqnarray}
  8\pi \rho
  =
  \frac{2\partial_{R}m_{1}(R)}{(\partial_{R}r)r^{2}}
  =
  \frac{\partial_{R}m_{1}(R)}{4\pi r^{2}} (1-f)^{-1/2}
  .
  \label{eq:LTB-energy-density-perturbation-3}
\end{eqnarray}
On the other hand, substituting
Eq.~(\ref{eq:dRa-f=0-tau0=R-with-Killing-time-sum-tau}) into
Eq.~(\ref{eq:dust-energy-momentum-tesor-general}), we obtain
\begin{eqnarray}
  T_{ab}
  &=&
      \rho (d\tau)_{a} (d\tau)_{b}
      \nonumber\\
  &=&
      \rho \left(
      (dt)_{a}
      +
      f^{-1}
      (1-f)^{1/2}
      (dr)_{a}
      \right)
      \left(
      (dt)_{b}
      +
      f^{-1}
      (1-f)^{1/2}
      (dr)_{b}
      \right)
      \nonumber\\
  &=&
      \rho
      (dt)_{a}
      (dt)_{b}
      +
      \rho
      \frac{(1-f)^{1/2}}{f}
      2
      (dt)_{(a} (dr)_{b)}
      +
      \rho
      \frac{1-f}{f^{2}}
      (dr)_{a}
      (dr)_{b}
      .
      \label{eq:dust-energy-momentum-tesor-general-static}
\end{eqnarray}
Then, we obtain the components of the energy-momentum tensor for the
static coordinate $(t,r)$ as
\begin{eqnarray}
  \tilde{T}_{tt}
  =
  \rho
  ,
  \quad
  \tilde{T}_{tr}
  =
  \frac{(1-f)^{1/2}}{f}
  \rho
  ,
  \quad
  \tilde{T}_{rr}
  =
  \frac{1-f}{f^{2}}
  \rho
  ,
  \quad
  \tilde{T}_{(e0)}
  =
  0
  .
  \label{eq:LTB-dust-Ttt-Ttr-Trr-def}
\end{eqnarray}
From this component, we can confirm the continuity equations (\ref{eq:div-barTab-linear-vac-back-u-A-mode-dec-l=0-t-sum-f})
and (\ref{eq:div-barTab-linear-vac-back-u-A-mode-dec-l=0-r-sum-f}).


As derived in Ref.~\cite{K.Nakamura-2021-PartIII}, the linearized
Tolman-Bondi solution (\ref{eq:LTB-linear-pert-with-Schwarzschild-BG})
with the Schwarzschild background is given by
\begin{eqnarray}
  {}_{\scrX}\!h_{ab}
  &=&
      \frac{2m_{1}(R)}{r}
      \left[
      (dt)_{a} (dt)_{b} + \frac{1}{f^{2}} (dr)_{a} (dr)_{b}
      \right]
      +
      \frac{2-f}{f (1-f)^{1/2}}
      \frac{m_{1}(R)}{r}
      2 (dt)_{(a} (dr)_{b)}
      \nonumber\\
  &&
     + {\pounds}_{V_{(LTB)}}g_{ab}
     ,
     \label{eq:LTB-linear-Schw-BG-sc-LieV1gab-LieV2gab-fnal}
\end{eqnarray}
where $V_{(LTB)a}$ are given by
\begin{eqnarray}
  \label{eq:V1a-V2a-summary}
  V_{(LTB)a}
  :=
  \left[(1-f)^{1/2} r_{1} + \frac{1}{2} f \int dt f_{1}(R)\right] (dt)_{a}
  +
  \frac{r_{1}}{f} (dr)_{a}
  .
\end{eqnarray}
As shown in Ref.~\cite{K.Nakamura-2021-PartIII}, the $l=0$ solution
(\ref{eq:l=0-componentHFAB-solution-complete-gauga-fixing-final}) with
the components (\ref{eq:LTB-dust-Ttt-Ttr-Trr-def}) of the linearized
energy-momentum tensor realize the first line of
Eq.~(\ref{eq:LTB-linear-Schw-BG-sc-LieV1gab-LieV2gab-fnal}).
In this sense, the solutions
(\ref{eq:l=0-componentHFAB-solution-complete-gauga-fixing-final}) of
the $l=0$ mode perturbations are justified by the LTB solutions.


However, we emphasize that the linearized LTB solution
(\ref{eq:LTB-linear-Schw-BG-sc-LieV1gab-LieV2gab-fnal}) does have the
term ${\pounds}_{V_{LTB}}g_{ab}$.
As noted by Eqs.~(\ref{eq:second-gauge-fixed-LTB-2}) and
(\ref{eq:general-second-kind-gauge-trans-expression}), we can always
eliminate the terms of the Lie derivative of the background metric as
the second-kind gauge-degree of freedom.
Therefore, according to our rule of the comparison, the term
${\pounds}_{V_{(LTB)}}g_{ab}$ should be regarded as the second-kind
gauge-degree of freedom in the ``conventional gauge-fixing approach''
discussed in this paper.
In this case, we have to regard that the perturbation $f_{1}(R)$ of
the initial distribution $f(R)$ of the energy of dust field in
Eq.~(\ref{eq:LTB-Hubble-equation}) is an ``unphysical degree of
freedom'' in spite that the behavior of the LTB solution crucially
depends on the signature of the function $f(R)$ as shown in
Eqs.~(\ref{eq:Morita-Nakamura-Kasai-1998-f-positive})--(\ref{eq:Morita-Nakamura-Kasai-1998-f=0}).


\section{Summary and Discussions}
\label{sec:Summary_and_Discussions}


In this paper, we have discussed comparison of our gauge-invariant
formulation for $l=0,1$ perturbations on the Schwarzschild background
spacetime proposed in the series of papers~\cite{K.Nakamura-2021-CQG,K.Nakamura-2021-LHEP,K.Nakamura-2021-PartI,K.Nakamura-2021-PartII,K.Nakamura-2021-PartIII}
and a ``conventional complete gauge-fixing approach.''
It is well-known that we cannot construct gauge-invariant variables
for $l=0,1$ mode perturbations through a similar manner to the
construction of the gauge-invariant variables for $l\geq 2$ mode
perturbations if we use the decomposition formula
(\ref{eq:hAB-fourier})--(\ref{eq:hpq-fourier}) with the spherical
harmonic functions $Y_{lm}$ as the scalar harmonics $S_{\delta}$.
In our gauge-invariant formulation for $l=0,1$ perturbations on the
Schwarzschild background spacetime, we proposed the introduction of
the singular harmonic function at once.
Due to this, we can construct gauge-invariant variables for
$l=0,1$-mode perturbations in a similar manner to the $l\geq 2$
modes of perturbations.
After deriving the mode-by-mode perturbative Einstein equations in
terms of the gauge-invariant variables, we impose the regularity on
the introduced singular harmonics when we solve the derived Einstein
equations.
Based on this proposal, we derived formal solutions to the
$l=0,1$-mode linearized Einstein equations without the specification
of the components of the linear perturbation of the energy-momentum tensor~\cite{K.Nakamura-2021-CQG,K.Nakamura-2021-PartI,K.Nakamura-2021-PartII}.
Our proposal enables us to develop higher-order perturbations of the
Schwarzschild spacetime~\cite{K.Nakamura-2021-LHEP}.
Furthermore, we check that our derived solutions realized the
linearized version of the LTB solution and non-rotating
C-metric~\cite{K.Nakamura-2021-PartII}.
In this sense, we conclude that our proposal is physically
reasonable.
On the other hand, it is often said that ``gauge-invariant
formulations in general-relativistic perturbations are equivalent to
complete gauge-fixing approaches.''
For this reason, we check this statement through the comparison of our
gauge-invariant formulation and a ``conventional complete gauge-fixing
approach'' in which we use the spherical harmonic functions $Y_{lm}$ as
the scalar harmonics $S_{\delta}$ from the starting point.


After reviewing the concept of ``gauges'' in general relativistic
perturbation theories and our proposed gauge-invariant formulation for
the $l=0,1$-mode perturbations and our derived $l=0,1$-mode solutions,
we considered $l=1$ odd-mode perturbations, $l=1$ even-mode
perturbations, and $l=0$ even-mode perturbations, separately.
As a result, it is shown that we can actually derive similar solutions
even in the treatment of the ``conventional complete gauge-fixing
approach.''
However, we have to emphasize that the derived solutions are slightly
different from those derived based on our gauge-invariant formulation
from a conceptual point of view.


In the case of $l=1$ odd-mode perturbations, we obtained the
formal solution to the linearized Einstein equation through our proposed
gauge-invariant formulation.
This formal solution includes the term of the Lie derivative of the
background metric.
In our gauge-invariant formulation, we describe the solutions
only through gauge-invariant variables.
Therefore, we should regard the term of the Lie derivative of the
background metric is gauge invariant.
On the other hand, in the conventional gauge-fixing approach where we
use the spherical harmonics $Y_{lm}$ from the starting point, we
cannot construct gauge-invariant variable for $l=1$ odd-mode
perturbations in a similar manner to those for $l\geq 2$ mode
perturbations.
For this reason, we have to treat gauge-dependent variables for
perturbations.
However, the linearized Einstein equations and the continuity
equations of the linearized energy-momentum tensor for $l=1$ odd-mode
perturbations in terms of these gauge-dependent variables have the
completely same form that is derived through our gauge-invariant
formulation.
Therefore, the same formal solutions derived through our
gauge-invariant formulation should be the formal solutions to these
linearized Einstein equations and the continuity equation of the
linearized energy-momentum tensor in terms of gauge-dependent
variables.
As noted above, we treat gauge-dependent variables in the
conventional gauge-fixing approach.
Therefore, the above Lie derivative terms of the background metric may
include the gauge degree of freedom of the second kind, which should be
regarded as an ``unphysical degree of freedom.''
Checking the residual gauge degree of freedom, we conclude that the above
Lie derivative terms of the background metric include the second-kind
gauge degree of freedom.
However, in our formal solution, there is a variable that should be
obtained by solving the $l=1$ Regge-Wheeler equation.
We conclude that the solution to this $l=1$ Regge-Wheeler equation is
not the gauge degree of freedom of the second kind but a physical
degree of freedom.
Furthermore, we have to impose appropriate boundary conditions to
solve this $l=1$ Regge-Wheeler equation.
Since the $l=1$ Regge-Wheeler equation is an inhomogeneous linear
second-order partial differential equation, the boundary conditions
for an inhomogeneous linear second-order partial differential equation
are adjusted by the homogeneous solutions to this linear second-order
partial differential equation.
According to the check of the residual gauge degree of freedom, we
have to conclude that a part of homogeneous solutions to this equation
is the gauge degree of freedom of the second kind.
We have to eliminate this part from our consideration because this
gauge degree of freedom is ``unphysical.''
This is the restriction of the boundary conditions of the linearized
Einstein equations.


In the case of $l=1$ even-mode perturbations, the situation is
worse than the $l=1$ odd-mode case.
As in the $l=1$ odd-mode case, we obtained the
formal solution to the linearized Einstein equation through our proposed
gauge-invariant formulation.
This formal solution includes the term of the Lie derivative of the
background metric, which is gauge-invariant within our proposed
gauge-invariant formulation.
On the other hand, in the conventional gauge-fixing approach where we
use the spherical harmonics $Y_{lm}$ from the starting point, we
cannot construct gauge-invariant variable for $l=1$ even-mode
perturbations in a similar manner to those for $l\geq 2$ mode
perturbations.
For this reason, we have to treat gauge-dependent variables for
perturbations.
As in the $l=1$ odd-mode case, the linearized Einstein equations and
the continuity equations of the linearized energy-momentum tensor for
$l=1$ even-mode perturbations in terms of these gauge-dependent
variables have the completely same form that is derived through our
gauge-invariant formulation.
Therefore, the same formal solutions derived through our
gauge-invariant formulation should be the formal solutions to these
linearized Einstein equations and the continuity equation of the
linearized energy-momentum tensor in terms of gauge-dependent
variables.
As noted above, we treat gauge-dependent variables in the
conventional gauge-fixing approach as in the $l=1$ odd-mode case.
Therefore, the above Lie derivative terms of the background metric may
include the second-kind gauge degree of freedom, which should be
regarded as an ``unphysical degree of freedom.''
However, in our formal solution, there is a variable that should be
obtained by solving the $l=1$ Zerilli equation.
We conclude that the solution to this $l=1$ Zerilli equation is
not the gauge degree of freedom of the second kind but the physical
degree of freedom in the non-vacuum case.
Furthermore, we have to impose appropriate boundary conditions to
solve this $l=1$ Zerilli equation in the non-vacuum case.
Since the $l=1$ Zerilli equation is an inhomogeneous linear
second-order partial differential equation.
The boundary conditions for an inhomogeneous linear second-order
partial differential equation are adjusted by the homogeneous
solutions to this linear second-order partial differential equation.
According to the check of the residual gauge degree of freedom, we
have to conclude that all homogeneous solutions to this equation
are the gauge degree of freedom of the second kind, i.e., all
homogeneous solution to the $l=1$ Zerilli equation is ``unphysical.''
We have to eliminate these homogeneous solutions from our
consideration because these are regarded as ``unphysical.''
Due to this situation, we have to say that we have to impose the
boundary conditions using these ``unphysical degrees of freedom'' to
obtain the ``physical solution'' through $l=1$ Zerilli equation.
This is a dilemma.


In the case of $l=0$ even-mode perturbations, we obtain the complete
gauge fixed solution.
This solution does not include any term of the Lie derivative of the
background metric.
All terms of the Lie derivative of the background metric are regarded
as the gauge degree of freedom of the second kind.
On the other hand, the solution that is derived through our proposed
gauge-invariant formulation includes the terms of the Lie derivatives
of the background metric which is regarded as the first-kind gauge in
our gauge-invariant formulation, i.e., these are ``physical.''
This difference leads a problem when we compare the derived solution
with the linearized LTB solution.
In the linearized LTB solution, the initial energy distribution
$f_{1}(R)$ of the dust field in the Newtonian sense is included in the
terms of the Lie derivative of the background metric.
When we realize this linearized exact solution by the solution
obtained by our gauge-invariant formulation, this initial energy
distribution is regarded as a physical degree of freedom.
On the other hand, when we realize this linearized exact solution
through the solution obtained by a conventional complete gauge-fixing
approach, we have to regard this initial energy distribution of the dust
field as the gauge degree of freedom of the second kind and have to
regard it ``unphysical.''
Since the behavior of the exact LTB solution crucially depends on this
initial energy distribution of the dust field, it is not natural to
regard that this initial degree of freedom is unphysical.


In summary, we have to conclude that there is a case where the
boundary conditions and initial conditions are restricted in the
conventional complete gauge-fixing approach where we use the
decomposition of the metric perturbation by the spherical harmonics
$Y_{lm}$ from the starting point.
On the other hand, such a situation does not occur in our proposed
gauge-invariant formulation.
As a theory of physics, this point should be regarded as the
incompleteness of the conventional complete gauge-fixing approach where
we use the decomposition of the metric perturbation by the spherical
harmonics $Y_{lm}$ from the starting point.
This is the main result of this paper.


Let us discuss why such a difference occurs between our proposed
gauge-invariant formulation and a conventional complete gauge-fixed
approach where we use the decomposition of the metric perturbation by
the spherical harmonics $Y_{lm}$ from the starting point.
The aim of the introduction of the singular harmonic functions in our
proposed gauge-invariant formulation is to increase the degree of
freedom to clarify the distinction between the gauge degree of freedom
of the second kind and the physical degree of freedom.
As emphasized in Sec.~\ref{sec:Rule_of_comparison}, if we choose
$S=Y_{lm}$, we have $\hat{D}_{p}S$ $=$ $\epsilon_{pq}\hat{D}^{q}S$ $=$
$0$,
$\left(\hat{D}_{p}\hat{D}_{q}-\frac{1}{2}\gamma_{pq}\hat{D}^{r}\hat{D}_{r}\right)S$
$=$ $2 \epsilon_{r(p}\hat{D}_{q)}\hat{D}^{r}S$ $=$ $0$ for $l=0$
modes and
$\left(\hat{D}_{p}\hat{D}_{q}-\frac{1}{2}\gamma_{pq}\hat{D}^{r}\hat{D}_{r}\right)S$
$=$ $2 \epsilon_{r(p}\hat{D}_{q)}\hat{D}^{r}S$ $=$ $0$ for $l=1$ modes.
This is the essential reason for the fact that we cannot construct
gauge-invariant variables for $l=0,1$ mode metric perturbation in a
similar manner to the derivation of gauge-invariant variables for
$l\geq 2$ modes.
Due to these vanishing vector- or tensor-harmonics, the associated
mode coefficients do not appear, and we cannot construct
gauge-invariant variables for $l=0,1$-modes in a similar manner to the
case of $l\geq 2$ modes.
In our series of
papers~\cite{K.Nakamura-2021-CQG,K.Nakamura-2021-LHEP,K.Nakamura-2021-PartI,K.Nakamura-2021-PartII,K.Nakamura-2021-PartIII},
we regarded that this is due to the lack of the degree of freedom.
For this reason, in our proposed gauge-invariant formulation, we
introduced singular harmonic functions $k_{\hat{\Delta}}$ and
$k_{\hat{\Delta+2}m}$ for $l=0$ and $l=1$ modes, respectively.
Owing to the introduction of these singular harmonic functions, we have
$\hat{D}_{p}S$ $\neq$ $0$ $\neq$ $\epsilon_{pq}\hat{D}^{q}S$ and
$\left(\hat{D}_{p}\hat{D}_{q}-\frac{1}{2}\gamma_{pq}\hat{D}^{r}\hat{D}_{r}\right)S$
$\neq$ $0$ $\neq$ $2 \epsilon_{r(p}\hat{D}_{q)}\hat{D}^{r}S$ for $l=0$
modes and $\left(\hat{D}_{p}\hat{D}_{q}-\frac{1}{2}\gamma_{pq}\hat{D}^{r}\hat{D}_{r}\right)S$
$\neq$ $0$ $\neq$ $2 \epsilon_{r(p}\hat{D}_{q)}\hat{D}^{r}S$ for $l=1$
modes.
Thanks to these non-vanishing vector- or tensor-harmonics, the
associated mode coefficients do appear, and we can construct
gauge-invariant variables for $l=0,1$-modes in a similar manner
to the case of $l\geq 2$ modes.
This leads to the development of the gauge-invariant perturbation
theory for all
modes~\cite{K.Nakamura-2021-CQG,K.Nakamura-2021-PartI,K.Nakamura-2021-PartII,K.Nakamura-2021-PartIII}
and the development of the higher-order gauge-invariant
perturbations~\cite{K.Nakamura-2021-LHEP}.


As noted in the Part I paper~\cite{K.Nakamura-2021-PartI}, the
decomposition using the spherical harmonics $Y_{lm}$ from the starting
point corresponds to the imposition of the regular boundary condition
on $S^{2}$ on the metric perturbations from the starting point.
In this sense, our introduction of the singular harmonic functions
corresponds to the change of the boundary conditions on $S^{2}$.
As shown in the Part I paper~\cite{K.Nakamura-2021-PartI}, if we
introduce this change of boundary conditions on $S^{2}$, we can
clearly distinguish the gauge degree of freedom of the second kind and
the physical degree of freedom, i.e., we can easily construct
gauge-invariant variables for $l=0,1$-mode perturbations.
This indicates that the imposition of the boundary conditions on
$S^{2}$ and the construction of the gauge-invariant variables does not
commute as the procedure of calculations.
This is the appearance of the non-locality of $l=0,1$-mode
perturbations as pointed out in Ref.~\cite{K.Nakamura-2013}.
Due to these reasons, we reached the conceptual difference between
the $l=0,1$-mode solutions discussed in this paper.
Namely, in the conventional complete gauge-fixing approach where we use
the decomposition by the harmonic function $Y_{lm}$ from the starting
point, the degree of freedom of the metric perturbations is not
sufficient so that we can distinguish the gauge degree of freedom of
the second kind, i.e., ``unphysical modes'' and ``physical modes.''
In spite of this situation of the lack of degree of freedom, if we
dare to carry out the ``complete gauge-fixing'' as the ``elimination
of unphysical modes,'' this ``complete gauge-fixing'' leads to the
restriction of the boundary conditions and initial conditions as
shown in this paper.


On the other hand, in our proposed gauge-invariant formulation, there
are no conceptual difficulties, such as the restriction of the boundary
conditions and initial conditions pointed out in this paper, due to the
sufficient degree of freedom of the metric perturbations.
Incidentally, due to this sufficient degree of freedom of the metric
perturbations through the introduction of the singular harmonic
functions, our proposed gauge-invariant variables are equivalent to
variables of the complete gauge-fixing within our proposed formulation
in which the degree of freedom of the metric perturbations is
sufficiently extended.
Furthermore, we can develop higher-order perturbation theory without
gauge-ambiguities if we apply our proposal and there are wide
applications of the higher-order gauge-invariant perturbations on
the Schwarzschild background spacetime as briefly discussed in
Ref.~\cite{K.Nakamura-2021-LHEP}.
We leave these further developments of the application of our
formulation to concrete problems of perturbations on the Schwarzschild
background spacetime as future works.


\appendix

\section{Linearized Einstein tensor}
\label{sec:Linearized_Einstein_tensor}


In this Appendix, we derive components of the linearized Einstein
tensor.
Although similar formulae are derived in Appendix C in the Paper
I~\cite{K.Nakamura-2021-PartI} in the gauge-invariant form.
However, in this paper, we have to derive the components of the
Einstein tensor with the spherically background spacetime without any
gauge fixing.
In this case, the formulation of the ``gauge-ready formulation''
proposed in Ref.~\cite{Noh-Hwang-2004} in the context of cosmological
perturbation theories is an appropriate formulation.
Therefore, in this Appendix, we derive the components of the
linear-order Einstein tensor without any gauge fixing following the
philosophy of the ``gauge-ready formulation'' 'in
Ref.~\cite{Noh-Hwang-2004}.
The derived formulae in the philosophy of ``gauge ready formulation''
is useful in the ingredient of this manuscript.


Here, we consider the metric perturbation as
\begin{eqnarray}
  {}_{{\cal X}}\!\bar{g}_{ab}
  =
  g_{ab} + \epsilon {}_{{\cal X}}\!h_{ab} + O(\epsilon^{2}).
\end{eqnarray}
The connection $C^{c}_{\;\;ab}$ between the covariant derivative
$\bar{\nabla}_{a}$ associated with the metric $\bar{g}_{ab}$ and the
covariant derivative $\nabla_{a}$ associated with the metric $g_{ab}$
is given by
\begin{eqnarray}
  \label{eq:connection-def-full-order}
  C^{c}_{\;\;ab} = \frac{1}{2} \bar{g}^{cd} \left(
  \nabla_{a} \bar{g}_{db} + \nabla_{b} \bar{g}_{da} - \nabla_{d}\bar{g}_{ab}
  \right)
  ,
\end{eqnarray}
where $\bar{g}^{ab}$ is the inverse of $\bar{g}_{ab}$.
Here, we expand the connection $C^{c}_{\;\;ab}$ with respect to
$\epsilon$ as
\begin{eqnarray}
  C^{c}_{\;\;ab} =
  \epsilon {}^{(1)}\!C^{c}_{\;\;ab}
  +
  O(\epsilon^{2})
  .
\end{eqnarray}
Then, we have
\begin{eqnarray}
  \label{eq:connection-def-1st-order}
  {}^{(1)}\!C^{c}_{\;\;ab}
  =
  \frac{1}{2} g^{cd} \left(
  \nabla_{a}h_{db} + \nabla_{b}h_{da} - \nabla_{d}h_{ab}
  \right)
  .
\end{eqnarray}
The relation between the Riemann curvature
$\bar{R}_{abc}^{\;\;\;\;\;\;d}$ associated with the metric
$\bar{g}_{ab}$ and the curvature $R_{abc}^{\;\;\;\;\;\;d}$ associated
with the background metric $g_{ab}$ is given by
\begin{eqnarray}
  \label{eq:Riemann-Curvature-abcd-full-order}
  \bar{R}_{abc}^{\;\;\;\;\;\;d}
  =
  R_{abc}^{\;\;\;\;\;\;d}
  - 2 \nabla_{[a}^{}C^{d}_{\;\;b]c}
  + 2 C^{e}_{\;\;c[a}C^{d}_{\;\;b]e},
\end{eqnarray}
Then, we have
\begin{eqnarray}
  \label{eq:Riemann-Curvature-abcd-full-order-perurvative}
  \bar{R}_{abc}^{\;\;\;\;\;\;d}
  =
  R_{abc}^{\;\;\;\;\;\;d}
  - 2 \epsilon \nabla_{[a}^{} {}^{(1)}\!C^{d}_{\;\;b]c}
  +
  O(\epsilon^{2})
  .
\end{eqnarray}
The perturbative expansion of the Ricci tensor $\bar{R}_{ac}$ is given
by
\begin{eqnarray}
  \label{eq:Ricci-Curvature-ac-full-order-perurvative}
  \bar{R}_{ac}
  =
  R_{ac}
  - 2 \epsilon \nabla_{[a}^{} {}^{(1)}\!C^{b}_{\;\;b]c}
  +
  O(\epsilon^{2})
  .
\end{eqnarray}
The perturbative expansion of the curvature $\bar{R}_{a}^{\;\;c}$ is
given by
\begin{eqnarray}
  \bar{R}_{a}^{\;\;d}
  &=&
      \bar{g}^{cd} \bar{R}_{ac}
      \nonumber\\
  &=&
      \bar{g}^{cd} \bar{R}_{ac}
      + \epsilon \left(
      - 2 g^{cd} \nabla_{[a}^{} {}^{(1)}\!C^{b}_{\;\;b]c}
      -  h^{cd} R_{ac}
      \right)
      +
      O(\epsilon^{2})
      .
      \label{eq:Ricci-Curvature-ac-mix-full-order-perurvative}
\end{eqnarray}
The perturbation of the scalar curvature is given by
\begin{eqnarray}
  \bar{R}
  &=&
      \bar{g}^{ac}\bar{R}_{ac}
      \nonumber\\
  &=&
      R
      + \epsilon \left(
      - h^{ac} R_{ac}
      - 2 g^{ac} \nabla_{[a}^{} {}^{(1)}\!C^{b}_{\;\;b]c}
      \right)
      +
      O(\epsilon^{2})
      .
      \label{eq:Scalar-Curvature-ac-full-order-perurvative}
\end{eqnarray}
Then, the perturbative expansion of the Einstein tensor
$\bar{G}_{a}^{\;\;d}$ is given by
\begin{eqnarray}
  \bar{G}_{a}^{\;\;d}
  &=&
      \bar{R}_{a}^{\;\;d} - \frac{1}{2} \delta_{a}^{\;\;d} \bar{R}
      \nonumber\\
  &=&
      G_{a}^{\;\;d}
      + \epsilon \left(
      - 2 g^{cd} \nabla_{[a}^{} {}^{(1)}\!C^{b}_{\;\;b]c}
      +  \delta_{a}^{\;\;d}  g^{ec} \nabla_{[e}^{} {}^{(1)}\!C^{b}_{\;\;b]c}
      -  h^{cd} R_{ac}
      +  \frac{1}{2} \delta_{a}^{\;\;d} h^{ec} R_{ec}
      \right)
      +
      O(\epsilon^{2})
      \nonumber\\
  &=:&
      G_{a}^{\;\;d}
      + \epsilon {}^{(1)}\!G_{a}^{\;\;d}
      +
      O(\epsilon^{2})
      .
      \label{eq:Einstein-tensor-ad-mix-full-order-perurvative}
\end{eqnarray}
Namely, the 1st-order perturbation of the Einstein tensor is given by
\begin{eqnarray}
  {}^{(1)}\!G_{a}^{\;\;d}
  &=&
      -  2 g^{cd} \nabla_{[a}^{} {}^{(1)}\!C^{b}_{\;\;b]c}
      + \delta_{a}^{\;\;d}  g^{ec} \nabla_{[e}^{} {}^{(1)}\!C^{b}_{\;\;b]c}
      -  h^{cd} R_{ac}
      + \frac{1}{2} \delta_{a}^{\;\;d} h^{ec} R_{ec}
      .
      \label{eq:Einstein-tensor-ad-mix-1st-order-perturvative}
\end{eqnarray}
Substituting Eq.~(\ref{eq:connection-def-1st-order}) into
Eq.~(\ref{eq:Einstein-tensor-ad-mix-1st-order-perturvative}), we obtain
\begin{eqnarray}
  {}^{(1)}\!G_{a}^{\;\;d}
  &=&
      -  \frac{1}{2} \nabla_{b}\nabla^{b}h_{a}^{\;\;d}
      + R_{abc}^{\;\;\;\;\;\;d} h^{bc}
      \nonumber\\
  &&
      + \frac{1}{2} \left(
      g_{ac} \nabla^{d}_{}
      - 2 \delta_{[a}^{\;\;d} \nabla_{c]}^{}
      \right) \nabla_{b}h^{cb}
      - \frac{1}{2} \left(
      \nabla_{a}\nabla^{d} - \delta_{a}^{\;\;d} \nabla_{c}\nabla^{c}
      \right) h_{b}^{\;\;b}
      \nonumber\\
  &&
      -  \frac{1}{2} R_{ac} h^{dc}
      + \frac{1}{2} R^{df} h_{af}
      + \frac{1}{2} \delta_{a}^{\;\;d} h^{ec} R_{ec}
      .
      \label{eq:Einstein-tensor-ad-mix-1st-order-perturvative-metric}
\end{eqnarray}
Since we consider the vacuum background spacetime $R_{ab}=0$, we
obtain
\begin{eqnarray}
  {}^{(1)}\!G_{a}^{\;\;d}
  &=&
      -  \frac{1}{2} \nabla_{b}\nabla^{b}h_{a}^{\;\;d}
      + R_{abc}^{\;\;\;\;\;\;d} h^{bc}
      \nonumber\\
  &&
      + \frac{1}{2} \left(
      g_{ac} \nabla^{d}_{}
      - 2 \delta_{[a}^{\;\;d} \nabla_{c]}^{}
      \right) \nabla_{b}h^{cb}
      - \frac{1}{2} \left(
      \nabla_{a}\nabla^{d} - \delta_{a}^{\;\;d} \nabla_{c}\nabla^{c}
      \right) h_{b}^{\;\;b}
      .
      \label{eq:Einstein-tensor-ad-mix-1st-order-perturvative-metric-vac-BG}
\end{eqnarray}


To derive the component of ${}^{(1)}\!G_{a}^{\;\;d}$, we denote the
components of the metric perturbation and the derivative operator as
\begin{eqnarray}
  &&
     \bar{h}_{AB} := h_{AB}, \quad
     \bar{h}_{A}^{\;\;D} := y^{DE} h_{AE}, \quad
     \bar{h}^{pD} := \gamma^{pq} y^{DE} h_{qE}, \quad
     \nonumber\\
  &&
     \bar{h}_{p}^{\;\;q} := \gamma^{qr} h_{pr}, \quad
     \bar{h}^{pq} := \gamma^{pr}\gamma^{qs} h_{rs}, \quad
     \label{eq:barh-defs}
\end{eqnarray}
and
\begin{eqnarray}
  \bar{D}^{C} = y^{CE}\bar{D}_{E}, \quad
  \hat{D}^{p} = \gamma^{pq} \hat{D}_{q}.
  \label{eq:barDhatD-defs}
\end{eqnarray}


Now, the components of ${}^{(1)}\!G_{a}^{\;\;d}$ in terms of the
variable defined by Eq.~(\ref{eq:barh-defs}) as follows:
\begin{eqnarray}
  {}^{(1)}\!G_{A}^{\;\;B}
  &=&
      -  \frac{1}{2} \bar{D}_{C}\bar{D}^{C}\bar{h}_{A}^{\;\;B}
     -  \frac{1}{2r^{2}} \hat{D}_{p}\hat{D}^{p}\bar{h}_{A}^{\;\;B}
     -  \frac{2}{r^{2}} (\bar{D}^{C}r) (\bar{D}_{C}r) \bar{h}_{A}^{\;\;B}
     + \frac{2}{r^{2}} \bar{h}_{A}^{\;\;B}
     \nonumber\\
  &&
      + \frac{1}{2} \bar{D}^{B}\bar{D}^{C}\bar{h}_{AC}
      + \frac{1}{2} \bar{D}_{A}\bar{D}_{C}\bar{h}^{BC}
     -  \frac{1}{r} (\bar{D}^{C}r) \bar{D}_{C}\bar{h}_{A}^{\;\;B}
      -  \frac{1}{2} \bar{D}_{A}\bar{D}^{B}\bar{h}_{C}^{\;\;C}
      \nonumber\\
  &&
     + \frac{1}{r} (\bar{D}^{C}r) \bar{D}^{B}\bar{h}_{AC}
     +  \frac{1}{r} (\bar{D}_{C}r) \bar{D}_{A}\bar{h}^{BC}
     \nonumber\\
  &&
     + \frac{1}{2r^{2}} \bar{D}^{B}\hat{D}^{p}\bar{h}_{Ap}
     + \frac{1}{2r^{2}} \bar{D}_{A}\hat{D}_{p}\bar{h}^{Bp}
      \nonumber\\
  &&
     -  \frac{1}{2r^{2}} \bar{D}_{A}\bar{D}^{B}\bar{h}_{r}^{\;\;r}
     + \frac{1}{2r^{3}} (\bar{D}_{A}r) \bar{D}^{B}\bar{h}_{r}^{\;\;r}
     + \frac{1}{2r^{3}} (\bar{D}^{B}r) \bar{D}_{A}\bar{h}_{r}^{\;\;r}
     -  \frac{1}{r^{4}} (\bar{D}_{A}r) (\bar{D}^{B}r) \bar{h}_{r}^{\;\;r}
     \nonumber\\
  &&
     + y_{A}^{\;\;B} \left(
      -  \frac{1}{2} \bar{D}_{C}\bar{D}_{D}\bar{h}^{CD}
     + \frac{1}{2} \bar{D}_{C}\bar{D}^{C}\bar{h}_{D}^{\;\;D}
     + \frac{1}{2r^{2}} \hat{D}_{p}\hat{D}^{p}\bar{h}_{C}^{\;\;C}
     -  \frac{2}{r} (\bar{D}_{D}r) \bar{D}_{C}\bar{h}^{CD}
     \right.
     \nonumber\\
  && \quad\quad\quad
     \left.
     + \frac{1}{r} (\bar{D}^{C}r) \bar{D}_{C}\bar{h}_{D}^{\;\;D}
     -  \frac{1}{r^{2}} (\bar{D}_{C}r) (\bar{D}_{D}r) \bar{h}^{CD}
     + \frac{3}{2r^{2}} (\bar{D}^{D}r) (\bar{D}_{D}r) \bar{h}_{C}^{\;\;C}
     \right.
     \nonumber\\
  && \quad\quad\quad
     \left.
     -  \frac{1}{r^{2}} \bar{D}_{C}\hat{D}_{p}\bar{h}^{pC}
     -  \frac{1}{r^{3}} (\bar{D}_{C}r) \hat{D}_{p}\bar{h}^{Cp}
     \right.
     \nonumber\\
  && \quad\quad\quad
     \left.
     + \frac{1}{2r^{2}} \bar{D}_{C}\bar{D}^{C}\bar{h}_{r}^{\;\;r}
     + \frac{1}{2r^{4}} \hat{D}_{p}\hat{D}^{p}\bar{h}_{r}^{\;\;r}
     -  \frac{1}{2r^{3}} (\bar{D}^{C}r) \bar{D}_{C}\bar{h}_{r}^{\;\;r}
     -  \frac{1}{2r^{4}} \hat{D}_{p}\hat{D}_{s}\bar{h}^{ps}
     \right.
     \nonumber\\
  && \quad\quad\quad
     \left.
     + \frac{1}{2r^{4}} (\bar{D}_{C}r) (\bar{D}^{C}r) \bar{h}_{r}^{\;\;r}
     -  \frac{3}{2r^{2}} \bar{h}_{C}^{\;\;C}
     \right)
     .
     \label{eq:linearized-Einstein-tensor-AB-result-sum}
\end{eqnarray}
\begin{eqnarray}
  {}^{(1)}\!G_{A}^{\;\;q}
  &=&
      \frac{1}{2r^{2}} \hat{D}^{q}\bar{D}^{C}\bar{h}_{AC}
      -  \frac{1}{2r^{2}} \bar{D}_{A}\hat{D}^{q}\bar{h}_{C}^{\;\;C}
      + \frac{1}{2r^{3}} (\bar{D}_{A}r) \hat{D}^{q}\bar{h}_{C}^{\;\;C}
      \nonumber\\
  &&
      -  \frac{1}{2r^{2}} \bar{D}_{C}\bar{D}^{C}\bar{h}_{A}^{\;\;q}
      -  \frac{1}{2r^{4}} \hat{D}_{r}\hat{D}^{r}\bar{h}_{A}^{\;\;q}
      + \frac{1}{2r^{4}} \bar{h}_{A}^{\;\;q}
      + \frac{1}{2r^{4}} \hat{D}^{q}\hat{D}^{p}\bar{h}_{Ap}
      \nonumber\\
  &&
      + \frac{1}{2r^{2}} \bar{D}_{A}\bar{D}_{D}\bar{h}^{qD}
      -  \frac{1}{r^{3}} (\bar{D}_{A}r) \bar{D}_{D}\bar{h}^{qD}
      -  \frac{1}{r^{4}} (\bar{D}_{A}r) (\bar{D}_{D}r) \bar{h}^{qD}
      + \frac{1}{r^{3}} (\bar{D}_{D}r) \bar{D}_{A}\bar{h}^{qD}
      \nonumber\\
  &&
      + \frac{1}{2r^{4}} \bar{D}_{A}\hat{D}_{s}\bar{h}^{qs}
      -  \frac{1}{r^{5}} (\bar{D}_{A}r) \hat{D}_{s}\bar{h}^{qs}
      -  \frac{1}{2r^{4}} \bar{D}_{A}\hat{D}^{q}\bar{h}_{r}^{\;\;r}
      + \frac{1}{r^{5}} (\bar{D}_{A}r) \hat{D}^{q}\bar{h}_{r}^{\;\;r}
      ,
      \label{eq:linearized-Einstein-tensor-Aq-result-sum}
\end{eqnarray}
\begin{eqnarray}
  {}^{(1)}\!G_{p}^{\;\;q}
  &=&
      -  \frac{1}{2r^{2}} \bar{D}_{C}\bar{D}^{C}\bar{h}_{p}^{\;\;q}
      -  \frac{1}{2r^{4}} \hat{D}_{s}\hat{D}^{s}\bar{h}_{p}^{\;\;q}
      + \frac{1}{r^{3}} (\bar{D}_{C}r) \bar{D}^{C}\bar{h}_{p}^{\;\;q}
      -  \frac{2}{r^{4}} (\bar{D}_{C}r) (\bar{D}^{C}r) \bar{h}_{p}^{\;\;q}
      + \frac{2}{r^{4}} \bar{h}_{p}^{\;\;q}
      \nonumber\\
  &&
     -  \frac{1}{2r^{2}} \hat{D}_{p}\hat{D}^{q}\bar{h}_{C}^{\;\;C}
      + \frac{1}{2r^{2}} \hat{D}_{p}\bar{D}_{C}\bar{h}^{qC}
      + \frac{1}{2r^{2}} \hat{D}^{q}\bar{D}^{C}\bar{h}_{pC}
      \nonumber\\
  &&
     + \frac{1}{2r^{4}} \hat{D}_{p}\hat{D}_{r}\bar{h}^{qr}
     + \frac{1}{2r^{4}} \hat{D}^{q}\hat{D}^{r}\bar{h}_{pr}
     -  \frac{1}{2r^{4}} \hat{D}_{p}\hat{D}^{q}\bar{h}_{r}^{\;\;r}
      \nonumber\\
  &&
      + \gamma_{p}^{\;\;q} \left(
      -  \frac{1}{2} \bar{D}_{C}\bar{D}_{D}\bar{h}^{CD}
      -  \frac{1}{r} (\bar{D}_{C}r) \bar{D}_{D}\bar{h}^{CD}
      + \frac{1}{2} \bar{D}_{C}\bar{D}^{C}\bar{h}_{D}^{\;\;D}
      + \frac{1}{2r} (\bar{D}^{C}r) \bar{D}_{C}\bar{h}_{D}^{\;\;D}
      \right.
      \nonumber\\
  && \quad\quad\quad
     \left.
      + \frac{1}{2r^{2}} \hat{D}_{s}\hat{D}^{s}\bar{h}_{C}^{\;\;C}
      -  \frac{1}{r^{2}} \bar{D}_{C}\hat{D}_{r}\bar{h}^{Cr}
      \right.
      \nonumber\\
  && \quad\quad\quad
     \left.
      -  \frac{1}{2r^{4}} \hat{D}_{r}\hat{D}_{s}\bar{h}^{sr}
      -  \frac{3}{2r^{4}} \bar{h}_{r}^{\;\;r}
      + \frac{1}{2r^{2}} \bar{D}_{C}\bar{D}^{C}\bar{h}_{r}^{\;\;r}
      -  \frac{1}{r^{3}} (\bar{D}^{C}r) \bar{D}_{C}\bar{h}_{r}^{\;\;r}
      \right.
      \nonumber\\
  && \quad\quad\quad
     \left.
      + \frac{2}{r^{4}} (\bar{D}_{C}r) (\bar{D}^{C}r) \bar{h}_{r}^{\;\;r}
     + \frac{1}{2r^{4}} \hat{D}_{s}\hat{D}^{s}\bar{h}_{r}^{\;\;r}
      \right)
      .
      \label{eq:linearized-Einstein-tensor-pq-result-sum}
\end{eqnarray}




\begin{thebibliography}{99}
\bibitem{LIGO-home-page}
  LIGO Scientific Collaboration 2024 home page: https://www.ligo.org/
\bibitem{Virgo-home-page}
  Virgo 2024 home page : https://www.virgo-gw.eu/
\bibitem{KAGRA-home-page}
  KAGRA 2024 home page: https://gwcenter.icrr.u-tokyo.ac.jp/en/
\bibitem{LIGO-INDIA-home-page}
  LIGO INDIA 2024 home page : https://www.ligo-india.in/
\bibitem{ET-home-page}
  Einstein Telescope 2024 home page : https://www.et-gw.eu/
\bibitem{CosmicExplorer-home-page}
  Cosmic Explorer 2024 home page : https://cosmicexplorer.org/
\bibitem{LISA-home-page}
  LISA 2024 home page : https://lisa.nasa.gov/
\bibitem{DECIGO-PTEP-2021}
  S.~Kawamura, et al., Prog. Theor. Exp. Phys. 2021 (2021), 05A105.
\bibitem{TianQin-PTEP-2021}
  J.~Mei, et al., Prog. Theor. Exp. Phys. 2020 (2020), 05A107.
\bibitem{Taiji-PTEP-2021}
  Z.~Luo, et al., Prog. Theor. Exp. Phys. 2020 (2020),05A108.
\bibitem{L.Barack-A.Pound-2019}
  L.~Barack and A.~Pound, Rep. Prog. Phys. {\bf 82} (2019) 016904.
\bibitem{K.Nakamura-2021-CQG}
  K.~Nakamura, ``Proposal of a gauge-invariant treatment of
  $l=0,1$-mode perturbations on Schwarzschild background spacetime'',
  Class. Quantum Grav. {\bf 38} (2021), 145010.
\bibitem{K.Nakamura-2021-LHEP}
  K.~Nakamura, ``Formal Solutions of Any-Order Mass, Angular-Momentum,
  anda Dipole Perturbations on the Schwarzschild Background
  Spacetime'', Letters in High Energy Physics {\bf 2021} (2021), 215.
\bibitem{K.Nakamura-2021-PartI}
  K.~Nakamura, ``Gauge-invariant perturbation theory on the
  Schwarzschild background spacetime Part I: --- Formulation and
  odd-mode perturbations ---'', arXiv:2110.13408v8 [gr-qc].
\bibitem{K.Nakamura-2021-PartII}
  K.~Nakamura, ``Gauge-invariant perturbation theory on the
  Schwarzschild background spacetime Part II: --- Even-mode
  perturbations ---'', arXiv:2110.13512v5 [gr-qc].
\bibitem{K.Nakamura-2021-PartIII}
  K.~Nakamura, ``Gauge-invariant perturbation theory on the
  Schwarzschild background spacetime Part III: --- Realization of
  exact solutions ---'', arXiv:2110.13519v5 [gr-qc].
\bibitem{T.Regge-J.A.Wheeler-1957}
  T.~Regge and J.~A.~Wheeler, Phys. Rev. {\bf 108} (1957), 1063.
\bibitem{F.Zerilli-1970-PRL}
  F.~Zerilli, Phys. Rev. Lett. {\bf 24} (1970), 737.
\bibitem{F.Zerilli-1970-PRD}
  F.~Zerilli, Phys. Rev. D {\bf 2} (1970), 2141.
\bibitem{H.Nakano-2019}
  H.~Nakano, Private note on ``Regge-Wheeler-Zerilli formalism'' (2019).
\bibitem{V.Moncrief-1974a}
  V.~Moncrief, Ann. Phys. (N.Y.) {\bf 88} (1974), 323.
\bibitem{V.Moncrief-1974b}
  V.~Moncrief, Ann. Phys. (N.Y.) {\bf 88} (1974), 343.
\bibitem{C.T.Cunningham-R.H.Price-V.Moncrief-1978}
  C.~T.~Cunningham, R.~H.~Price, and V.~Moncrief, Astrophys. J. {\bf
    224} (1978), 643.
\bibitem{Chandrasekhar-1983}
  S.~Chandrasekhar, {\it The mathematical theory of black holes} (Oxford:
  Clarendon Press, 1983).
\bibitem{Gerlach_Sengupta-1979a}
  U.H.~Gerlach and U.K.~Sengupta, Phys.~Rev.\ D\ {\bf 19} (1979), 2268.
\bibitem{Gerlach_Sengupta-1979b}
  U.H.~Gerlach and U.K.~Sengupta, Phys.~Rev.\ D\ {\bf 20} (1979), 3009.
\bibitem{Gerlach_Sengupta-1979c}
  U.H.~Gerlach and U.K.~Sengupta, J. Math. Phys.\ {\bf 20} (1979), 2540.
\bibitem{Gerlach_Sengupta-1980}
  U.H.~Gerlach and U.K.~Sengupta, Phys.~Rev.\ D\ {\bf 22} (1980), 1300.
\bibitem{T.Nakamura-K.Oohara-Y.Kojima-1987}
  T.~Nakamura, K.~Oohara, Y.~Kojima,
  Prog. Theor. Phys. Suppl. No. {\bf 90} (1987), 1.
\bibitem{Gundlach-Martine-Garcia-2000}
  C.~Gundlach and J.M.~Mart\'in-Garc\'ia, Phys.~Rev.\ D{\bf 61}
  (2000), 084024.
\bibitem{Gundlach-Martine-Garcia-2001}
  J.M.~Mart\'in-Garc\'ia and C.~Gundlach, Phys.~Rev.\ D{\bf 64}
  (2001), 024012.
\bibitem{A.Nagar-L.Rezzolla-2005-2006}
  A.~Nagar and L.~Rezzolla, Class. Quantum Grav. {\bf 22} (2005),
  R167; Erratum {\it ibid}. {\bf 23} (2006), 4297.
\bibitem{K.Martel-E.Poisson-2005}
  K.~Martel and E.~Poisson, Phys. Rev. D {\bf 71} (2005), 104003.
\bibitem{K.Nakamura-2003}
  K.~Nakamura, Prog.~Theor.~Phys. {\bf 110}, (2003), 723.
\bibitem{K.Nakamura-2005}
  K.~Nakamura, Prog. Theor. Phys. {\bf 113} (2005), 481.
\bibitem{K.Nakamura-2011}
  K.~Nakamura, Class.~Quantum~Grav.\ {\bf 28} (2011), 122001.
\bibitem{K.Nakamura-IJMPD-2012}
  K.~Nakamura, Int. J. Mod. Phys. D {\bf 21} (2012), 124004.
\bibitem{K.Nakamura-2013}
  K.~Nakamura, Prog.~Theor.~Exp.~Phys. {\bf 2013} (2013), 043E02.
\bibitem{K.Nakamura-2014}
  K.~Nakamura, Class.~Quantum Grav. {\bf 31} (2014), 135013.
\bibitem{K.Nakamura-2010}
  K.~Nakamura, Advances in Astronomy, {\bf 2010} (2010), 576273.
\bibitem{K.Nakamura-2020}
  K.~Nakamura, ``Second-order Gauge-invariant Cosmological
  Perturbation Theory: Current Status Updated in 2019'', Chapter 1 in
  {\it Theory and Applications of Physical Science, Vol.3}
  Ed. M.~Rafatullah, (Book Publisher International, 2020), ISBN
  978-93-89816-24-2 (Print); ISBN 978-93-89816-25-9 (eBook),
  Preprint arXiv:1912.12805 [gr-qc].
\bibitem{A.J.Christopherson-K.A.Malik-D.R.Matravers-K.Nakamura-2011}
  A.~J.~Christopherson, K.~A.~Malik, D.~R.~-Matravers,
  K.~Nakamura, Class.~Quantum~Grav.\ {\bf 28} (2011), 225024.
\bibitem{K.Nakamura-2006}
  K.~Nakamura, Phys. Rev. D {\bf 74} (2006), 101301(R).
\bibitem{K.Nakamura-2007}
  K.~Nakamura, Prog. Theor. Phys. {\bf 117} (2007), 17.
\bibitem{K.Nakamura-2008}
  K.~Nakamura, ````Gauge'' in General Relativity: -- Second-order
  general relativistic gauge-invariant perturbation theory --'',
  in {\it Lie Theory and its Applications in Physics VII}
  ed. V.~K.~Dobrev et al, (Heron Press, Sofia, 2008)
\bibitem{K.Nakamura-2009a}
  K.~Nakamura, Phys. Rev. D {\bf 80} (2009), 124021.
\bibitem{K.Nakamura-2009b}
  K.~Nakamura, Prog. Theor. Phys. {\bf 121} (2009), 1321.
\bibitem{L.Landau-E.Lifshitz-1962}
  L.~Landau and E.~Lifshitz, {\it The Classical Theory of Fields}
  (Addison-Wesley, Reading, Mass., 1962).
\bibitem{W.Kinnersley-M.Walker-1970}
  W.~Kinnersley and M.~Walker, Phys. Rev. D {\bf 2} (1970), 1359.
\bibitem{J.B.Griffiths-P.Krtous-J.Podolsky-2006}
  J.~B.~Griffiths, P.~Krtous, and J.~Podolsky, Class. and Quantum
  Grav. {\bf 23} (2006), 6745.
\bibitem{R.K.Sachs-1964}
  R.~K.~Sachs, ``Gravitational Radiation'', in {\it Relativity,
    Groups and Topology} ed. C.~DeWitt and B.~DeWitt, (New York:
  Gordon and Breach, 1964).
\bibitem{J.M.Stewart-M.Walker-1974}
  J.~M.~Stewart and M.~Walker, Proc.~R.~Soc.~London\ A {\bf 341}
  (1974), 49.
\bibitem{J.M.Stewart-M.Walker-1990}
  J.~M.~Stewart, Class.~Quantum~Grav.\ {\bf 7} (1990), 1169.
\bibitem{J.M.Stewart-1990}
  J.~M.~Stewart, {\it Advanced General Relativity} (Cambridge
  University Press, Cambridge, 1991).
\bibitem{M.Bruni-S.Matarrese-S.Mollerach-S.Sonego-1997}
  M.~Bruni, S.~Matarrese, S.~Mollerach and S.~Sonego,
  Class. Quantum Grav.\ {\bf 14} (1997), 2585.
\bibitem{M.Bruni-S.Sonego-CQG1999}
  M.~Bruni and S.~Sonego, Class.~Quantum~Grav.\ {\bf 16} (1999), L29.
\bibitem{S.Sonego-M.Bruni-1998}
  S.~Sonego and M.~Bruni, Commun.~Math.~Phys.\ {\bf 193} (1998), 209.
\bibitem{Noh-Hwang-2004}
  H.~Noh and J.~Hwang, Phys. Rev. D {\bf 69} (2004), 104011.
\end{thebibliography}
\end{document}